\documentclass[epjc3,5p,times,twocolumn,a4paper,showpacs,showkeys,amsmath,amssymb,superscriptaddress,nofootinbib]{svjour3} 
\usepackage{graphicx}
\usepackage{fix-cm}
\usepackage{import}
\usepackage{txfonts}
\usepackage{epsfig}
\usepackage{listings}
\usepackage{mdwlist}
\usepackage{subfigure}
\usepackage{pennames}
\usepackage[english]{babel}

\graphicspath{{./}{../00_introduction/}{../01_beamline/}{../02_COBRA/}{../03_DC/}{../04_TC/}{../05_XEC/}{../06_trigger/}{../07_DAQ/}{../08_slow_control/}{../09_calibrations/}{../10_simulation/}{../11_operation/}{../12_summary/}}
\DeclareGraphicsExtensions{.eps,.pdf,.jpeg,.png}

\newcommand{\meg}{\ifmmode\mu^+ \to \mathrm{e}^+ \gamma\else$\mu^+ \to \mathrm{e}^+ \gamma$\fi}
\newcommand{\michel}{\ifmmode\mu^+ \to \mathrm{e}^+ \nu\nu\else$\mu^+ \to \mathrm{e}^+ \nu\nu$\fi}
\newcommand{\radiative}{\ifmmode\mu^+ \to \mathrm{e}^+\gamma \nu\nu\else$\mu^+ \to \mathrm{e}^+\gamma \nu\nu$\fi}

\journalname{Eur. Phys. J. C} 

\begin{document}


\title{The MEG detector for \meg\ decay search }

\newcommand*{\INFNPi}{INFN Sezione di Pisa$^{a}$; Dipartimento di Fisica$^{b}$ dell'Universit\`a, Largo B.~Pontecorvo~3, 56127 Pisa, Italy}
\newcommand*{\INFNGe}{INFN Sezione di Genova$^{a}$; Dipartimento di Fisica$^{b}$ dell'Universit\`a, Via Dodecaneso 33, 16146 Genova, Italy}
\newcommand*{\INFNPv}{INFN Sezione di Pavia$^{a}$; Dipartimento di Fisica$^{b}$ dell'Universit\`a, Via Bassi 6, 27100 Pavia, Italy}
\newcommand*{\INFNRm}{INFN Sezione di Roma$^{a}$; Dipartimento di Fisica$^{b}$ dell'Universit\`a ``Sapienza'', Piazzale A.~Moro, 00185 Roma, Italy}
\newcommand*{\INFNLe}{INFN Sezione di Lecce$^{a}$; Dipartimento di Matematica e Fisica$^{b}$ dell'Universit\`a, Via per Arnesano, 73100 Lecce, Italy}
\newcommand*{\ICEPP} {ICEPP, University of Tokyo 7-3-1 Hongo, Bunkyo-ku, Tokyo 113-0033, Japan }
\newcommand*{\UCI}   {University of California, Irvine, CA 92697, USA}
\newcommand*{\KEK}   {KEK, High Energy Accelerator Research Organization 1-1 Oho, Tsukuba, Ibaraki, 305-0801, Japan}
\newcommand*{\PSI}   {Paul Scherrer Institut PSI, CH-5232, Villigen, Switzerland}
\newcommand*{\Waseda}{Research Institute for Science and Engineering, Waseda~University, 3-4-1 Okubo, Shinjuku-ku, Tokyo 169-8555, Japan}
\newcommand*{\BINP}  {Budker Institute of Nuclear Physics of Siberian Branch of Russian Academy of Sciences, 630090, Novosibirsk, Russia}
\newcommand*{\JINR}  {Joint Institute for Nuclear Research, 141980, Dubna, Russia}
\newcommand*{\ETHZ}  {Swiss Federal Institute of Technology ETH, CH-8093 Z\" urich, Switzerland}
\newcommand*{\NOVST}  {Novosibirsk State Technical University, 630092, Novosibirsk, Russia}

\date{Received: date / Accepted: date}

\author{J.~Adam~\thanksref{addr1,addr2} \and
        X.~Bai~\thanksref{addr3}  \and
        A.~M.~Baldini~\thanksref{addr4}$^a$ \and
        E.~Baracchini~\thanksref{addr3,addr9,addr11} \and
        C.~Bemporad~\thanksref{addr4}$^{ab}$ \and
        G.~Boca~\thanksref{addr7}$^{ab}$ \and
        P.W.~Cattaneo~\thanksref{e1,addr7}$^{a}$  \and
        G.~Cavoto~\thanksref{addr8}$^{a}$ \and
        F.~Cei~\thanksref{addr4}$^{ab}$ \and
        C.~Cerri~\thanksref{addr4}$^{a}$ \and
        M.~Corbo~\thanksref{addr4}$^{ab}$ \and
        N.~Curalli~\thanksref{addr4}$^{ab}$ \and
        A.~De Bari~\thanksref{addr7}$^{ab}$ \and
        M.~De Gerone~\thanksref{addr1,addr8}$^{ab,}$\thanksref{addr5}$^{ab}$ \and
        L.~Del Frate~\thanksref{addr4}$^{a}$ \and
        S.~Doke~\thanksref{addr10} \and
        S.~Dussoni~\thanksref{addr5}$^{ab,}$\thanksref{addr4}$^{ab}$\and
        J.~Egger~\thanksref{addr1} \and
        K.~Fratini~\thanksref{addr5}$^{ab}$ \and
        Y.~Fujii~\thanksref{addr3}  \and
        L.~Galli~\thanksref{addr1,addr4}$^{ab}$ \and
        S.~Galeotti~\thanksref{addr4}$^{a}$ \and
        G.~Gallucci~\thanksref{addr4}$^{ab}$ \and
        F.~Gatti~\thanksref{addr5}$^{ab}$ \and
        B.~Golden~\thanksref{addr11} \and
        M.~Grassi~\thanksref{addr4}$^{a}$ \and
        A.~Graziosi~\thanksref{addr8}$^{a}$ \and
        D.N.~Grigoriev~\thanksref{addr12,addr14} \and
        T.~Haruyama~\thanksref{addr9} \and
        M.~Hildebrandt~\thanksref{addr1} \and
        Y.~Hisamatsu~\thanksref{addr3}  \and
        F.~Ignatov~\thanksref{addr12} \and
        T.~Iwamoto~\thanksref{addr3}  \and
        D.~Kaneko~\thanksref{addr3}  \and
        K.~Kasami~\thanksref{addr9}  \and
        P.-R.~Kettle~\thanksref{addr1} \and
        B.I.~Khazin~\thanksref{addr12} \and
        O.~Kiselev~\thanksref{addr1} \and 
        A.~Korenchenko~\thanksref{addr13}  \and
        N.~Kravchuk~\thanksref{addr13}  \and
        G.~Lim~\thanksref{addr11} \and
        A.~Maki~\thanksref{addr9}  \and
        S.~Mihara~\thanksref{addr9}  \and
        W.~Molzon~\thanksref{addr11} \and
        T.~Mori~\thanksref{addr3}  \and
        F.~Morsani~\thanksref{addr4}$^{a}$ \and
        D.~Mzavia$^\dagger$~\thanksref{addr13}  \and
        R.~Nard\`o~\thanksref{addr7}$^{ab}$ \and
        H.~Natori~\thanksref{addr3,addr9}  \and
        D.~Nicol\`o~\thanksref{addr4}$^{ab}$ \and
        H.~Nishiguchi~\thanksref{addr9}  \and
        Y.~Nishimura~\thanksref{addr3}  \and 
        W.~Ootani~\thanksref{addr3}  \and
        K.~Ozone~\thanksref{addr3}  \and
        M.~Panareo~\thanksref{addr6}$^{ab}$ \and
        A.~Papa~\thanksref{addr1,addr4}$^{ab}$ \and
        R.~Pazzi$^\dagger$~\thanksref{addr4}$^{ab}$ \and
        G.~Piredda~\thanksref{addr8}$^{a}$ \and
        A.~Popov~\thanksref{addr12} \and
        F.~Raffaelli~\thanksref{addr4}$^{a}$ \and
        F.~Renga~\thanksref{addr1,addr8}$^{ab}$ \and
        E.~Ripiccini~\thanksref{addr8}$^{ab}$ \and
        S.~Ritt~\thanksref{addr1} \and
        M.~Rossella~\thanksref{addr7}$^{a}$ \and
        R.~Sawada~\thanksref{addr3}  \and
        M.~Schneebeli~\thanksref{addr1,addr2} \and
        F.~Sergiampietri~\thanksref{addr4}$^{a}$ \and
        G.~Signorelli~\thanksref{addr4}$^{a}$ \and
        S.~Suzuki~\thanksref{addr10}  \and
        F.~Tenchini~\thanksref{addr4}$^{ab}$ \and
        C.~Topchyan~\thanksref{addr11} \and
        Y.~Uchiyama~\thanksref{addr1, addr3} \and
        R.~Valle~\thanksref{addr5}$^{ab}$ \and 
        C.~Voena~\thanksref{addr8}$^{a}$ \and
        F.~Xiao~\thanksref{addr11} \and
        S.~Yamada~\thanksref{addr9} \and
        S.~Yamamoto~\thanksref{addr9} \and
        S.~Yamashita~\thanksref{addr3}  \and
        Yu.V.~Yudin~\thanksref{addr12} \and
        D.~Zanello~\thanksref{addr8}$^{a}$ 
}

\institute{ \PSI \label{addr1} 
           \and
              \ETHZ \label{addr2}
           \and
              \ICEPP \label{addr3}
           \and
             \INFNPi \label{addr4}
           \and
             \UCI    \label{addr11}
           \and
             \INFNPv \label{addr7}
           \and
             \INFNRm \label{addr8}
           \and
             \INFNGe \label{addr5}
           \and
             \Waseda \label{addr10}
           \and
             \BINP   \label{addr12}
           \and
             \KEK    \label{addr9}
           \and
             \JINR   \label{addr13}
           \and
             \INFNLe \label{addr6}
           \and
             \NOVST  \label{addr14}
}


\thankstext[*]{e1}{Corresponding author: paolo.cattaneo@pv.infn.it } 
\thankstext[$\dagger$]{}{Deceased } 
\maketitle 
 
\begin{abstract}
The MEG (Mu to Electron Gamma) experiment has been running at the Paul Scherrer Institut (PSI), 
Switzerland since 2008 to search for the decay \meg\ by using 
one of the most intense continuous $\mu^+$ beams in the world.
This paper presents the MEG components: the positron 
spectrometer, including a thin target, a superconducting magnet, a set of drift chambers for measuring
the muon decay vertex and the positron momentum, a 
timing counter for measuring the positron time, and a liquid xenon detector for measuring 
the photon energy, position and time. 
The trigger system, the read-out electronics and the data acquisition system are also
presented in detail. The paper is completed with a description of the equipment 
and techniques developed for the calibration in time and energy and the simulation of the whole apparatus.

\end{abstract}


\tableofcontents 


%

\section{Introduction}
\label{sec:introduction}
\begin{figure*}[t]
\centering
  \includegraphics[width=1.\textwidth,angle=0] {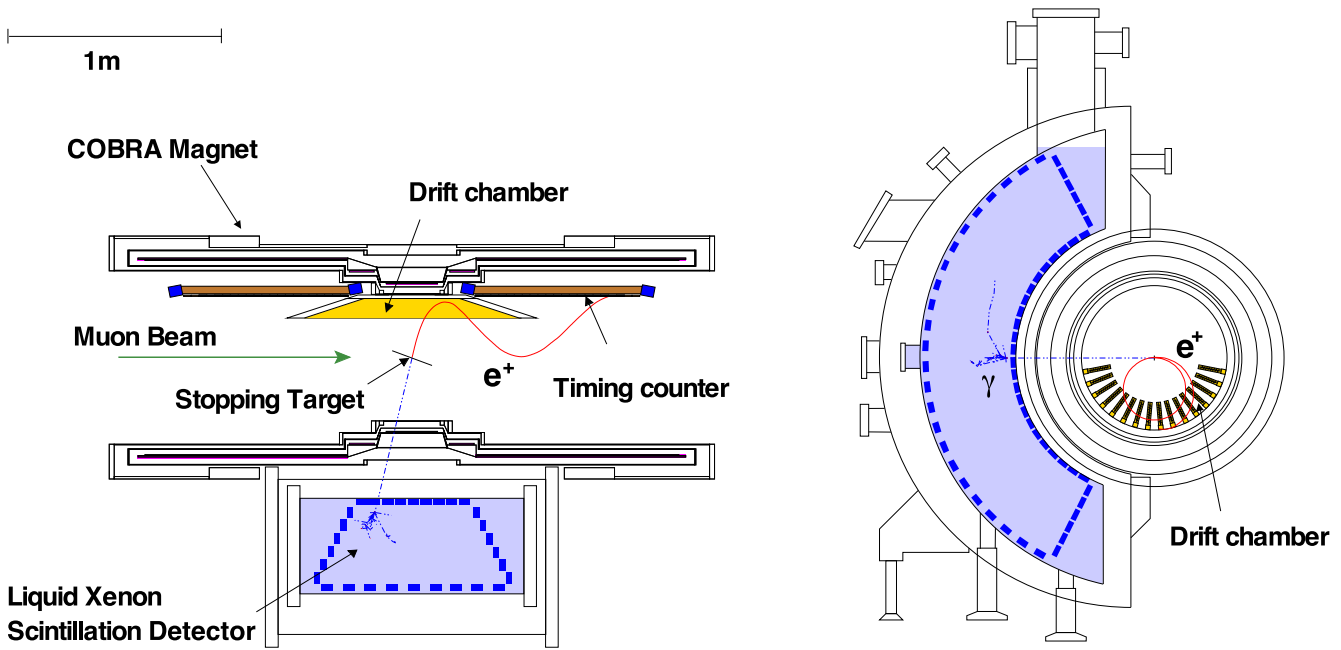}
 \caption{Schematic view of the MEG detector showing one simulated signal event emitted 
from the target. 
}
 \label{introduction:megdet}
\end{figure*}

A search for the Charged Lepton Flavour Violating (CLFV) decay \meg, 
the MEG experiment (see \cite{hisamatsu_2007} and references therein for a 
detailed report of the experiment motivation, design criteria and goals)
is in progress at the Paul Scherrer Institut (PSI) in Switzerland. 
Preliminary results have already been published \cite{meg2009,meg2010}.
The goal is to push the sensitivity to this decay down to $\sim 5\times$10$^{-13}$ 
improving the previous limit set by the MEGA experiment,
$1.2\times 10^{-11}$ \cite{brooks_1999_prd}, by a factor 20.

CLFV processes are practically forbidden in the Standard Model (SM), which,
even in presence of neutrino masses and mixing, predicts tiny branching
ratios (BR~$\ll 10^{-50}$) for CLFV decays.
Detecting such decays would be a clear indication of new physics beyond the SM,
as predicted by many extensions such as supersymmetry \cite{barbieri}.
Hence, CLFV searches with improved sensitivity either 
reveal new physics or constrain the allowed parameter space of SM extensions.

In MEG positive muons stop and decay in a thin target located at the centre of 
a magnetic spectrometer. The signal has the simple kinematics of a two-body decay from
a particle at rest: 
one monochromatic positron and one monochromatic photon moving in opposite directions
each with an energy of $52.83$~MeV (half of the muon mass) and being coincident in time.

This signature needs to be extracted from a background induced by Michel (\michel)
and radiative (\radiative) muon decays.
The background is dominated by accidental coincidence events where a positron 
and a photon from different muon decays with energies close to the kinematic 
limit overlap within the direction and time 
resolution of the detector.
Because the rate of accidental coincidence events is proportional to the square 
of the $\mu^+$ decay rate, while the rate of signal events is proportional only to 
the $\mu^+$ decay rate, direct-current beams allow a better signal 
to background ratio to be achieved than for pulsed beams. Hence we use the PSI continuous surface muon beam
with intensity $\sim 3\times 10^7\,\mu^+$/s (see Sect.~\ref{sec:beamline}).

A schematic of the MEG apparatus is shown in Fig.~\ref{introduction:megdet}.
A magnet, COBRA (COnstant Bending RAdius), generates a gradient 
magnetic field, for the first time among particle physics experiments, with the 
field strength gradually decreasing at increasing distance along the magnet axis 
from the centre.

This configuration is optimised to sweep low-momentum positrons from Michel decays rapidly out 
of the magnet, and to keep the bending radius of the positron trajectories 
only weakly dependent on their emission angle within the acceptance region
(see Sect.~\ref{sec:cobra}).

The positron track parameters are measured by a set of very low mass Drift CHambers
(DCH) designed to minimise the multiple scattering (see Sect.~\ref{sec:dch}).
The positron time is measured by a Timing Counter (TC) consisting of scintillator 
bars read out by PhotoMultiplier Tubes (PMT) (see Sect.~\ref{timingcounter}).

For $\gamma$-ray detection, we have developed an innovative detector using
Liquid Xenon (LXe) as a scintillation material viewed by PMTs submersed
in the liquid. This detector provides accurate measurements of the $\gamma$-ray
energy and of the time and position of the interaction point (see Sect.~\ref{sec:LXe}).

We monitor the performance of the experiment continuously by introducing a variety
of calibration methods (see Sect.~\ref{sec:calsec}). 

We employ a flexible and sophisticated electronics system for triggering the
\meg\ candidate events along with calibration data (see Sect.~\ref{trigger}).

Special considerations influenced the development of the data acquisition system.
We record signals from all detectors with a custom-designed waveform digitiser,
the Domino Ring Sampler (DRS) (see Sects.~\ref{sec:daq} and \ref{slowcontrol}).

We evaluate detector response functions using data with help
of detector simulations (see Sect.~\ref{sec:simul}).

In this paper, we mostly refer to a cylindrical coordinate system ${\it (r,\phi,z)}$ with origin 
in the centre of COBRA.
The z-axis is parallel to the COBRA axis and directed along the incoming muon beam.
The axis defining $\phi=0$ (the x-axis of the corresponding Cartesian coordinate system) 
is directed opposite to the centre of the LXe detector
(as a consequence the y-axis is directed upwards).
Positrons move along trajectories of decreasing $\phi$ coordinate.
When required, the polar angle $\theta$ with respect to the z-axis is also used.
The region with $z<0$ is referred as upstream, that with $z>0$ as downstream.

The angular acceptance of the experiment is defined by the size of the 
LXe fiducial volume, corresponding approximately to $\phi\in (\frac{2}{3}\pi,\frac{4}{3}\pi)$
and $|\cos\theta|<0.35$, for a total acceptance of $\sim 11\%$.
The spectrometer is designed to fully accept positrons from 
\meg\ decays when the $\gamma$-ray falls in the acceptance region.

%
%
\section{Beam Line}
\label{sec:beamline}
\begin{figure*}[t]
\centering
\includegraphics[width=1.0\linewidth]{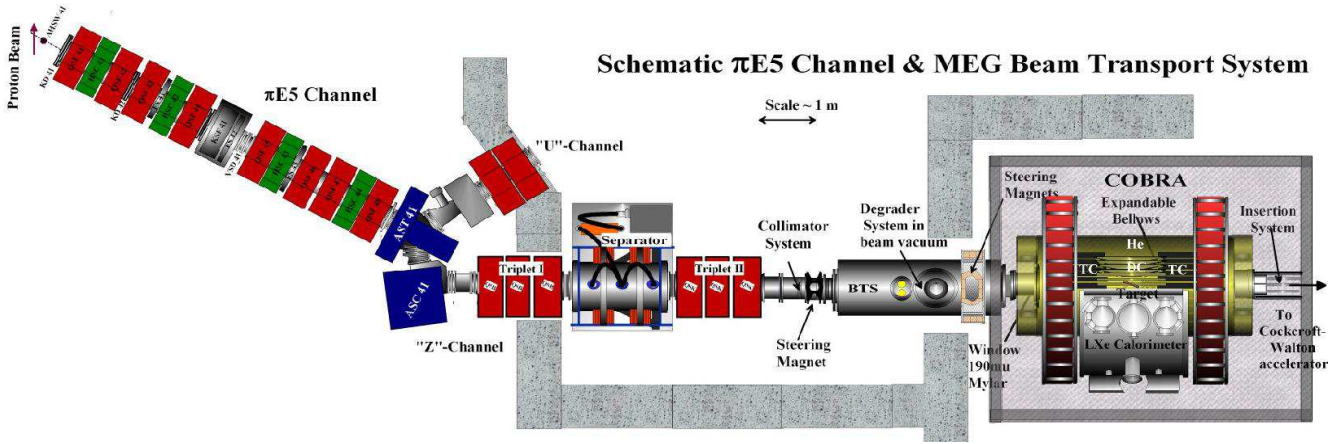}
\caption{Schematic of the $\pi$E5 Channel and the MEG  beam transport system showing the beam 
elements and detectors: Wien-filter, collimator system and the superconducting Beam Transport Solenoid (BTS) 
and central degrader system. Coupled to these are the He-filled COBRA spectrometer, consisting 
of a gradient-field magnet and central target system. The downstream insertion system allowing 
the remote insertion of the Cockcroft--Walton beam pipe, for calibration purposes, is also shown.}
\label{fig:PiE5}      
\end{figure*}
\subsection{Experimental Requirements}
\label{sec:expe}
In order to meet the stringent background requirements necessary for a high-sensitivity, 
high-rate coincidence experiment such as MEG, one of the world's highest intensity continuous 
muon beams, the $\pi$E5 channel at PSI, is used in 
conjunction with the high-performance MEG beam transport system which couples the $\pi$E5 
channel to the COBRA spectrometer (see Sect.~\ref{sec:cobra}). This allows more than 
$10^{8}\,\mu^+$/s, 
with an optimal suppression of beam-correlated background, to be transported to the ultra-thin 
stopping-target at the centre of COBRA. The main requirements of such a low-energy surface 
muon beam are: high-intensity with high-transmission optics at 28~MeV/c, close to the kinematic 
edge of stopped muon decay and to the maximum intensity of the beam momentum spectrum; 
small beam emittance to minimise the stopping-target size; momentum-bite 
\footnote{Selected range of momenta transmitted by the channel.} of less than $10\%$ with 
an achromatic final focus to provide an almost monochromatic beam with a high 
stopping density \footnote{A measure of the muon stopping rate per unit target 
volume, influenced by both the beam momentum and bite as well as the target 
material properties.} and enabling muons to be stopped 
in a minimally thick target; minimisation and separation of beam-related background, such 
as beam $e^{+}$ originating from $\pi^{0}$-decay in the production target or decay particles 
produced along the beam line; minimisation of material interactions along the beam line, 
requiring both vacuum and helium environments to reduce the multiple scattering and 
the photon-production probability for annihilation-in-flight \footnote{The reaction 
$\mathrm{e}^+\mathrm{e}^-\to \gamma\gamma$ with $\mathrm{e}^+$ from $\mu^+$ Michel decay 
and $\mathrm{e}^-$ in the material.} or Brems\-strahlung. 

Furthermore, 
the experiment requires an arsenal of calibration and monitoring tools, including, on 
the beam side, the possibility of sources of monochromatic photons and positrons for 
calibration purposes. This in turn requires a 70.5~MeV/c stopped negative-pion beam tune 
to produce photons from the charge-exchange ($\pi^{-} p\to \pi^{0} n$) 
and radiative capture reactions ($\pi^{-} p\to \gamma n$) on a liquid hydrogen target
(see Sect.~\ref{sec:cex}), as well as a 53~MeV/c positron beam tune to produce monochromatic 
Mott-scattered positrons 
(see Sect.~\ref{sec:mott}).

Finally, the MEG beam transport system also allows the introduction of external targets and 
detectors, as well as a proton beam line, from a dedicated Cockcroft--Walton {(C--W)} accelerator 
(see Sect.~\ref{sec:accelerator}), 
to the centre of the spectrometer helium volume. This is achieved without affecting the gas environment, 
by means of specially designed end-caps. The protons from C--W accelerator interact 
with a dedicated target to produce $\gamma$-ray lines induced by nuclear reactions.

\subsection{$\pi$E5 Beam line and MEG Beam Transport System} 
\begin{figure}[h]
\centering
\includegraphics[width=1.0\linewidth, height=80mm,clip=true]{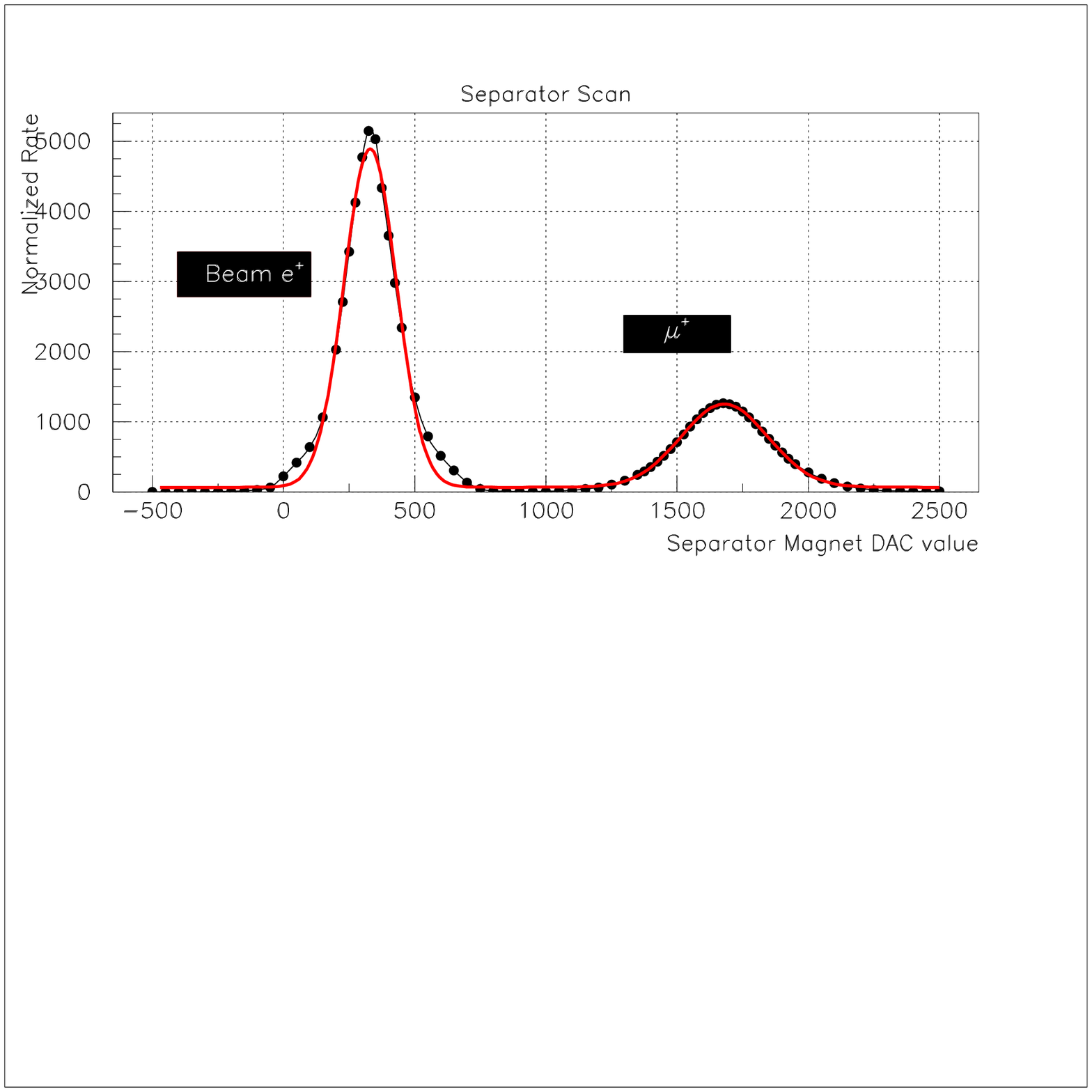}
\caption{Separator scan plot, measured post collimator system. The black dots represent 
on-axis, low discriminator threshold intensity measurements at a given separator 
magnetic field value during 
the scan, for a fixed electric field value of -195~kV. The red curve represents a double 
Gaussian fit to the data points, with a constant background. A separation of 8.1 muon beam 
$\sigma_\mu$ is found, corresponding to more than 12~cm separation of beam-spot centres 
at the collimator. The raw muon peak measured at low discriminator threshold also contains 
a contribution from Michel positrons due to decays in the detector. These can easily 
be distinguished by also measuring at high threshold, where only muons are visible on 
account of their higher energy loss and hence higher pulse-height. }
\label{fig:sepmu_pos}
\end{figure}

The $\pi$E5 channel is a 165$^{\circ}$ backwards-oriented, windowless, high-acceptance 
(150~msr), low-momentum ($< 120$~MeV/c), dual-port $\pi^-$, $\mu^-$ or $ e$-channel. 
For the MEG experiment, the channel is tuned to +28~MeV/c with a momentum-bite of 
between $5-7\%$ FWHM, depending on the opening of the momentum selecting slits placed in 
the front-part of the channel. This constitutes an optimal high-intensity surface muon tune 
(stopped pion-decay at the surface of the production target) at backward pion production 
angles \cite{Frosch,Kettle}. 
A schematic of the $\pi$E5 channel and the connecting MEG beam line is shown in Fig.~\ref{fig:PiE5}.
The main production target is a rotating radiation-cooled graphite wheel of 4~cm thickness 
in the beam direction.
It is connected to the $\pi$E5 area by a quadrupole and sextupole channel that
couples to the initial extraction and focussing elements of the MEG beam line, quadrupole 
triplet I, which exits the shielding wall and provides optimal high transmission through 
the Wien-filter ($E\times B$ field separator). 
With an effective length of about 83~cm for both the electric and magnetic fields, together 
with a 200~kV potential across the 19~cm gap of the electrodes, a mass separation equivalent 
to an angular separation of +88~mrad and $-25$~mrad between muons and  positrons and between muons and 
pions is achieved, respectively. This together with Triplet II and a collimator system 
(equivalent to 11~X$_{0}$ of Pb), placed 
after the second triplet, separates the muons from the eight-times higher beam positron 
contamination coming from the production target. A separation quality between muons and 
beam positrons of 8.1~$\sigma_{\mu}$ is so achieved, corresponding to a 12$\,$cm physical 
separation at the collimator system as shown in Fig.~\ref{fig:sepmu_pos}. 
This allows an almost pure muon beam \footnote{Positron suppression factor $\sim 850$ 
under the muon peak.} to propagate to the Beam Transport Solenoid (BTS) (see Sect.~\ref{bts}), 
which couples the injection to the superconducting COBRA spectrometer in an iron-free 
way. By means of a 300~$\mu$m thick Mylar~\textsuperscript{\textregistered} degrader system placed at the intermediate focus, 
the BTS also minimises the multiple scattering contribution to the beam and adjusts the muon range for a 
maximum stop-density at the centre of the muon target, placed inside a helium atmosphere, at the 
centre of the COBRA spectrometer. 

\subsection{The Beam Transport Solenoid (BTS)} 
\label{bts}
\begin{figure}[h!]
\centering
\includegraphics[width=0.8\linewidth]{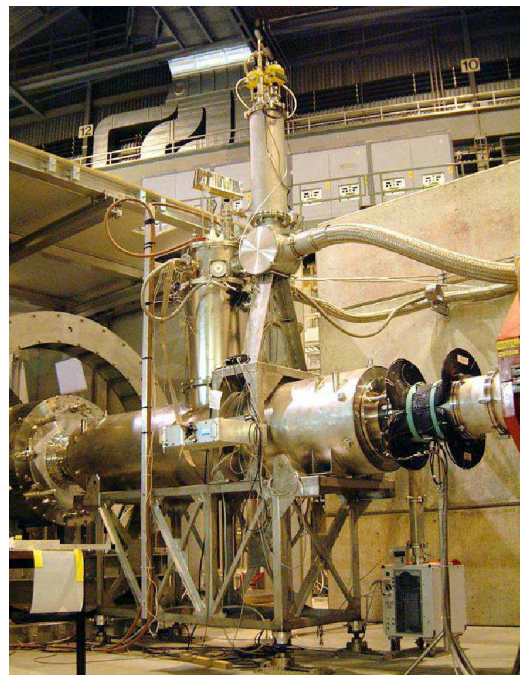}
\caption{The Beam Transport Solenoid (BTS), shown coupled to the beam line between Triplet II 
and the COBRA spectrometer. The main cryogenic tower with the current leads and the Joule--Thompson 
valve tower with the LHe transfer line from the cryoplant coupled to it as well as the smaller 
LN$_{2}$ transfer line in the background can be seen. The two correction dipole magnets on the downstream side are not shown here.}
\label{fig:bts}
\end{figure}

The BTS,
a 2.8~m long iron-free superconducting solenoid, with a 38~cm 
warm-bore aperture, coupling directly to the beam line vacuum, is shown in Fig.~\ref{fig:bts}. 
With a maximal current of 300~A a field of 0.54~T can be reached, higher than the 
0.36~T required for an intermediate focus at the central degrader system for muons. The double-layered 2.63~m long coil is 
made up of a $45\%$ Ni/Ti,
$55\%$ Cu matrix conductor with an inductance of 1.0~H so yielding a maximum stored 
energy of 45~kJ. 

On the cryogenic side, a dedicated liquid helium (LHe) port to the central cryoplant is used 
to transport the supercritical helium at $\sim 10\,$bar to two Joule--Thompson valves (in the upper 
tower in Fig.~\ref{fig:bts}), which are used to liquefy and automatically regulate the level of 
the 150~$\ell$ of LHe in the cryostat. A separate liquid nitrogen line cools an outer heat-shield 
used to minimise the LHe losses.

\begin{figure*}[t]
\centering
\includegraphics[width=0.8\linewidth, height=70mm]{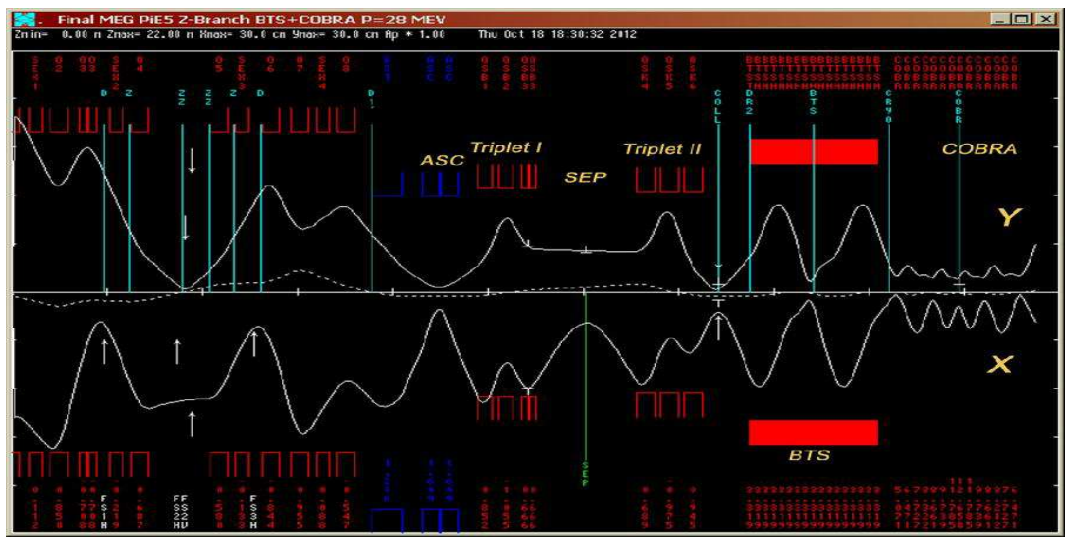}
\caption{2$^{nd}$-order TRANSPORT calculation showing the vertical (top-half) and 
horizontal (bottom-half) beam envelopes plotted along the beam-axis from close to target E up 
to the centre of COBRA, a distance of 22~m. The vertical and horizontal scales differ 33:1. The 
coloured elements show the vacuum chamber apertures of the various beam elements, while the 
dotted line represents the central dispersion trajectory for particles with a $1\%$ higher momentum.}
\label{fig:transport}
\end{figure*}

Furthermore, three iron-free correction dipole magnets (one $\cos\theta$-type placed upstream 
of the BTS) one horizontal and one vertical conventional-dipole type, both mounted on the 
downstream side of the BTS, are employed to ensure an axial injection into both solenoids. 
The three dipole magnets compensate for 
a radial asymmetry in the fringe fields induced by the interaction of the stray field from the 
large-aperture, iron-free COBRA magnet with an iron component of the hall-floor foundations.  
An example of the optics in Fig.~\ref{fig:transport} shows the $2\sigma$ beam envelopes both vertically and horizontally along the 
22~m long beam line, together with the apertures of the various elements and the dispersion 
trajectory of particles having a $1\%$ higher momentum (dotted line).
 
\subsection{Spectrometer Coupling System}
The COBRA spectrometer is sealed at both ends by end-caps which constitute the final part 
of the beam line, separating the vacuum section from the helium environment inside. 
These specially designed 1160~mm diameter light-weight (35--40~kg), only 4~mm-thick aluminium 
constructs minimise the material available for possible photon background production, crucial to 
a background-sensitive experiment. They also integrate the beam vacuum window on the upstream side 
and an automated thin-windowed (20~$\mu$m) bellows insertion system, capable of an extension-length 
of about 1.7~m. The bellows insertion system is used to introduce various types of targets such 
as a liquid hydrogen target used in conjunction with the pion beam, or a 
lithium tetraborate photon production target used together with the C--W accelerator, on 
the downstream side.
 
\subsection{Commissioning}
The commissioning of the MEG beam line was completed with the first ``engineering run" 
in late 2007. During commissioning, several beam optical solutions were studied, 
including the investigation of both $\pi$E5 branches (``U"-branch and ``Z"-branch) as well as different 
beam-correlated background separation methods: a combined degrader energy-loss/magnetic-field 
separation method or a crossed $(E \times B)$-field separator (Wien filter). 
The optimum solution favoured the ``Z"-channel with a Wien filter, both from the experimental 
layout and the rate and background-suppression point of view.

In order to maximise the beam transmission rate and the stopping density in the target, an 
understanding of the momentum spectrum and channel momentum acceptance is necessary. This was carefully 
studied by measuring the muon momentum spectrum in the vicinity of the kinematic edge of 
stopped pion decay (29.79~MeV/c). As with all phase-space measurements performed at high rate, 
a scanning technique was used: either one using a small cylindrical ``pill" scintillator of 
2~mm diameter 
and 2~mm thickness coupled to a miniature PMT \cite{hamamatsu-beam}, or 
one using a thick depletion layer Avalanche PhotoDiode (APD), without a scintillator, in the 
case of measurements in a magnetic field. Both these solutions gave an optimum 
pulse-height separation between positrons and muons. This allows integrated beam-spot intensity 
measurements up to 2~GHz to be made.

\begin{figure}[h!]
\centering
\includegraphics[width=1.0\linewidth]{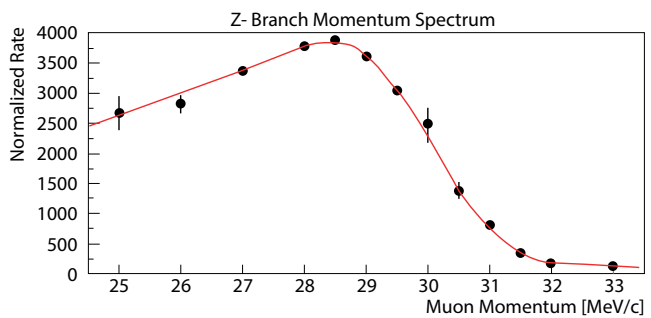}
\caption{Measured MEG muon momentum spectrum with fully open slits. Each point is obtained by
optimising the whole beam line for the corresponding central momentum and measuring the full beam-spot intensity.
The red line is a fit to the data, based on a theoretical P$^{3.5}$ behaviour, folded with a Gaussian resolution
function corresponding to the momentum-bite plus a constant cloud-muon background.}
\label{fig:spect}
\end{figure}

For the momentum spectrum, taken with the full $7\%$ FWHM momentum-bite 
and shown in Fig.~\ref{fig:spect}, the whole
beam line was tuned to successive momentum values and the integral muon rate was then determined by 
scanning the beam-spot using the beam line as a spectrometer. This allows to determine both the 
central momentum and the momentum-bite 
by fitting a theoretical distribution, folded with a Gaussian resolution 
function corresponding to the momentum-bite, plus a constant cloud-muon background over this 
limited momentum range. The red curve corresponding  to the fitted function shows the expected 
P$^{3.5}$ empirical behaviour below the kinematic edge. The cloud-muon content below the 
surface-muon peak, which has its origin in pion decay-in-flight in the vicinity of the production 
target, was determined to be $3.3\%$ of the surface-muon rate. This was determined from negative 
muon measurements at 28~MeV/c and using the known pion production cross-section differences 
\cite{Crawford}, since negative surface-muons do not exist due to the formation of pionic atoms 
and hence the small cloud-muon contribution is much easier to measure in the absence of a 
dominating surface-muon signal.

Another important feature investigated was the muon polarisation on decaying: since the muons 
are born in a $100\%$ polarised state, with their spin vector opposing their momentum vector, 
the resultant angular distribution on decay is an important issue in understanding the angular 
distribution of the Michel positrons and the influence of detector acceptance. The depolarisation 
along the beam line, influenced by divergence, materials, magnetic fields and beam 
contaminants was studied and incorporated into the MEG Monte Carlo, which was used to 
compare results with the measured angular distribution asymmetry (upstream/downstream events) 
of Michel positrons. The predicted muon polarisation of $91\%$ was confirmed by these 
measurements, which yielded a value of $89\pm4$~$\%$ \cite{polarization,polarization-note}. 
The main single depolarising contribution is due to the mean polarization of the 
$3.3\%$ cloud-muon contamination, which opposes that of the surface muons. However, because 
the depolarisation in the polyethylene target is totally quenched by the strength of the 
longitudinal magnetic field in COBRA, a high polarization can be maintained \cite{polarization}.

\begin{figure}[h!]
\centering
\includegraphics[width=1.0\linewidth,clip=true]{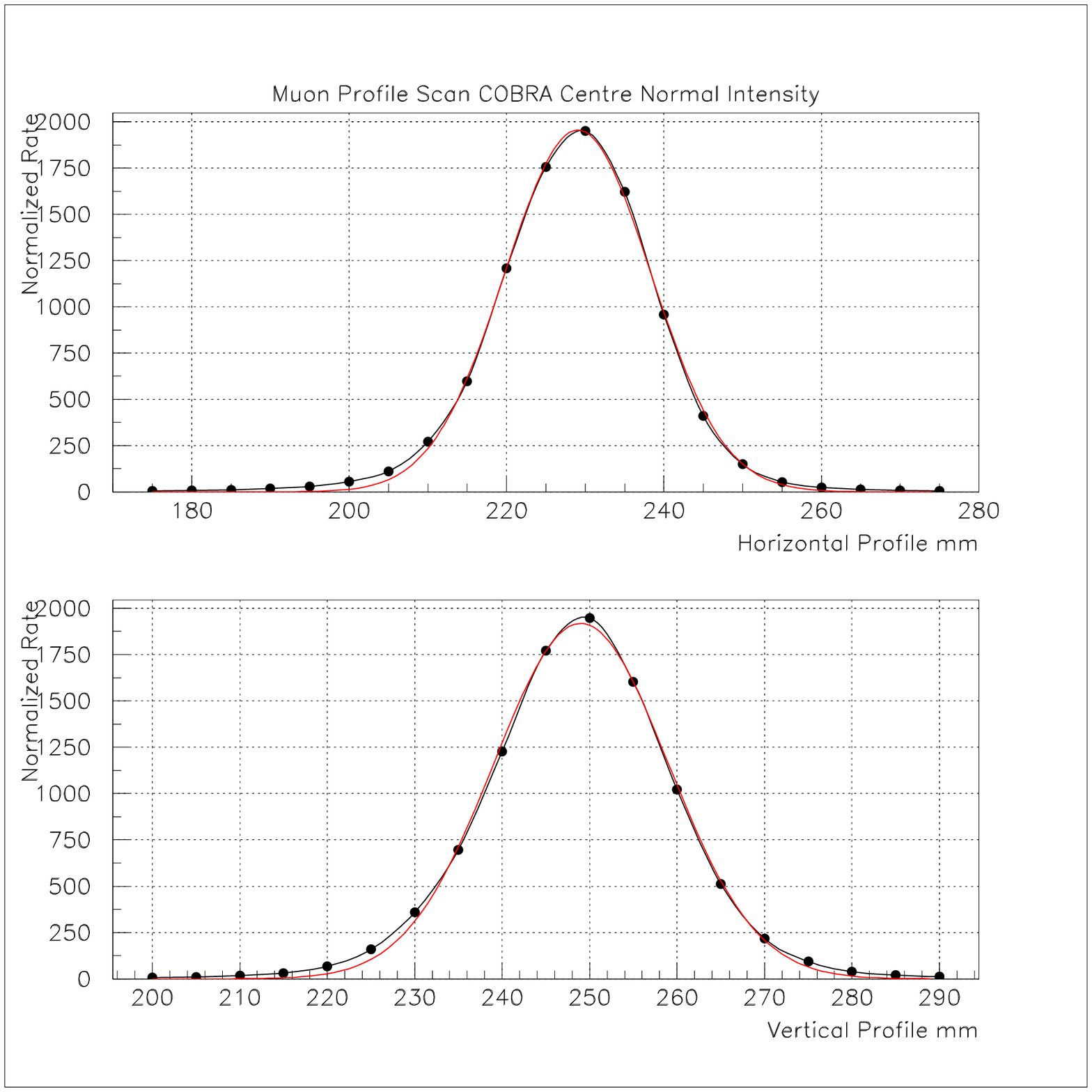}
\caption{Scanned horizontal and vertical muon beam-spot profiles measured at the centre of COBRA. 
The red lines show the Gaussian fits to the 
distributions that yield $\sigma_{x,y} \sim 10$ mm and a total stopping intensity of $\sim 3\times 10^7\,\mu^+/$s
at a proton beam intensity of 2.2$\,$mA and momentum-bite of $5.0\%$ FWHM.}
\label{fig:spot}
\end{figure}

In order to achieve the smallest beam-spot size at the target, the position of the final 
degrader as well as the polarity combination of COBRA and the BTS were studied. The smallest 
beam divergences and hence the smallest beam-spot size at the target were achieved by having 
a reverse polarity magnetic field of the BTS with respect to COBRA. This is due to the fact 
that the radial field is enhanced at the intermediate focus between the two magnets, thus 
effectively increasing the radial focussing power and minimising the growth of the beam 
divergences through the COBRA volume. The final round beam-spot at the target, with 
$\sigma_{x,y}\sim  10$~mm is shown in Fig.~\ref{fig:spot}.

\subsection{Beam line Performance}

Although the beam line was tuned for maximal rate with minimal positron beam contamination, 
the final 
beam stopping intensity was determined by a full analysis of the detector performance and 
an evaluation of its sensitivity to the $\meg$ decay. This yielded an optimal 
muon stopping rate in the target of $3\times 10^7\,\mu^+/$s.

\begin{table} [!htb]
\caption{\label{tab:beamline}\emph{MEG Final Beam Line Characteristics}}
\begin{center}
\begin{tabular}{ll}
\hline
\hline
Parameter & Value  \\
\hline
\\
Solid Angle $\Delta\Omega$ & 150$\,$msr \\
Beam Line length L & 22$\,$m \\
Muon Central Momentum P$_{\mu}$ & 28~MeV/c \\
Pion Central Momentum P$_{\pi}$ & -70.5~MeV/c \\
Positron Central Momentum P$_{e}$ & 53~MeV/c \\
Muon Momentum-bite $\Delta$P/P& $5.0\%$ (full-width) \\
Beam-spot size $\sigma_{X,Y}$& 9--10~mm  \\
beam $e^{+}/\mu^{+}$ ratio (Sep. off)  & 8\\
$\mu^{+}/e^{+}$ separation (Sep. on) & 8.1$\sigma_{\mu}$, 12~cm \\
Optimal $\mu^{+}$ stopping intensity & $3\times 10^7\,\mu^+/$s\\
\\
\hline
\hline
\end{tabular}
\end{center}
\end{table}

Furthermore, for both the pion beam and Mott positron beam calibration tunes, mentioned in 
Sect.~\ref{sec:expe}, which involve prompt 
particle production at the proton target, the central momentum was chosen to give a maximal 
time-of-flight separation between the wanted particles and the unwanted background particles at the 
centre of COBRA, based on the use of the accelerator modulo 19.75~ns 
radio-frequency signal as a clock, which has a fixed time relationship to the production 
at the proton target. In the case of the Mott positron beam the final momentum should also 
be close to the signal energy of 52.83~MeV/c. A summary of the final beam parameters is 
given in Table~\ref{tab:beamline}.

%
\subsection{Target}
\label{target}
\subsubsection{Requirements}
Positive muons are stopped in a thin target at the centre of COBRA. 
The target is designed to satisfy the following criteria:
\begin{itemize}
\item[1] Stop $>80$~\%\ of the muons in a limited axial region at the magnet centre.
\item[2] Minimise conversions of $\gamma$-rays from radiative muon decay in the target. 
\item[3] Minimise positron annihilation-in-flight with $\gamma$-rays in the acceptance region.
\item[4] Minimise multiple scattering and Bremsstrahlung of positrons from muon decay.  
\item[5] Allow the determination of the decay vertex position and the positron direction at the vertex by projecting the reconstructed positron trajectory to the target plane.
\item[6] Be dimensionally stable and be remotely removable to allow frequent insertion of ancillary calibration targets. 
\end{itemize}
Criterion 1 requires a path-length along the incident muon axial trajectory of $\sim 0.054$ g/cm$^2$ 
for low-Z materials and for the average beam momentum of $19.8$ MeV/c (corresponding to a kinetic 
energy of 1.85 MeV) at the centre of COBRA. 
To reduce scattering and conversion of outgoing particles, the target is constructed 
in the form of a thin sheet in a vertical plane, with the direction perpendicular to the target at 
$\theta\sim 70^\circ$, oriented so that the muon beam is incident on the target side 
facing the LXe detector (see Fig.~\ref{introduction:megdet}). The requirement that we be able to deduce the positron angle 
at the vertex sets the required precision on the knowledge of the position of the target in the direction 
perpendicular to its plane.  This derives from the effect of the track curvature on the 
inferred angle; an error in the projected track distance results in an error in the positron angle 
at the target due to the curvature over the distance between the assumed and actual target 
plane position. This angle error is larger when the positrons are incident at a large angle 
with respect to the normal vector; for an incidence angle of $45^\circ$, an error in the target plane 
position of $500\,\mu$m results in an angle error of $\sim 5$ mrad. 
This sets the requirements that the target 
thickness be below $500\,\mu$m (eliminating low-density foam as a target material) and that its 
position with respect to the DCH spectrometer be measured to $\sim 100\,\mu$m.

\subsubsection{Realisation}
To meet the above criteria, the target was implemented as a $205\,\mu$m thick sheet of a low-density 
(0.895 g/cm$^3$) layered-structure of a polyethylene and polyester with an elliptical shape and semi-major 
and semi-minor axes of 10 cm and 4 cm. The target was captured between
two rings of Rohacell~\textsuperscript{\textregistered} foam of density 0.053 g/cm$^3$ and hung by a Delrin~\textsuperscript{\textregistered} and Rohacell  
stem from a remotely actuated positioning device attached to the DCH support frame. To allow 
a software verification of the position of the target plane, as described in Sect.~\ref{softalig}, eight holes of 
radius 0.5 cm were punched in the foil.


\subsubsection{Automated insertion }

The positioning device consists of a pneumatically operated, bi-directional cylinder that 
inserts and retracts the target. Physical constraints on the position of the cylinder required 
the extraction direction to not be parallel to the target plane; hence the position of the plane 
when inserted was sensitive to the distance by which the piston travelled. This distance was 
controlled by both a mechanical stop on the piston travel and a sensor that indicated when the 
target was fully inserted.

\subsubsection{Optical alignment}

The position of the target plane is measured every run period with an optical survey that images crosses inscribed 
on the target plane; this survey is done simultaneously with that of the DCH. 
The relative accuracy of the target and DCH positions was estimated to be $\pm$(0.5, 0.5, 0.8) mm 
in the ($x$, $y$, $z$) directions. 

\subsubsection{Software Alignment}
\label{softalig}

\begin{figure}[htb]
 \centerline{\hbox{
  \includegraphics[width=.45\textwidth,angle=0] {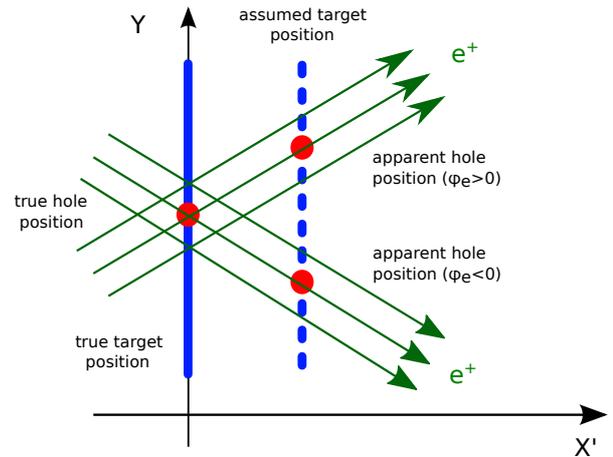}
 }}
 \caption[]{An illustration of the target hole imaging technique. If there is an offset in the $x'$ direction 
between the assumed target position and the true target position, the apparent $y$ position of a target hole will 
depend on the angle of emission at the target $\phi_{e}$. The shift in $y$ is approximately proportional to the 
shift in the $x'$ direction by $\tan(\phi_{e})$.}
 \label{targetholereco.eps}
\end{figure}

\begin{figure}[htb]
 \centerline{\hbox{
  \includegraphics[width=.45\textwidth,angle=0] {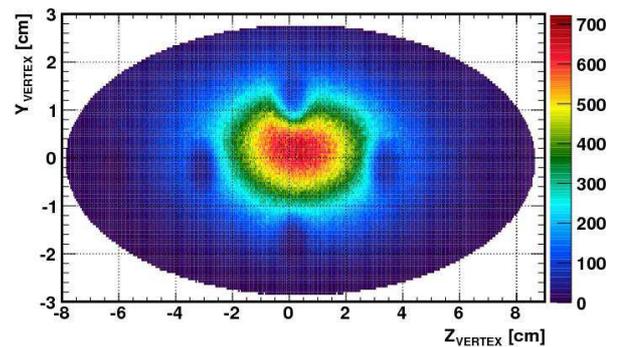}
 }}
 \caption[]{The target illumination plot for the 2011 data sample.}
 \label{target2011.eps}
\end{figure}

\begin{figure}[htb]
 \centerline{\hbox{
  \includegraphics[width=.45\textwidth,angle=0] {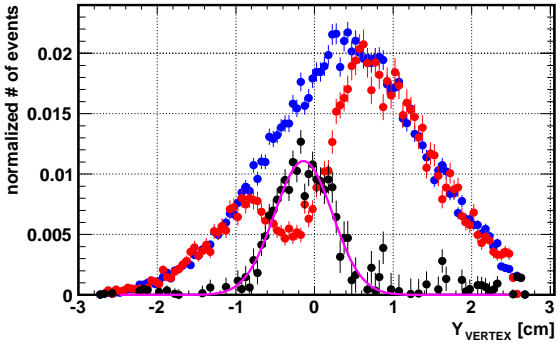}
 }}
 \caption[]{The normalised $y$ distribution of events with $20^\circ <\phi_{e}<30^\circ$ in a slice 
at fixed $z'$ centred on a target hole (in red) and in adjacent slices (in blue). The difference (in black) 
is fitted with a Gaussian.}
 \label{targethole2_3_zslicefit2.eps}
\end{figure}

\begin{figure}[htb]
 \centerline{\hbox{
  \includegraphics[width=.45\textwidth,angle=0] {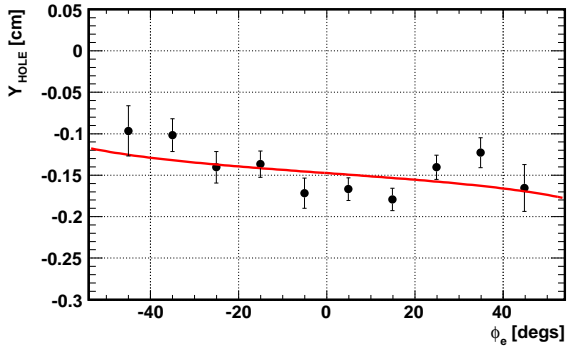}
 }}
 \caption[]{The apparent $y$ position of a target hole versus $\phi_{e}$ fitted with a $\tan(\phi_{e})$ function, 
from which we infer that the $x'$ offset with respect to the optical survey is 0.2$\pm$0.1 mm.}
 \label{targethole2_3_phi.eps}
\end{figure}

The optical survey was checked by imaging the holes in the 
target. We define $z'$ as the position parallel to the target plane and perpendicular to the $y$ 
coordinate and $x'$ as the position perpendicular to the target plane. The $y$ and $z'$ positions are 
easily checked by projecting particle trajectories to the nominal target plane and determining 
the $z'$ and $y$ hole coordinates by fitting the illuminations to holes superimposed on a beam profile 
of approximately Gaussian shape. We note that these target coordinates are not important to the 
analysis, since a translation of the target in the target plane has no effect on the inferred 
vertex position. The $x'$ coordinate was checked by imaging the $y$ positions of the holes as a 
function of the track direction at the target $\phi_{e}$. Figure~\ref{targetholereco.eps} illustrates the technique. 
The $x'$ target position is determined by requiring that the target hole images be insensitive 
to $\phi_{e}$. Figure~\ref{target2011.eps} shows the illumination plot for the 2011 data sample. 
Figure~\ref{targethole2_3_zslicefit2.eps} shows the normalised $y$ distribution of events with $20^\circ<\phi_{e}<30^\circ$
in a slice at fixed $z'$ centred on a target hole (in red) and in adjacent slices (in blue). 
The difference (in black) is fitted with a Gaussian. Figure~\ref{targethole2_3_phi.eps} shows the apparent $y$ position 
of one of the target holes versus $\phi_{e}$, fitted with a $\tan(\phi_{e})$ function, from which we infer that 
the $x'$ offset with respect to the optical survey is 0.2$\pm$0.1 mm. By imaging all the holes 
(i.e. finding $x'$($y$,$z'$)) both the planarity and the orientation of the target plane was determined. 

\begin{figure}[htb]
 \centerline{\hbox{
  \includegraphics[width=.45\textwidth,angle=0] {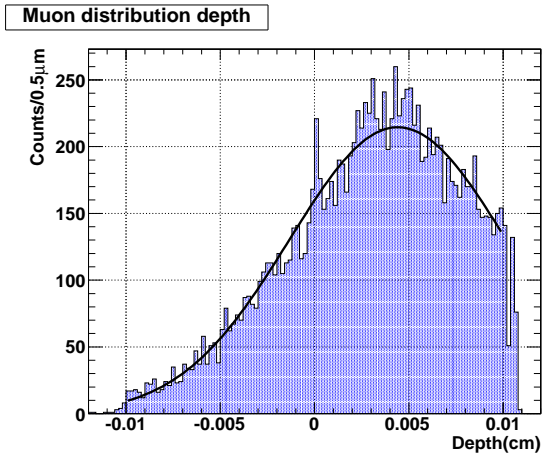}
 }}
 \caption[]{Distribution of the $\mu^+$ decay vertex along $x'$.}
 \label{tardep}
\end{figure}


\subsubsection{Multiple scattering contribution from the target}

The target itself constitutes the most relevant irreducible contribution to the positron
multiple scattering and therefore to the angular resolution of the spectrometer.

The simulated distribution of the $\mu^+$ decay vertex along $x'$ 
is shown in Fig.~\ref{tardep}.
The average is slightly beyond the centre of the target. Averaging over the decay vertex position 
and over the positron direction in the MEG acceptance region, the difference between the positron 
direction at the $\mu^+$ decay vertex and that at the target boundary follows a Gaussian distribution 
with $\sigma=5.2$ mrad.



%
%

%
 
\section{COBRA magnet}
\label{sec:cobra}
\subsection{Concept}
\begin{figure}[htb]
   \centerline{\hbox{
   \includegraphics[width=.45\textwidth,angle=0] {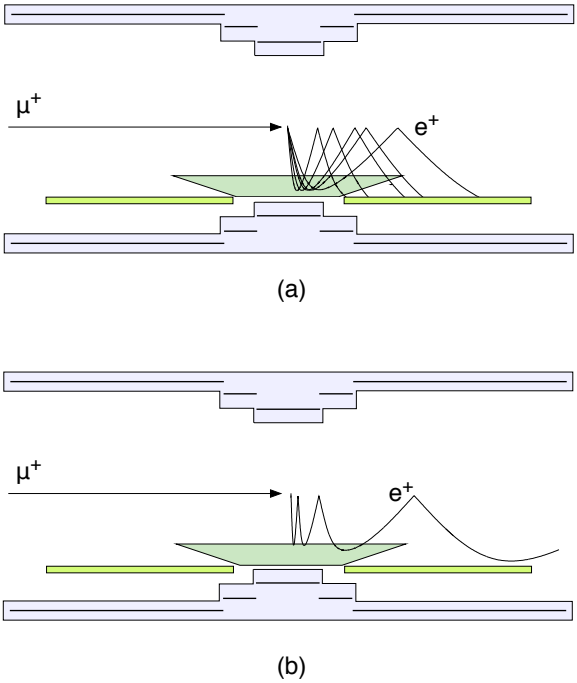}
   }}
   \caption[]{Concept of the gradient magnetic field of COBRA.
  The positrons follow trajectories at a constant bending radius weakly dependent on the emission angle $\theta$ 
  (a) and those transversely emitted from the target ($\cos \theta\sim 0$) are quickly swept away from the DCH (b).}
   \label{fig:COBRA concept}
\end{figure}

\begin{figure}[htb]
   \centerline{\hbox{
   \includegraphics[width=.50\textwidth,angle=0] {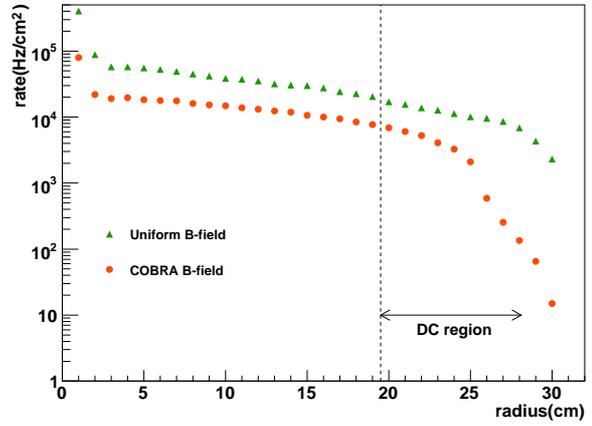}
   }}
   \caption[]{Hit rate of the Michel positrons as a function of the radial distance from the target in both the gradient and uniform field cases.}
   \label{fig:DCH hit rate} 
\end{figure}

The COBRA (COnstant Bending RAdius) magnet is a thin-wall superconducting magnet generating a gradient magnetic field 
which allows stable operation of the positron spectrometer in a high-rate environment. 
The gradient magnetic field is specially designed so that the positron emitted from the target 
follows a trajectory with an almost constant projected bending radius weakly dependent on the 
emission polar angle $\theta$ (Fig.~\ref{fig:COBRA concept}(a)).
Only high-momentum positrons can therefore reach the DCH placed at the outer radius of the inner bore of COBRA.
Another good feature of the gradient field is that the positrons emitted at $\cos \theta\sim 0$ 
are quickly swept away (Fig.~\ref{fig:COBRA concept}(b)).
This can be contrasted to a uniform solenoidal field where positrons emitted transversely would turn 
repeatedly in the spectrometer.
The hit rates of Michel positrons expected for the gradient and uniform fields are compared in 
Fig.~\ref{fig:DCH hit rate} and indicate a significant reduction for the gradient field.

\begin{figure}
       \centerline{\hbox{
       \includegraphics[width=0.45\textwidth]{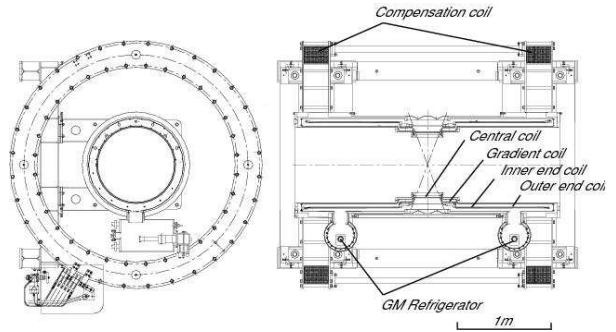}
       }}
       \caption{Cross-sectional view of the COBRA magnet.}
       \label{fig:COBRA cross section}
    \end{figure}

The central part of the superconducting magnet is as thin as $0.197\,\mathrm{X}_0$ so that only a fraction 
of the $\gamma$-rays from the target interacts before reaching
the LXe detector placed outside COBRA.
The COBRA magnet (see Fig.~\ref{fig:COBRA cross section})
is equipped with a pair of compensation coils to reduce the stray field around the LXe detector 
for the operation of the PMTs.

The parameters of COBRA are listed in Table\,\ref{table:COBRA parameters}.

\subsection{Design}

   \begin{table*}
      \caption{Parameters of the COBRA magnet.}
      \label{table:COBRA parameters}
      \begin{center}
         \begin{tabular}{c|ccccc}
            Coil      & Central & Gradient & Inner end & Outer end & Compensation \\\hline\hline
            Conductivity   & Super  & Super   & Super   & Super   & Resistive \\
            Inner dia. (mm)& 699.1 & 809.1 &919.1 &919.1  &2210  \\
            Outer dia. (mm)& 711.6 & 820.6 &  929.5&929.5  &  2590\\
            Length (mm) & 240.3 & 110.4 &189.9  & 749.2 &265  \\
            $z$-coordinate of coil centre(mm) & 0.8 & $\pm235$ &  $\pm405.4$& $\pm874.95$ &  $\pm1190$\\
            Layers     & 4 & 4 & 3 & 3 & 14 \\
              Turns (total) & 1068 & 399 & 240 & 1548 & 280 \\
              Winding density(Turns/m) & 4444.4 & 3614.1 & 1263.8 & 2066.2 & 1056.6 \\
             Inductance(H) & 1.64 & 0.62 & 0.35 & 2.29 & 0.54 \\
             Current (A) & 359.1 & 359.1 & 359.1 & 359.1 & 319.2 \\
             Energy $E$ (kJ)& 106 & 40 & 23 & 148 & 35 \\
             Weight $M$ (kg)& 9 & 4 & 7 & 28 & 1620 \\
             $E/M$ (kJ/kg) & 11.8 & 10.0  & 3.3 & 5.3 & 0.02 \\
         \end{tabular}
      \end{center}
    \end{table*}

\subsubsection{Superconducting magnet}
\begin{figure}[htb]
   \centerline{\hbox{
   \includegraphics[width=.45\textwidth,height=.2\textwidth,angle=0] {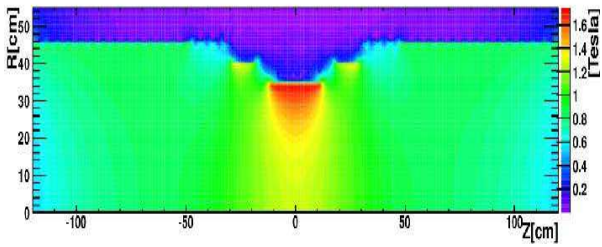}
   }}
   \caption[]{
   Gradient magnetic field generated by the COBRA magnet.}
   \label{fig:COBRA gradient field}
\end{figure}

\begin{figure}[htb]
   \centerline{\hbox{
   \includegraphics[width=.45\textwidth,angle=0] {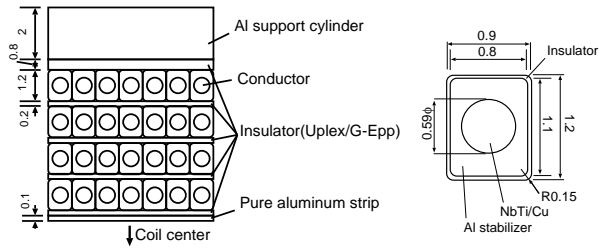}
   }}
   \caption[]{
   Layer structure of the central coil (left) and the cross-sectional view of the conductor (right). Units in mm.}
   \label{fig:Coil layer structure}
\end{figure}

The gradient magnetic field ranging from 1.27~T at the centre to 0.49~T at either end of the magnet cryostat is generated by a step structure of five coils with three different radii (one central coil, two gradient coils and two end coils) (Fig.~\ref{fig:COBRA gradient field}).
The field gradient is optimised by adjusting the radii of the coils and the winding densities of the conductor.
The coil structure is conductively cooled by two mechanical refrigerators attached to the end coils; 
each is a two-stage Gifford--McMahon (GM) refrigerator \cite{COBRA:GM-refrigerator} with a cooling power of 1~W at 4.2~K.
The thin support structure of the coil is carefully designed in terms of the mechanical strength and the thermal conductivity.
The basic idea is to make the coil structure as thin as possible by using a high-strength conductor on a thin aluminium mechanical support cylinder.
Figure~\ref{fig:Coil layer structure} shows the layer structure of the central coil, which has the highest current density.
A high-strength conductor is wound in four layers inside the 2\,mm-thick aluminium support cylinder.
Pure aluminium strips with a thickness of $100\,\mu$m are attached on the inner surface of the coil structure in order to increase the thermal conductivity.
Several quench protection heaters are attached to all coils in order to avoid a local energy dump in case of a quench. 
The high thermal conductivity of the coil structure and the uniform quench induced by the protection heaters are important to protect the magnet. 


\begin{figure}[htb]
   \centerline{\hbox{
   \includegraphics[width=.45\textwidth,angle=0] {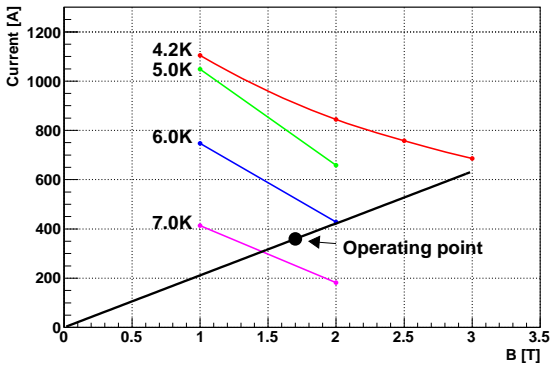}
   }}
   \caption[]{
   Measured critical current of the conductor as a function of the applied magnetic field. The operating condition of COBRA is also shown.}
   \label{fig:conductor performance}
\end{figure}

A high-strength aluminium stabilised conductor is used for the superconducting coils to minimise the thickness of the support cylinder.
The cross-sectional view of the conductor is shown in Fig.~\ref{fig:Coil layer structure}.
 A copper matrix NbTi multi-filamentary core wire is clad with aluminium stabiliser which is 
mechanically reinforced by means of ``micro-alloying'' and ``cold work hardening" \cite{micro-alloying,thin-cable}.
The overall yield strength of the conductor is measured to be higher than 220~MPa at 4.4~K.
The measured critical current of the conductor as a function of applied magnetic field is shown in Fig.~\ref{fig:conductor performance}
where the operating condition of COBRA is also shown (the highest magnetic field of 1.7\,T is reached in the coils with the operating current of 359.1\,A).
This indicates that COBRA operates at 4.2~K with a safety margin of $40\%$ for the conductor.

\subsubsection{Compensation coils}
\begin{figure}[htb]
   \centerline{\hbox{
   \includegraphics[width=.45\textwidth,angle=0] {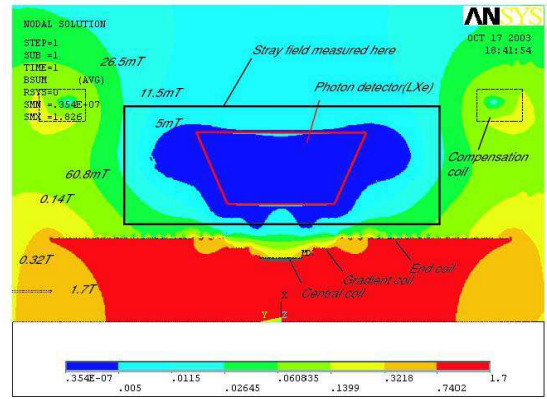}
   }}
   \caption[]{
   Distribution of the residual magnetic field around the LXe detector. The PMTs of the LXe detector are placed along the trapezoidal box shown in this figure.}
   \label{fig:field cancellation}
\end{figure}

The COBRA magnet is equipped with a pair of resistive compensation coils in order to reduce 
the stray field from the superconducting magnet around the LXe detector
(Fig.~\ref{fig:COBRA cross section}).
The stray field should be reduced below $5\times 10^{-3}$~T for the operation of the PMTs of the LXe detector.
The stray field from the superconducting magnet is efficiently cancelled around the LXe detector
since the distribution of the magnetic field generated by the compensation coils is similar to that of the superconducting magnet around that region but the sign of the field is opposite.
Figure~\ref{fig:field cancellation} shows that the expected distribution of the residual magnetic field in the vicinity of the LXe detector is always below $5\times 10^{-3}$~T.

\subsection{Performance}

\subsubsection{Excitation tests}
The superconducting magnet was successfully tested up to 379\,A coil current, which is $5.6\%$ higher than the 
operating current.  The temperature and stress distribution are monitored by using resistive temperature sensors and strain gauges attached to the coils, respectively.
The temperatures of all the coils are found to stay lower than 4.5\,K even at the highest coil current. 

The strains of the coils are measured by the strain gauges up to 379\,A coil current at the central coils and the support cylinder
where the strongest axial electromagnetic force is expected.
We observe a linear relation between the strains and the square of the coil current, which is proportional to the electromagnetic force acting on the coils.
This indicates that the mechanical strength of the coils and the support strength are sufficient up to 379\,A.

The effect of the compensation coils is also measured.
The gain of the PMTs of the LXe detector, measured using LEDs (see Sect.~\ref{sec:led}), is suppressed by a factor of 50 on average in the stray magnetic field from the superconducting magnet,
whereas it recovers up to $93\%$ of the zero field value with the compensation coils.


\subsubsection{Mapping of magnetic field}
The magnetic field of COBRA was measured prior to the experiment start with a commercial three-axis Hall probe 
mounted on a wagon moving three-dimensionally along $z$, $r$ and $\phi$.
The ranges of the measurement are
$|z|<110\,\mathrm{cm}$ with 111 steps,
$0^{\circ}<\phi<360^{\circ}$ with 12 steps and
$0\,\mathrm{cm} < r < 29\,\mathrm{cm}$ with 17 steps, 
which mostly cover the positron tracking volume.
The probe contains three Hall sensors orthogonally aligned and is aligned on the moving wagon such that the three Hall sensors can measure $B_z, B_r$ and $B_\phi$ individually.
The field measuring machine, the moving wagon and the Hall probe are aligned at a precision of a few mrad using a laser tracker.
However, even small misalignments could cause a relatively large effect on the secondary components, $B_r$ and $B_\phi$ due to the large main component $B_z$,
while the measurement of $B_z$ is less sensitive to misalignments.
Only the measured $B_z$ is, therefore, used in the analysis to minimise the influence of misalignments
and the secondary components $B_r$ and $B_\phi$ are computed from the measured $B_z$ using the Maxwell equations as
%
%

\begin{eqnarray}
   B_\phi(z, r, \phi) &=& B_\phi(z_0, r, \phi) + \frac{1}{r}\int_{z_0}^{z}\frac{\partial B_z(z', r, \phi)}{\partial \phi}dz' \label{eq:reconstructed Bphi}\\
   B_r(z, r, \phi) &=& B_r(z_0, r, \phi) + \int_{z_0}^{z}\frac{\partial B_z(z', r, \phi)}{\partial r}dz'. \label{eq:reconstructed Br}
\end{eqnarray}

The computations require the measured values for $B_r$ and $B_\phi$ only at the plane defined by $z=z_0$.
This plane is chosen at $z_0 = 1\,\mathrm{mm}$ 
near the symmetry plane at the magnet centre where $B_r$ is measured to be small 
($|B_r|<2\times 10^{-3}$~T) as expected.
The effect of the misalignment of the $B_\phi$-measuring sensor on $B_\phi(z_0, r, \phi)$ is estimated 
by requiring the reconstructed $B_r$ and $B_\phi$ be consistent with all the other Maxwell equations.

Since $r = 0$ is a singularity of Eqs.~(\ref{eq:reconstructed Bphi}) and (\ref{eq:reconstructed Br})
and the effect of the misalignment of the $B_z$-measuring sensor is not negligible at large $r$ due to large $B_r$,
the secondary components are computed only for $2\,\mathrm{cm}<r<26\,\mathrm{cm}$.
The magnetic field calculated with a detailed coil model is used outside this region 
and also at $z < -106\,\mathrm{cm}$ and $z > 109\,\mathrm{cm}$ where the magnetic field was not measured.

The magnetic field is defined at the measuring grid points as described above 
and a continuous magnetic field map to be used in the analysis is obtained by interpolating the grid points by a B-spline fit.

%
%
\section{Drift chamber system}
\label{sec:dch}

\subsection{Introduction}
The Drift CHamber (DCH) system of the MEG experiment \cite{Hildebrandt2010111} 
is designed to ensure precision measurement of positrons from \meg\ decays.
It must fulfil several stringent requirements: 
cope with a huge number of Michel positrons due to a very high muon stopping rate 
up to $3\times 10^7\,\mu^+$/s, be a low-mass tracker as the momentum 
resolution is limited by multiple Coulomb scattering and in order to minimise the 
accidental $\gamma$-ray background by positron annihilation-in-flight, and finally 
provide excellent resolution in the measurement of the radial coordinate as well 
as in the $z$ coordinate. 

\begin{figure}[htb]
 \centerline{\hbox{
  \includegraphics[width=.45\textwidth,angle=0] {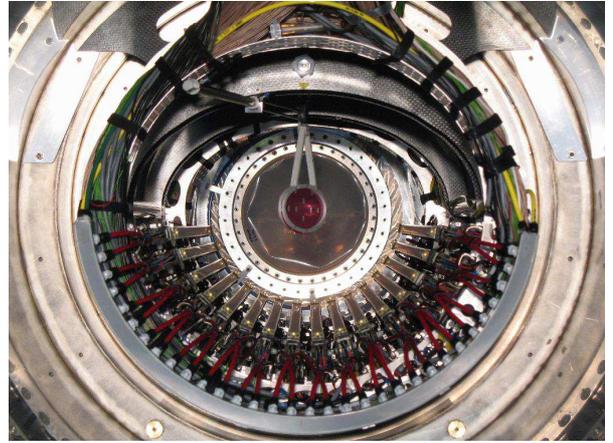}
 }}
 \caption[]{View of the DCH system from the downstream side of the MEG detector. The muon 
stopping target is placed in the centre, the 16 DCH chamber modules are mounted in a half circle.}
 \label{fig_dc_1.eps}
\end{figure}

The DCH system consists of 16 independent modules, 
placed inside the bore of COBRA (see Sect.~\ref{sec:cobra})
aligned in a half circle with 10.5$^{\circ}$ intervals covering
the azimuthal region $\phi\in (191.25^\circ,348.75^\circ)$ and the 
radial region between 19.3~cm and 27.9~cm (see Fig.~\ref{fig_dc_1.eps}). 

\subsection{Design of DCH module}
\begin{figure}[htb]
 \centerline{\hbox{
  \includegraphics[width=.45\textwidth,angle=0] {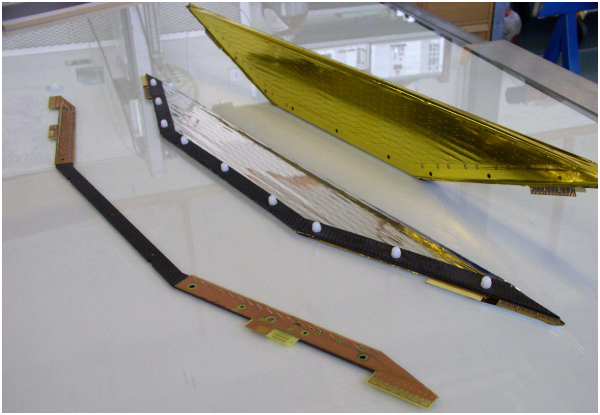}
 }}
 \caption[]{Main components of a DCH module: anode frame with wires (front), middle cathode and hood cathode 
(back).}
 \label{fig_dc_2.eps}
\end{figure}

All 16 DCH modules have the same trapezoidal shape 
(see Fig.~\ref{fig_dc_2.eps}) with base lengths of 40~cm and 104~cm, 
without any supporting structure on the long side (open frame geometry)
to reduce the amount of material.
The modules are mounted with the long side in the inner part of the spectrometer
(small radius) and the short one positioned on the central coil of the magnet (large radius).

\begin{figure}[htb]
 \centerline{\hbox{
  \includegraphics[width=.85\linewidth,angle=0] {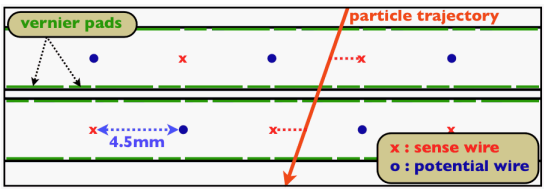}
 }}
 \caption[]{Schematic view of the cell structure of a DCH plane.}
 \label{cell}
\end{figure}

\begin{figure}[htb]
 \centerline{\hbox{
    \includegraphics[width=.95\linewidth] {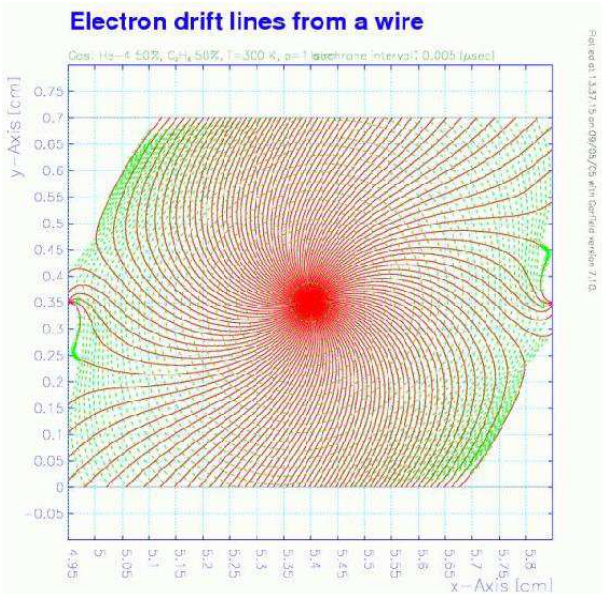}
 }}
 \caption{Isochrons (dashed green) and drift (solid red) lines of a cell with $B=1.60$~T.}
 \label{fieldline}
\end{figure}

Each module consists of two detector planes which are operated independently. The 
two planes are separated by a middle cathode made of two foils with a gap of 3~mm.
Each plane consists of two cathode foils with a gas gap of 7~mm. In between the foils
there is a wire array containing alternating anode and potential wires. They are 
stretched in the axial direction and are mounted with a pitch of 4.5~mm. The longest 
wire has a length of 82.8~cm, the shortest of 37.6~cm. The anode-cathode distance 
is 3.5~mm. The two wire arrays in the same module are staggered in radial direction by 
half a drift cell to resolve left-right ambiguities (see Fig.~\ref{cell}).
The isochronous and drift lines in a cell calculated with GARFIELD \cite{garfield} are 
shown in Fig.~\ref{fieldline} for $B=1.60$~T.

The two detector planes are enclosed by the so-called hood cathode.
The middle as well as the hood cathodes
are made of a 12.5~$\mu$m-thick polyamide foil with an aluminium deposition of 2500~{\AA}.

Thanks to such a low-mass construction and the use of a helium-based gas mixture, 
the average amount of material in one DCH module sums up to only 
\mbox{$2.6\times 10^{-4}\rm X_{0}$},
which totals \mbox{$2.0\times 10^{-3}\rm X_{0}$} along the positron track.

All frames of the DCH modules are made of carbon fibre. Despite being a very light material, it 
offers the advantage of being rather rigid. Consequently, there is only a negligible deformation of the 
frames due to the applied pretension compensating the mechanical tension of the wires (1.7~kg) or 
the cathode foils (12.0~kg).

\subsection{Charge division and vernier pads}
\label{ZDetermination}

A preliminary determination of the hit $z$ coordinate, $z_{\mathrm anode}$, is based on the principle of charge division.
To achieve this goal, the anodes are resistive Ni-Cr wires with a resistance per unit length of 
2.2~k$\Omega$/m. Firstly, the $z$ coordinate, is measured from the ratio of charges at both 
ends of the anode wire to an accuracy better than $2\%$ of the wire length.

\begin{figure}[htb]
 \centerline{\hbox{
  \includegraphics[width=.45\textwidth,angle=0] {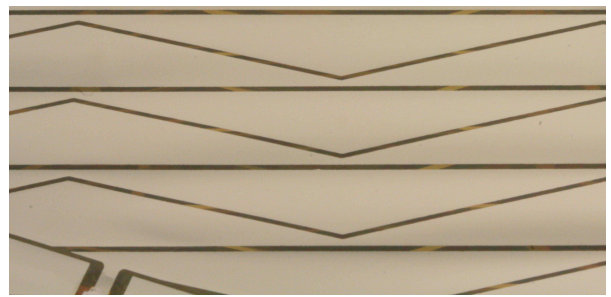}
 }}
 \caption[]{Double wedge or vernier pads etched on the aluminium deposition of the cathode foil.}
 \label{fig_dc_4.eps}
\end{figure}

Secondly the information from the cathodes is used to improve the measurement, by using a 
double wedge or vernier pad structure \cite{anderson84,green87,allison91} with pitch 
$\lambda =$ 5 cm, etched on the cathode planes on both sides of the anode wires (see Fig.~\ref{fig_dc_4.eps})
with a resistance per unit length of 50~$\Omega$/m. 

Due to the double wedge structure, the fraction of charge induced on each pad depends 
periodically on the $z$ coordinate. The period $\lambda$ is solved by means of $z_{\mathrm anode}$.
In total there are four cathode pad signals for each wire. To increase the capability of this method the 
vernier pads of one cathode plane is staggered by $\lambda$/4 in the axial direction with respect 
to its partner plane.

\subsection{Counting gas}
In order to reduce the amount of material along the positron trajectory, the helium-based 
gas mixture \mbox{$\rm{He:C_2H_6} = 50:50$} is adopted as a counting gas (\cite{uno93,hirano00}).
For this mixture the disadvantages of small primary ionisation and 
large diffusion of helium is balanced by the high primary ionisation, good quenching properties and 
high voltage (HV) stability 
of ethane. A reliable HV stability is very important as the DCHs are operated in a high-rate environment 
and with a very high gas gain of several times $\rm 10^5$. The main advantages of this mixture 
are a rather fast and saturated drift velocity, $\rm ~\sim4~cm/\mu s$, a Lorentz angle smaller 
than $8^{\circ}$ 
and a long radiation length ($\rm X_0/\rho = 640$~m).

\subsection{Pressure regulation system}
\label{sec:pressure}
A specially designed pressure control system manages the gas flows and gas pressures inside the DCH modules 
and inside the bore of COBRA, which is filled with almost pure He to also reduce multiple scattering 
along the positron path outside the DCH volume. The pressure regulation value, defined as the pressure 
difference between inside and outside the DCH, is on average $1.2\times 10^{-5}$~bar regulated to a precision better than $0.2\times 10^{-5}$~bar. 
This sensitivity is required as larger pressure differences would induce large distortions in the thin cathode 
foils and consequently lead to changes in the anode-cathode distances resulting in huge gain inhomogeneities 
along the anode wire as well as from wire to wire.

\subsection{HV system}
The DCHs are operated with a voltage of $\sim 1800$~V applied to the anode wires. 
In the first years of operation a HV system based on the MSCB system (see Sect.~\ref{High_voltage_system}) 
was used. In this case one power supply \cite{DC:iseg} was 
used as the primary HV device which fed 16 independent regulator units with 
2 channels each. The primary HV was daisy-chained along the 16 regulator modules.

In 2011 the MSCB.based HV system was replaced by a new commercial one. 
Two plug-in units \cite{DC:iseg2} with 16 independent channels each, operated 
in a Wiener Mpod mini crate, were chosen to reduce the high-frequency 
electronic noise contribution (mainly 14 MHz).

\subsection{Readout electronics}
Cathode and anode signals are conducted via printed circuit board tracks to the pre-amplifiers,
which are mounted at the frames of the DCH modules. The cathode signal is directly 
connected to the input of the pre-amplifiers, whereas the anode signal is 
decoupled by a 2.7~nF capacitor as the anode wire is on positive HV potential.

The pre-amplifier was custom-designed to match the geometrical boundary conditions and 
electrical requirements of the DCH system. The circuit is based on two operational amplifiers 
with feedback loops and protection diodes. The total gain of the pre-amplifier is $\sim$50 and 
the bandwidth is 140~MHz for the cathode signals and 190~MHz for the anode signals. 
In addition, the anode output is inverted to match the required positive input of the 
DRS read-out chip. The noise contribution of the pre-amplifier is 0.74~mV.

All output signals from the pre-amplifiers are individually transferred by non-magnetic, coaxial 
cables (Radiall MIL-C-17/93-RG178) to the back-end electronics. At the end-cap of the COBRA 
spectrometer there is a feedthrough patch panel which has resistive dividers splitting 
the anode signals into two outputs in the proportion 1:9. The larger signal goes to 
the DRS input, whereas the smaller one is amplified by a feedback amplifier and summed 
with several other anode outputs to form a DCH self-trigger signal.

\subsubsection{Alignment tools and optical Survey}

As the DCH is a position-sensitive detector, the position of the wires and 
the pads must be known very precisely. 
The alignment procedure consists of two parts, a geometrical 
alignment and a software alignment using cosmic rays and Michel positrons.

\begin{figure}[htb]
 \centerline{\hbox{
  \includegraphics[width=.45\textwidth,angle=0] {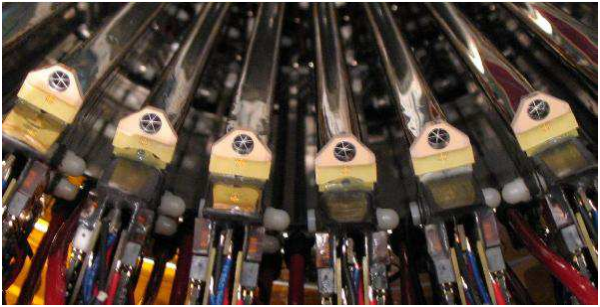}
 }}
 \caption[]{Corner cube reflectors and target marks on the downstream side of the DCH modules.}
 \label{fig_dc_3.eps}
\end{figure}

During the construction of each single frame the positions of the anode wires and of the cathode pads
are measured with respect to an alignment pin located at the bottom left edge of each frame.
Each cathode hood is equipped with two target marks (cross hairs) and two corner cube reflectors placed on the 
most upstream and most downstream upper edge of the cathode hood (see Fig.~\ref{fig_dc_3.eps}). After assembly 
of a DCH module the 
position of these identification marks was measured with respect to the alignment pin, which allows 
the alignment of the different frames within the assembled DCH module and acts as a reference for the wire 
as well as the pad positions.
Even though the chamber geometry is an open frame construction a geometrical precision of $\sim 50~\mu$m 
was achieved over the full length of the chamber for the position of the wires and the pads.

The support structure between two adjacent DCH modules is also 
equipped with target marks and corner cube reflectors on both the downstream and the upstream sides.

After the installation of the DCH system inside COBRA, an optical survey 
was performed. The positions of the target marks and of the corner cube reflectors on the cathode hoods and on the support 
structure were measured with respect to the beam axis and the downstream flange of the magnet using a 
theodolite (target marks) and a laser tracker system (corner cube reflectors) with a precision of 0.2~mm.

\subsubsection{Track-based alignment: the Millipede}
\label{sec:milli}
The optical survey measurements described above allow the determination of the wire positions in the 
absolute coordinate system of the experiment. 
To achieve a better accuracy, we developed a track-based alignment 
procedure consisting of three steps:
i) an internal alignment of DCH modules;
ii) the placement of the so-obtained DCH frame in the spectrometer, with
a fine adjustment of the relative DCH-target-COBRA position;
iii) the alignment of the spectrometer with respect to  the LXe detector
(see Sect.~\ref{sec:alixec}).

The first step is accomplished by recording cosmic ray tracks during the 
COBRA shutdown periods,  to make the internal DCH alignment independent of
the magnetic field map. Events are triggered by dedicated scintillation 
Cosmic Ray Counters (CRC), located around the magnet cryostat, which also provide 
a reference for the drift-time measurement. The alignment procedure utilises
the reconstructed position of hit DCH modules to minimise the residuals with
respect to straight muon tracks, according to the Millipede 
algorithm \cite{millipede}. Global parameters, associated with displacements of
each DCH module from its initial position, are obtained with an accuracy better
than $150\, \mu$m for each coordinate.

The second step relies on consistency checks
based on a sample of double-turn Michel positron tracks in the magnet volume.
The relative DCH-COBRA position is fine-tuned until the 
track parameters related to each single turn (which are reconstructed as 
independent tracks by the fitting algorithm) are consistent within the fit
uncertainties.

\subsubsection{Michel alignment}
The chamber positions from the optical survey are also cross-checked by using 
Michel positrons from $\mu^+$ decays. Both the radial chamber alignment and the 
alignment in $z$ are analysed by looking at the difference between the measured 
coordinate and that predicted by a fitted trajectory for each chamber plane. 
Because this diagnostic is insensitive to the absolute chamber positions, 
the overall location of the support structure in space is constrained by 
fixing the average radial and $z$ positions over all planes to those of the optical survey.  
Each chamber plane is then shifted to bring the mean of its pull distribution 
closer to the average over all planes.
The shifts are found to be consistent with the results in Sect.~\ref{sec:milli}.
Additional effects are investigated with Michel positrons such as 
tilts in the ($r$-$z$) plane, and displacement in the ($x$-$y$) plane. 
However no such effects have been found.


\subsubsection{Alignment with the LXe detector}
\label{sec:alixec}
Cosmic rays penetrating both the DCHs and the LXe detector are used to 
measure the position of the DCH frame relative to the LXe detector.
From the distribution of the difference between the position of 
incidence on the LXe detector reconstructed by the PMTs and that extrapolated 
from the DCHs, the two detectors can be aligned relative to each other with an accuracy of 1~mm 
in the horizontal and vertical directions.


%
\subsection{Calibrations}

\subsubsection{Time calibration}

Systematic time offsets between signals are calibrated by looking at the
distribution of the drift times. It is a broad distribution, whose width
is the maximum drift time in the cell, but it has a steep edge for
particles passing close to the sense wire. The position of this edge shows
systematic offsets for different channels (wires and pads), even in
the same cell, due to different delays in the electronics. The
determination of the edge position allows us to realign these offsets.

An essential ingredient of this procedure is the determination of the
track time, to be subtracted from the time of each hit for determining the
drift time. If positron tracks from muon decays are used, the time can be
provided by the track reconstruction itself (see Sect.~\ref{sec:performance}). Yet,
a better performance is obtained if cosmic rays are used, with the track
time provided with high precision by the CRC and corrected for
the time of flight from the reference counter to the cell. 
This approach disentangles the calibration from the positron tracking 
procedure, and guarantees a relative time alignment at the level of 
500~ps, comparable with the
counter time resolution. The track reconstruction uncertainty due to these
residual offset errors is negligible with respect to other contributions.

The same procedure can be used for both wire and pad signals.
Alternatively, to avoid problems due to the smaller size of pad signals,
wire offsets can be aligned first, and then the average time difference of
wire and pad signal peaks in the same cell can be used to align the pads
with respect to the wires.

\subsubsection{$z$ coordinate calibration}

A variety of electronic hardware properties influence the accuracy of 
the anode wire and cathode pad measurements of the $z$ coordinate.
As discussed in Sect.~\ref{ZDetermination}, the hit $z$ coordinate is 
first estimated by charge division using the anode wire signals and then 
refined using the charge distributions induced on the cathode pads.
The anode charge asymmetry is defined as 
\begin{equation}
A^{\mathrm anode} \equiv \frac{Q^{\mathrm anode}_{d}-Q^{\mathrm anode}_{u}}{Q^{\mathrm anode}_{d}+Q^{\mathrm anode}_{u}}, 
\end{equation}
where $Q^{\mathrm anode}_{u(d)}$ is the measured anode charge at the upstream (downstream) end.
The cathode pad charge asymmetry is defined similarly with $Q^{\mathrm cathode}_{u(d)}$ replacing $Q^{\mathrm anode}_{u(d)}$, 
where {\it cathode} is either the {\it middle} or the {\it hood} cathode of the cell.
Calculations show that the cathode charge asymmetry depends sinusoidally on $z$ to high precision. 
One group of calibrations involves adjusting the parameters entering into the 
anode $z$ calculation; this improve the probability of obtaining the correct pad period from the 
initial determination of $z$.

The anode $z$ coordinate is given by:
\begin{equation}
\label{anodeZ}
z_{\mathrm anode}=\left(\frac{L}{2}+\frac{R}{\rho}\right)A^{\mathrm anode}, 
\end{equation} 
where $L$ is the wire length, $R$ the pre-amplifier input impedance, and $\rho$ the wire resistivity. 

\begin{figure}[htb]
\centerline{\hbox{
  \includegraphics[width=.45\textwidth,angle=0] {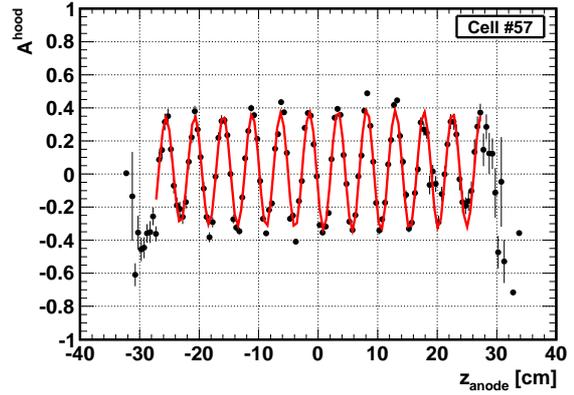}
 }}
\caption{A fit to the cathode pad asymmetry versus $z_{\mathrm anode}$ for the cathode hood of a particular drift cell.}
\label{hood112anodefit.eps}
\end{figure}

Figure~\ref{hood112anodefit.eps} demonstrates the fundamental calibration tool: the measured 
cathode charge asymmetry versus the measured $z_{\mathrm anode}$ coordinate for each cell. The sinusoid period 
is forced to the pad pitch ($\lambda =5$~cm) by adjusting the factor
\mbox{$(\frac{L}{2}+\frac{R}{\rho})$} and its phase is forced to be 0 for $z_{\mathrm anode}=0$ 
by calibrating the relative gain between anode charges $Q^{\mathrm anode}_{d}$ and $Q^{\mathrm anode}_{u}$. 

A calibration of the relative gain between cathode pads is performed to achieve optimal $z$ resolution.
This procedure uses plots similar to Fig.~\ref{hood112anodefit.eps}, except that $z_{\mathrm anode}$ is 
replaced with the more precisely determined coordinate from fitted tracks $z_{\mathrm track}$.  
A vertical offset indicates a relative difference between upstream and downstream cathode pad gains, 
which is zeroed by adjusting the cathode pad gains.
The systematic uncertainty in the cathode pad gains from this procedure is estimated by comparing the 
results using $z_{\mathrm track}$ and $z_{\mathrm anode}$ as the abscissa, 
to be $\sigma^{DCH}_{z}=70 \; \mu $m. 

The chambers, being operated at slight overpressure (see Sect.~\ref{sec:pressure}), experience bowing of the 
outer cathode hood foils. The bulge of the hood is largest near $z=0$ and at middle cell numbers 
because the foil is only attached at the edges of the chamber. This means that the two cathode foils are not 
equidistant from the anode wires, but instead the distance varies as a function of $z$ even within the same 
cell. The induced charge, as well as the magnitude of the charge asymmetry, also differ between the two 
planes depending on $z$.

\begin{figure}[htb]
\centerline{\hbox{
  \includegraphics[width=.45\textwidth,angle=0] {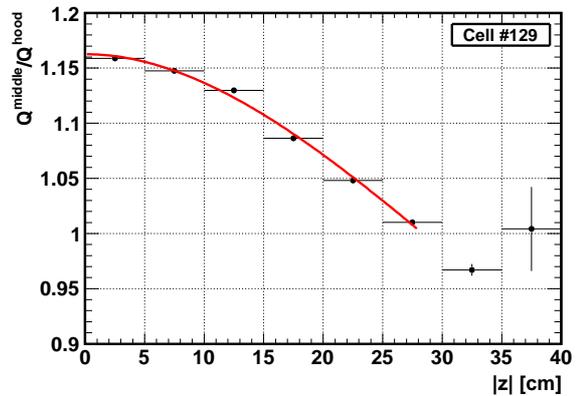}
 }}
\caption{A plot of the charge ratio versus $|z|$ of a particular drift cell. 
This is fit to a quadratic polynomial to correct for the relative cathode-to-hood gain.}
\label{Qratio129.eps}
\end{figure}

This effect is calibrated by correcting for the mismatch between asymmetry amplitudes.
For a given cell, 
\begin{equation}
\frac{Q^{middle}}{Q^{hood}}=\frac{Q^{middle}_{u}+Q^{middle}_{d}}{Q^{hood}_{u}+Q^{hood}_{d}}, 
\end{equation}
which is related to the ratio of asymmetry magnitudes for geometric reasons, exhibits a clear and 
expected dependence on $z$, as shown in Fig.~\ref{Qratio129.eps}. The dependence for each wire and a preliminary
evaluation of the pad $z$ coordinate are used to rescale the asymmetries on an event-by-event basis for 
a final evaluation of the pad $z$ coordinate. The charge ratio dependence of the charge asymmetry magnitude ratio 
is obtained by a scatter plot of those two variables summed over all wires and values of $z$.
 
\subsection{Tracking}
A chain of software algorithms first measures the arrival time and charge on each wire and pad associated 
with the passage of a charged particle through a cell. The arrival times and charges are converted into 
spatial coordinates.  
The information from each cell, referred to as a hit, is cross-checked with other hits in the same chamber, 
and groups of hits consistent with coming from the same particle are then collected into clusters.  
Patterns of clusters consistent with coming from the same particle are then collected into tracks.

Finally, a list of clusters associated with each track is passed to a 
Kalman filter to fit a trajectory. The fit incorporates average energy loss, multiple scattering, and the measured
magnetic field. It also attempts to link multiple turns consistent with belonging to a single trajectory.   
The Kalman filter provides the track momentum, decay vertex, positron emission angles, the predicted impact location 
at the TC and the total path length from the target to the TC. 
An event-by-event indicator of the resolution in each of these quantities is also given.

\subsection{Performance}
\label{sec:performance}
This section reports on the DCH performance regarding spatial, angular, momentum and vertex resolutions.

\subsubsection{Single Hit Resolution}
\begin{figure}[htb]
\centerline{\hbox{
  \includegraphics[width=.50\textwidth,angle=0] {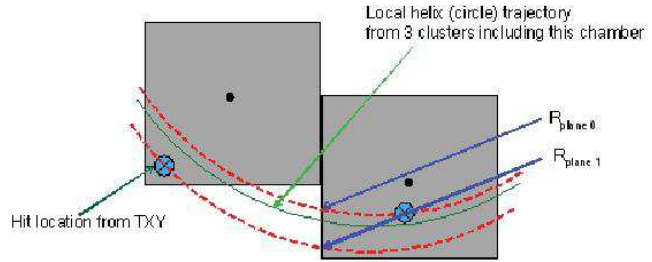}
 }}
\caption{A diagram of the technique for measuring the intrinsic radial coordinate resolution.}
\label{dRDCH.eps}
\end{figure}

\begin{figure}[htb]
\centerline{\hbox{
  \includegraphics[width=.50\textwidth,angle=0] {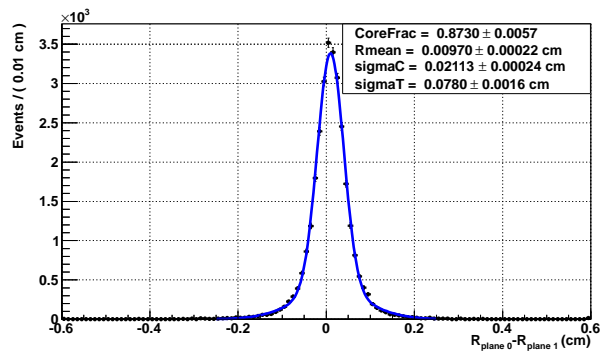}
 }}
\caption{A fit to the distribution of the difference of projected $r$ coordinates
 for two-hit clusters.}
\label{dR.eps}
\end{figure}

We refer to the uncertainty in a single hit position measurement as the intrinsic resolution, 
i.e. not affected by multiple scattering. A technique for measuring the intrinsic radial 
coordinate resolution is illustrated in Fig.~\ref{dRDCH.eps}. The method considers clusters 
with exactly one hit in each chamber plane, on adjacent cells.
A local track circle in the ($x$,$y$) plane is calculated from the position of this cluster 
and the two neighbouring clusters.
Each hit is then propagated to the central chamber plane using this trajectory and 
the difference in the two radial coordinates at the central plane is interpreted as a measure 
of the radial intrinsic resolution. Figure~\ref{dR.eps} 
shows the distribution of $R_{plane \; 0}-R_{plane \; 1}$. This distribution is fit to 
the convolution of two double Gaussian functions. 

The reason for using a double Gaussian function is that
the single hit resolution for drift chambers usually has large tails, which are due 
to the non-Gaussian fluctuations of the primary ionisation, to the large diffusion and 
the irregular shape of the time-to-distance relations for hits produced in the peripheral 
regions of drift cells, to wrongly left/right assignments, frequent for hits 
at small drift distances, to wrong drift time measurements in some unusually noisy
signals, etc. Along with the non-Gaussian material effects (multiple scattering and dE/dx), 
it produces tails in the track parameter resolutions (angles, momentum and vertex).

The result is a radial coordinate resolution of 
$\sigma^{DCH}_{r} = 210$ $\mu$m in the core (87\%) and $\sigma^{DCH}_{r} = 780$ $\mu$m in the tail.
The design resolution was parametrised by a single Gaussian with $\sigma^{DCH}_{r}=200$ $\mu$m.  

\begin{figure}[htb]
\subfigure[A view of a track segment in the ($x$,$y$) plane. 
See text for details.]{
\label{zresxyview.eps}
  \includegraphics[width=.225\textwidth,angle=0] {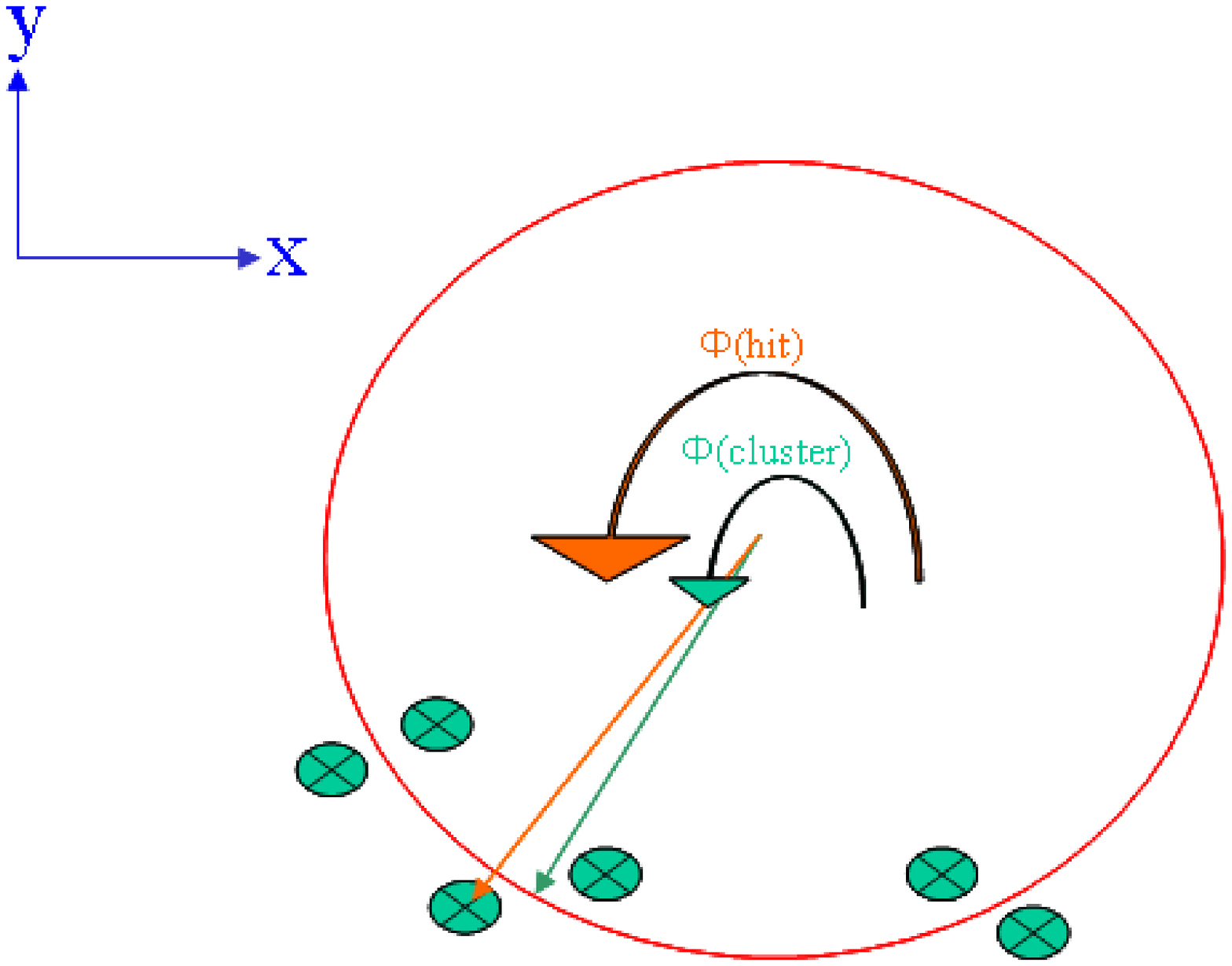}
}
\subfigure[A view of a track segment in the z$\phi$ plane.
See text for details.]{
\label{zreszphiview.eps}
  \includegraphics[width=.225\textwidth,angle=0] {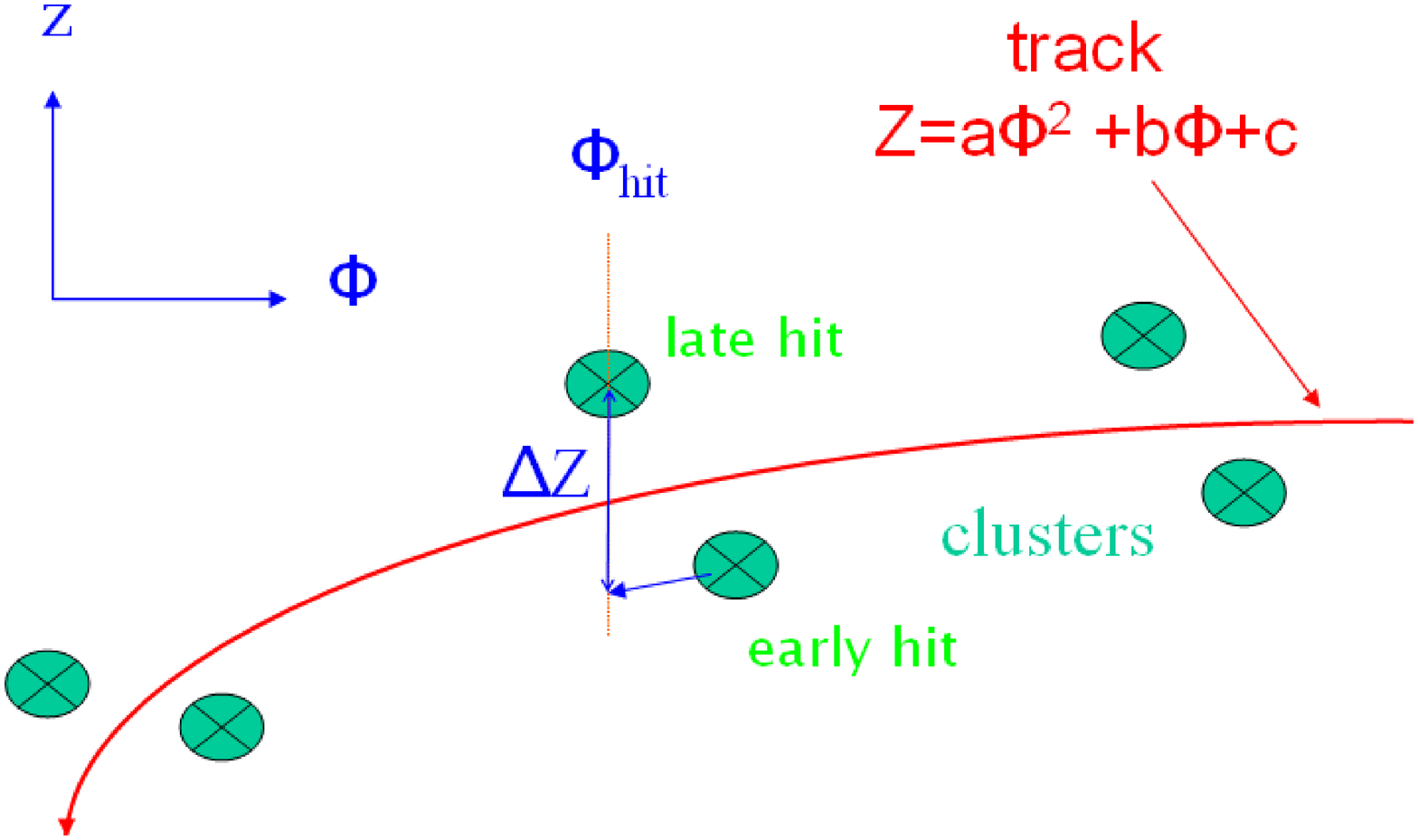}
}
\caption{A diagram of the technique for measuring the intrinsic $z$ coordinate resolution.}
\label{dZDCH}
\end{figure}

\begin{figure}[htb]
\centerline{\hbox{
  \includegraphics[width=.50\textwidth,angle=0] {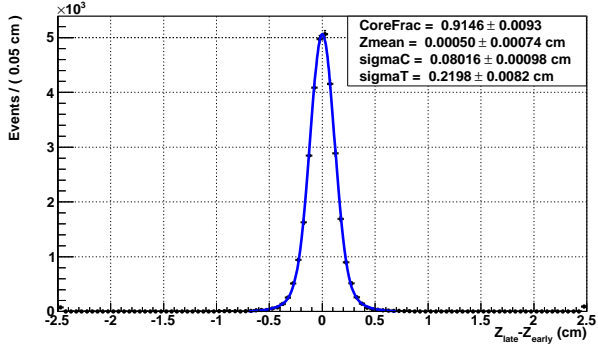}
 }}
\caption{A fit to the distribution of the difference of projected $z$ coordinates for two-hit clusters.}
\label{dZ.eps}
\end{figure}

A technique for measuring the intrinsic $z$ coordinate resolution is illustrated in Fig.~\ref{dZDCH}.  
Two-hit clusters belonging to a track segment are selected in the same way described for the radial coordinate 
resolution measurement to compare the two measured $z$ coordinates. The calculated track circle in the ($x$,$y$) 
plane is used to define a local coordinate system, with the centre of the circle defining the origin. 
Within this coordinate 
system, the polar angle $\phi$ is calculated for each overall cluster position and for each hit within the cluster, 
as shown in Fig.~\ref{zresxyview.eps}. From the three cluster coordinates, a quadratic trajectory 
for $z$ as a function of the local $\phi$ is computed. This trajectory is used to project the measured 
$z$ coordinates of the two hits in question to a common $\phi$ and the resulting distribution of the 
difference of the projected $z$ coordinates, $z_{late}-z_{early}$, is used to infer the intrinsic $z$ 
resolution, as pictured in Fig.~\ref{zreszphiview.eps}. This distribution, shown in Fig.~\ref{dZ.eps}, 
is fit with the convolution of two double Gaussian functions, resulting in a $z$ coordinate 
resolution of $\sigma^{DCH}_z = 800$ $\mu$m in the core (91\%) and $\sigma^{DCH}_z = 2.1$ mm in the tail.
The largest known contribution to the $z$ coordinate resolution comes from 
the stochastic fluctuations of the baseline in the presence of noise; this is estimated to be 
$\sigma^{DCH}_{z,noise}=550$ $\mu$m on average.  
The design $z$ coordinate resolution was $\sigma^{DCH}_{z}=300$ $\mu$m. 

\subsubsection{Angular Resolution}
\begin{figure}[htb]
\centerline{\hbox{
  \includegraphics[width=.5\textwidth,angle=0] {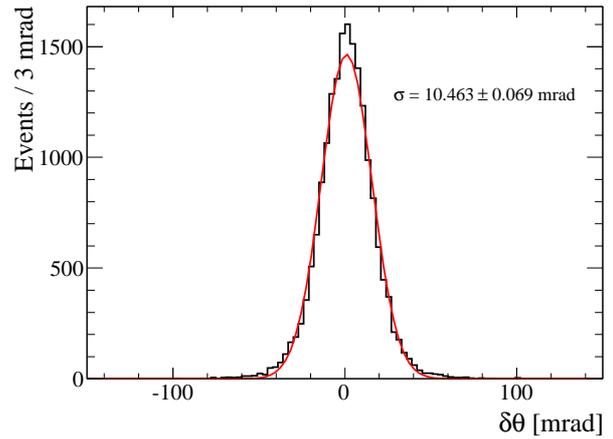}
 }}
\caption{A fit to the distribution of $\delta \theta_{e} \equiv \theta^{1st \; turn}_{e}-\theta^{2nd \; turn}_{e}$ 
on double-turn events. The distribution is fitted with a single Gaussian function convolved with itself, and 
the corresponding width is shown.}
\label{DCHThetaRes}
\end{figure}

\begin{figure}[htb]
\centerline{\hbox{
  \includegraphics[width=.5\textwidth,angle=0] {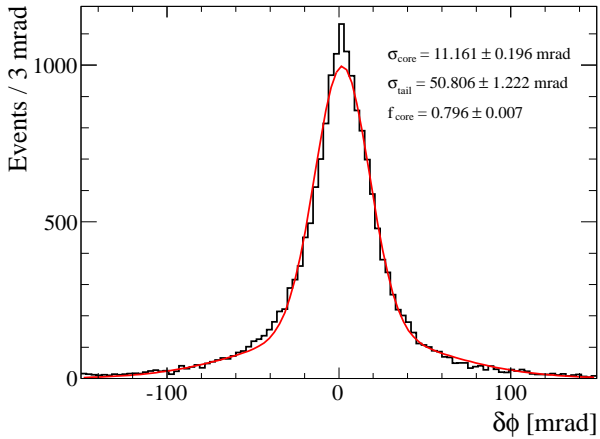}
 }}
\caption{A fit to the distribution of $\delta \phi_{e} \equiv \phi^{1st \; turn}_{e}-\phi^{2nd \; turn}_{e}$ 
on double-turn events. The distribution is fitted with a double Gaussian function convolved with itself, and 
the corresponding core and tail widths are shown, along with the fraction of events in the core component.}
\label{DCHPhiRes}
\end{figure}

The resolutions in the measurements of the positron angles at the target are measured from data by exploiting events
where the positron makes two turns in the DCHs. Each turn is treated as an independent track, fitted, 
and propagated to the beam line where the track angles are compared.  
The distributions of the difference 
of the two measured angles in double-turn events are shown in Fig.~\ref{DCHThetaRes} for $\theta_{e}$ and 
in Fig.~\ref{DCHPhiRes} for $\phi_{e}$.
The resolution in each turn is assumed to be the same and these distributions are fit to the convolution 
of a single ($\theta_{e}$) or double ($\phi_{e}$) Gaussian functions. These functions represent the resolution 
function of the positron angles. 

According to Monte Carlo studies, this method provides a significant overestimate of the true resolution. After correcting
for this, we obtain a single Gaussian resolution $\sigma_{\theta_{e}} = 9.4 \pm 0.5$ mrad and a
double Gaussian $\phi_{e}$ resolution of $\sigma_{\phi_{e}} = 8.4 \pm 1.4$ mrad in the core (80\%) and 
$\sigma_{\phi_{e}} = 38 \pm 6$ mrad in the tail, 
where the errors are dominated by the systematic uncertainty of the correction. 
The Monte Carlo resolutions are  $\sigma_{\theta_{e}}\sim 9$ mrad
$\sigma_{\phi_{e}}\sim 8$ mrad,
while the design resolutions were $\sigma_{\theta_{e},\phi_{e}}\sim 5$ mrad.

It is also important to stress that these resolutions are affected by correlations among the 
other positron observables, which can be treated on an event-by-event basis,
so that that the effective resolutions determining the experimental sensitivity are 
$\sigma_{\theta_{e}} = 8.5 \pm 0.5$ mrad and $\sigma_{\phi_{e}} = 7.7\pm 1.4$ mrad in the core.

The multiple scattering contribution to these resolution is $\sigma_{\theta_{e},\phi_{e}} \sim 6.0$ mrad,
the rest is due to the single hit resolution.

\subsubsection{Vertex Resolution}
\label{sec:vertex}
The resolution in the position of the decay vertex on the target is dominated by the positron angular resolution.
For a proper evaluation of the angular resolution function, a precise knowledge of the correlations 
between positron angle error and vertex position error is required. 
The average vertex position resolutions, however, can be measured directly by comparing the projected 
point of interception at the target plane on double-turn events. 
The difference in vertex $z$ coordinates of the two turns is fit to the convolution of a single Gaussian 
function with itself, while that of the vertex $y$ coordinates is fit to the convolution 
of a double Gaussian function with itself. 
After the Monte Carlo corrections are applied, the resolutions are 
$\sigma_{y_e} = 1.1 \pm 0.1$ mm in the core (86.7\%), $\sigma_{y_e} = 5.3 \pm 3.0$ mm 
in the tail and $\sigma_{z_e} =2.5 \pm 1.0$ mm. The resolutions for Monte Carlo events are 
$\sigma^{MC}_{y_e} = 1.0 \pm 0.1$ mm in the core and $\sigma^{MC}_{z_e} =2.9 \pm 0.3$ mm.
The values of $\sigma_{y_e}$ are corrected for the correlation with the positron energy
assumed to be the signal energy.

The design resolution was $\sigma_{y_e,z_e} \sim 1.0$ mm without correcting for correlation.

\subsubsection{Energy Resolution}
\begin{figure}[htb]
\centerline{\hbox{
  \includegraphics[width=.45\textwidth,angle=0] {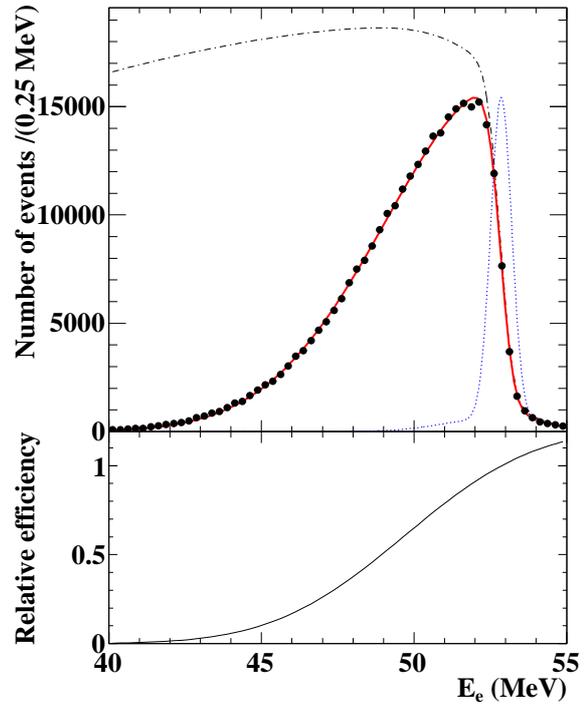}
 }}
\caption{A fit to the Michel positron energy spectrum. The theoretical spectrum (dashed black), the resolution
function (dashed blue) and the acceptance curve (in the bottom plot) are also shown.}
\label{EMichelAvg.eps}
\end{figure}

The positron energy resolution is measured with a fit of the energy distribution 
to the unpolarised Michel 
spectrum multiplied by an acceptance function and convolved with a resolution function:
\begin{eqnarray}
Probability \; density(E^{measured}_{e})= \nonumber \\
(Michel*Acceptance)(E^{true}_{e})\otimes Resolution.
\end{eqnarray}
Functional forms for both the acceptance and the 
resolution functions are based on the guidance provided by Monte Carlo simulation. The acceptance function is 
assumed to be:
\begin{equation}
Acceptance(E_{e}^{true}) = \frac{1+erf(\frac{E_{e}^{true}-\mu_{acc}}{\sqrt{2}\sigma_{acc}})}{2},
\end{equation}
and the resolution function is taken to be a double Gaussian.
The acceptance and the resolution parameters are extracted from the fit, as shown in Fig.~\ref{EMichelAvg.eps}. 
This gives an average resolution of $\sigma_{E_e} = 330 \pm 16$ keV in the core (82\%) and 
$\sigma_{E_e} = 1.13 \pm 0.12$ MeV in the tail. There is also a 60 keV systematic underestimation 
of the energy, to which we associate a conservative 25 keV systematic uncertainty from Monte Carlo studies. 
This is to be compared with the resolution goal of $\sigma_{E_{e}}=180$ keV (0.8\%\ FWHM).

\begin{figure}[htb]
\centerline{\hbox{
  \includegraphics[width=.45\textwidth,angle=0] {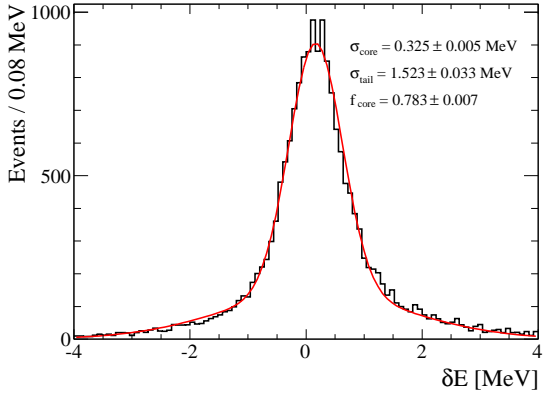}
 }}
\caption{A fit to the distribution of $\delta E_{e} \equiv E^{1st \; turn}_{e}-E^{2nd \; turn}_{e}$ on 
double-turn events. The distribution is fitted with a double Gaussian function convolved with itself, and 
the corresponding core and tail widths are shown, along with the fraction of events in the core component.}
\label{E2tAvg.eps}
\end{figure}

A complementary approach to determining the positron energy resolution is possible by using two-turn 
events as for the angular resolution. Figure~\ref{E2tAvg.eps} shows 
the distribution of the
energy difference between the two turns. This is fit to the convolution of a double Gaussian function with itself, 
the same shape assumed in the fit of the edge of the Michel spectrum. A disadvantage of this 
technique is its inability to detect a global shift in the positron energy scale.
This technique gives an average resolution of 
$\sigma_{E_e} = 330$ keV in the core (79\%) and $\sigma_{E_e} = 1.56$ MeV in the tail, in 
reasonable agreement with the results obtained from the fit of the Michel spectrum. A systematic offset of 108 keV 
between the energies of the two turns also appears; the energy of the first turn is systematically 
larger than the energy of the second turn. 
A related effect is the dependence of the measured Michel edge on $\theta_e$. 
These effects point to errors in the magnetic field mapping.

\subsubsection{Chamber detection efficiency}
The relative efficiency of each chamber plane is measured as the probability to 
have a reconstructed hit when its neighbouring plane in the same chamber has at least one hit
associated to a track. This probability is called the
``hardware'' efficiency, while the probability to have a hit associated to the track is called the ``software''
efficiency.
Low-efficiency planes can be traced to operation at voltages below the nominal voltage.
For the other planes the average hardware efficiency is $\sim 95\%$ and the average software efficiency is
$\sim 90\%$.

%
%

\subsubsection{DCH efficiency}
\label{sec:dchefficiency}

The absolute efficiency of the DCH system for signal and Michel positrons is difficult to measure.
Yet it is not used in the physics analysis, because 
the \meg\ Branching Ratio is normalised with respect to the number of reconstructed
Michel or radiative muon decays, so that the absolute positron reconstruction efficiency cancels. 
Nevertheless in Fig.~\ref{EMichelAvg.eps}, the fit to the Michel positron spectrum relative to tracks 
associated to a TC hit returns an estimate of the relative efficiency versus the positron momentum.
In particular the relative efficiency for Michel (48~MeV~$<E_e$) versus signal positrons is $\sim 0.70$.
This relative efficiency can also be estimated by Monte Carlo simulation to be $\sim 0.75$ in good agreement.

The absolute efficiency of DCH only for Michel positron (48~MeV~$<E_e$) as estimated from 
Monte Carlo is $\epsilon_{e,DCH}^{MC,Michel} \sim 90\%$, while for signal 
$\epsilon_{e,DCH}^{MC}\sim 83\%$. The largest source of inefficiency for the DCH alone 
are the tracks emitted at $\theta_e \sim 90^\circ$, where multiple turns generate inefficiency 
in the track-finding algorithm.

%

\section{Timing Counter}
\label{timingcounter}

The Timing Counter (TC) is dedicated to precisely measuring the impact time of positrons 
to infer their emission time at the decay vertex in the target
by correcting for the track length obtained from the DCH information.

The main requirements of the TC are:
\begin{itemize}
\item fast response to be used in the online trigger algorithms and to avoid rate effects;
\item online fast and approximate ($\sim 5$~cm) impact point position reconstruction for trigger purposes;
\item capability of determining the positron time with an accuracy $\sim 50$~ps 
\item good ($\sim 1$~cm) impact point position reconstruction in the offline event analysis; 
\item reliable operation in a harsh environment: high and non-uniform magnetic field, helium atmosphere, possibility of aging effects.
\item cover the full acceptance region for signal while 
matching the tight mechanical constraints dictated by the DCH and COBRA
\end{itemize}

These goals were achieved through extensive laboratory and beam tests 
\cite{Dussoni2010387,valle_2006,dussoni_2006,DeGerone2011}.


\subsection{Concept and design of the Timing Counter}
\label{tcdesign}

As mentioned in Sect.~\ref{sec:introduction} and visible in Fig.~\ref{introduction:megdet},
the TC matches the signal kinematics and is compatible with the mechanical constraints
by the placement of one module (sector) upstream and the other downstream.

\begin{figure}[!ht]
\centering
\includegraphics[width=.45\textwidth]{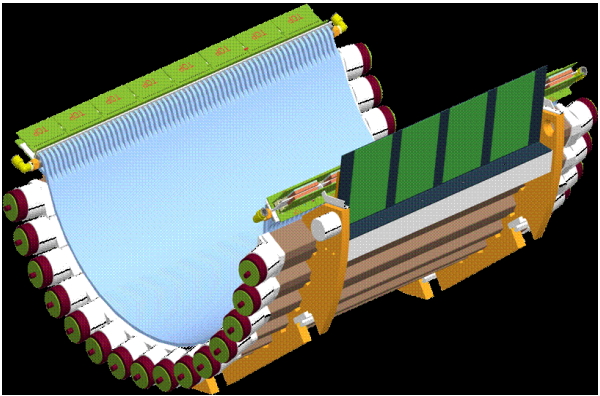}
\caption{Schematic picture of a TC sector. Scintillator bars are read out by PMTs.
Scintillation fibres
are placed on top of the scintillator bars and read out by APDs.}
\label{tcall}
\end{figure}

Each sector is barrel-shaped and consists of two sub-detectors, mounted on top of each other,
with full angular coverage for positrons from \meg\ decays
when the $\gamma$-ray points to the LXe detector (see Fig.~\ref{tcall} for a 
sketch of a sector). 

At the outer radius, the longitudinal detector, consisting of scintillating bars
read out at each end by PMTs, is dedicated mainly 
to time measurements and, thanks to its segmentation along $\phi$, also provides a 
measurement of the positron
impact $\phi$ coordinate. It also provides a measurement of the positron
impact $z$ coordinate by exploiting the separate time measurements
of both PMTs. An approximate estimate of these variables is obtained online 
with fast algorithms for trigger purposes.
This detector is therefore sufficient to reconstruct all positron variables
required to match a DCH track and recover the muon decay time by extrapolating the
measured time back to the target.

At the inner radius, the transverse detector, consisting of scintillating fibres,
is devoted to determining with high precision the impact $z$ coordinate to improve
the matching between the DCH track and TC point. 

Each sector is surrounded by a bag made of a 0.5~mm-thick EVAL \textsuperscript{\textregistered}
foil that separates 
the helium-filled volume of the magnet bore from the PMTs, which are sensitive to helium 
leakage through the borosilicate window \cite{hepmt,helat}. A buildup of helium inside the 
PMTs would lead to internal discharge making them unusable. 
The bag volume is continuously flushed with nitrogen to guarantee a stable operating condition.

\subsection{The longitudinal detector}
\label{tcbars}
Each sector of the longitudinal detector consists of an array of 15 scintillating bars
(Bicron BC404 \cite{saintgobain,Dussoni2010387}) 
with approximate square section and size $4.0\times
4.0\times 79.6\,\mathrm{cm}^3$; each bar is read out by a couple of 
fine-mesh 2" PMTs for high magnetic fields \cite{hamamatsu-tc} glued at the ends. 
The bars are arranged in a barrel-like shape to fit the COBRA 
inner profile with a $10.5^\circ$ gap between adjacent bars matching the DCH pitch. 
The scintillator type was chosen from among fast scintillator candidates 
through a series of beam tests by comparing their timing performances.
The PMTs were chosen to match the bar size taking into account the 
magnetic field strength and orientation \cite{bonesini_2006} as well as
the mechanical constraints.

The number of bars is defined by requirements on the size of the trigger regions
defined in the $(\phi,\theta)$ plane mapped onto the TC; the $\phi$ width corresponds
to the gap between adjacent bars.

The bar thickness was chosen to acquire sufficient energy deposition from positrons 
to obtain the design time resolution.

The bar width is dictated by the requirement of providing full efficiency for positrons
crossing the TC, leaving no gap between bars.
Full acceptance of positrons emitted in the experiment's angular range defines
the bar length.

\begin{figure}[!ht]
\centering
\includegraphics[width=3.0in]{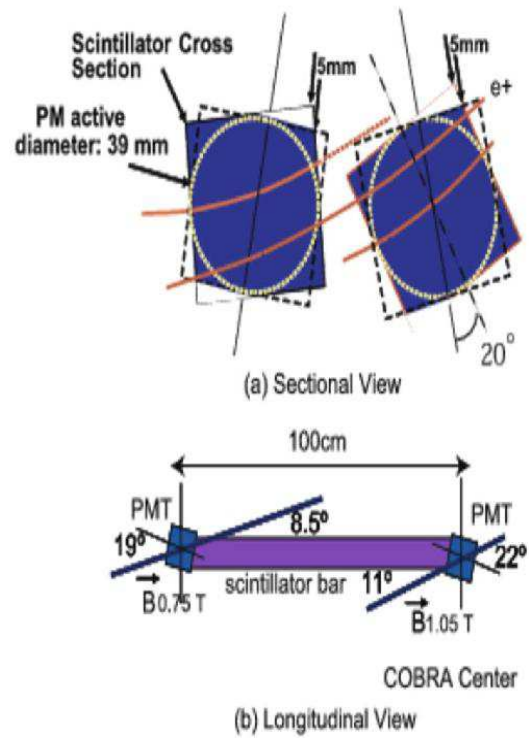}
\caption{The layout of a TC bar-PMT assembly.}
\label{tcbar}
\end{figure}

Furthermore, the details of the bar geometry are also dictated 
by the mechanical constraints coming from the TC transverse detector and the DCH system,
resulting in the bar geometry in Fig.~\ref{tcbar}. 

\subsubsection{Readout electronics}
\label{tcdevelopment}

\begin{figure}[!ht]
\centering
\includegraphics[width=.45\textwidth,height=2.in]{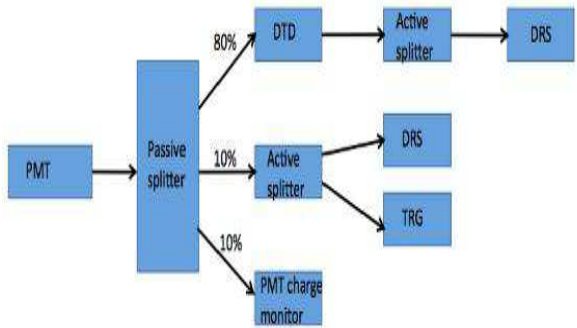}
\caption{Schematic picture of the read-out electronics of the TC longitudinal detector. 
Signals from each PMT are passively 
split and fed into dedicated electronics channels. 
Different splitting ratios are used to optimise time and pulse reconstruction.
The PMT charge monitor was deactivated in 2011 and its fraction is added to DRS/trigger input.}
\label{DTD}
\end{figure}

The signals from the PMTs are interfaced with the DRS chip fast 
digitiser (see Sect.~\ref{DAQ:drs4}) with the read-out scheme depicted in 
Fig.~\ref{DTD}. Each PMT signal is passively split into three channels 
with 80\%, 10\%\ and 10\%\ amplitude fractions, respectively.
The highest fraction is sent to a Double Threshold Discriminator (DTD)
specifically designed for time analysis purposes.

DTD is a high-bandwidth, low-noise discriminator with two different 
tunable thresholds: the lower one allows precise pulse-time measurements
(see Sect.~\ref{sec_timewalk}) while the higher one acts as a veto against 
low energy hits.
The value of the high threshold determines the detector
efficiency and is also related to the trigger threshold as
discussed in Sect.~\ref{tccommission}. When the DTD is fired, a
standard NIM waveform is generated and then digitised by the DRS
boards together with a copy of the PMT waveform. 
The NIM and PMT waveforms are analysed offline to extract 
the hit time. Several algorithms can be applied, the 
best one (used as default) fits the NIM waveform. 
This method reduces the
contribution from the DRS jitter and dynamic range to the overall time resolution.

The first 10\%\ copy of each signal is duplicated and fed to the
trigger system and the DRS: in this stage an active splitter described in 
Sect.~\ref{sec:splitter} is used
for the level translation needed to match the input ranges of
the trigger (see Sect.~\ref{trigger}) and DRS boards. 
The second 10\%\ signal copy is sent to a charge integrator to monitor the PMT ageing. 
Since 2011 we implemented a charge monitor based on pulse analysis. 
Consequently we feed 20\%\  of the PMT signals to the DRS/trigger inputs.

\subsection{The transverse detector}
\label{tcfibres}
The transverse detector consists of 256 scintillating
multi-clad fibres \cite{saintgobain-ticz} 
coupled to Avalanche PhotoDiodes (APD) \cite{hamamatsu-apd} covering the
inner surface of a TC sector. 

The small sectional area of the fibres ($5\times 5\,\mathrm{mm}^2$
plus $2\times 0.5\,\mathrm{mm}$ wrapping thickness totalling $6.0\,\mathrm{mm}$
in size along $z$) satisfies the required space resolution and 
matches the $5\times 5\,\mathrm{mm}^2$ APD 
sensitive area. In order to comply with the mechanical constraints, 
the fibre ends at the APD side are bent in two different ways: one set  has almost
straight termination while the other is ``S'' shaped with
small curvature radii ($\sim 2\, \mathrm{cm}$) and shows a significant light leakage, 
recovered with a high-reflectance wrapping \cite{degerone}. 

The APD advantages can be summarised as: small size, insensitivity to
magnetic field, and fast response. The main disadvantage is the low
gain, at most $\sim 10^{3}$ if biased near the
breakdown voltage, and the high capacitance that make them sensitive to
noise. For this reason a single APD current
pulse is read out as voltage across a load resistor and amplified. Bundles
of 8 APDs are mounted on distinct boards carrying amplifying stages, power
supplies, HV bias and ancillary control signals. 
Due to mechanical constraints, 
the fibre bundle mounted on a single front-end board is interleaved with the
fibre bundle connected to the adjacent board: the resulting assembly, 16
consecutive fibres read out by two adjacent boards, covers $9.6\times40.0\;\mathrm{cm^2}$ 
and is viewed in the trigger system as a single block.

Each APD has an 
independent discriminated output, acquired via a FPGA based board and stored. 
Therefore the transverse detector delivers the list of fired fibres for each event.

\begin{figure}[!ht]
\centering
\includegraphics[width=0.40\textwidth,angle=-90]{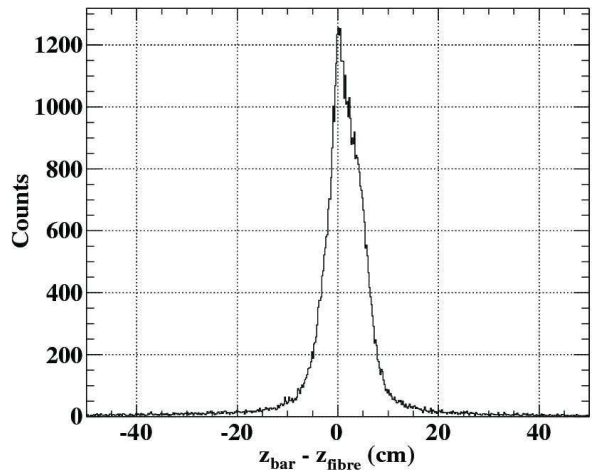}
\caption{Distribution of $z_{bar}-z_{fibre}$ for the downstream sector.}
\label{z_APD}
\end{figure}

The reconstructed coordinate of the hit fibre $z_{fibre}$ can be matched to the 
$z_{bar}$ measured by the longitudinal detector.
The distribution of $z_{bar}-z_{fibre}$ 
for the downstream sector obtained with cosmic rays
is shown in Fig.~\ref{z_APD}. Some tails are related to a $\mathcal{O}(10\%)$
inefficiency of the APD detector due to its geometry: 
incomplete coverage and $1~\mathrm{mm}$ gap between fibres.

%

\subsection{Commissioning and \em{in-situ} performance}
\label{tccommission}

The integration of the TC in MEG requires a set of
calibrations to determine the detector parameters
optimising its performance. 
Hereinafter only those involving the TC alone are described. 
Additional calibrations, involving other sub-detectors (e.g. time 
alignments with the DCH and LXe detectors) are described in Sect.~\ref{sec:accelerator} 
and in \cite{tccalib}.

\subsubsection{Gain equalisation}
Due to the non-solenoidal configuration of the magnetic field,
the inner and outer PMTs are subject to different field intensities and
orientations and therefore operate in different working conditions
resulting in different gains. Gains need to be equalised for their use
in triggering and analysis.

\begin{figure}[htb]
\begin{center}
\includegraphics[width=.5\textwidth]{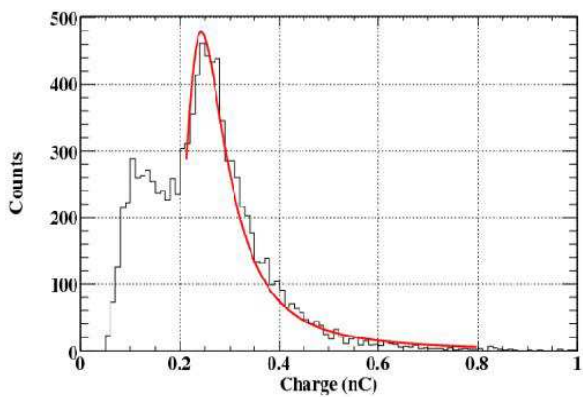}
\caption{Example of charge spectrum acquired with a TC bar, 
superimposed with a Landau function fit.}
\label{fig_landau}
\end{center}
\end{figure}

The PMT gain equalisation procedure exploits cosmic rays hitting the TC:
due to the uniformity of cosmic ray hit distribution along the bars
the charge and amplitude spectra of the inner and outer PMTs are
only marginally affected by geometrical effects.
It is therefore possible to tune the PMT gains acting on the bias high voltages
and equalise the peaks of the Landau distributions (Fig.~\ref{fig_landau})
within $15\%$ or better. 
In the trigger algorithm a further software equalisation is performed online to keep a uniform threshold on the whole detector. 
By means of dedicated weights for each PMT a percent level uniformity is obtained.

\subsubsection{Time Walk effect}
\label{sec_timewalk}
\begin{figure}[!ht]
\centering
\includegraphics[width=3.4in]{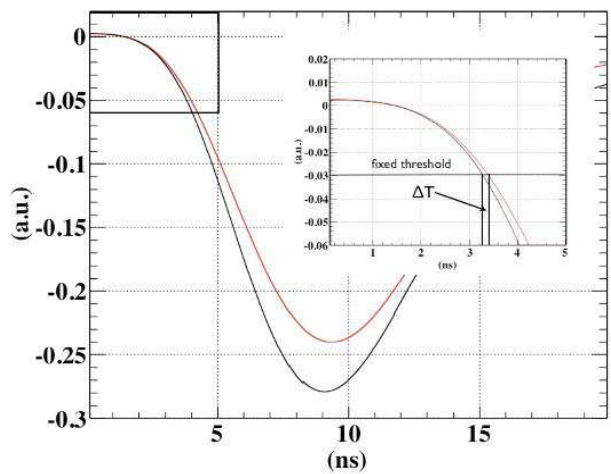}
\caption{Graphical representation of the Time Walk effect.}
\label{fig_timewalk}
\end{figure}

The Time Walk effect is the dependence on the pulse amplitude 
of the threshold-crossing time delay as shown in Fig.~\ref{fig_timewalk}.
The analog waveforms sampled at $1.6\,\mathrm{GHz}$ are averaged and
interpolated separately for each PMT over many events to obtain a
template waveform, used to evaluate the correction
required in time reconstruction. In the leading edge region 
of each template waveform, the dependence of the delay time versus 
the pulse amplitude is fitted with the empirical function: 
\begin{equation}
\label{twx}
TW(x)=A+B\sqrt{x}+C\log{x}
\end{equation}
where $x$ is the ratio of the DTD Low Level Threshold (LLT)
value to the PMT pulse amplitude and {\it A, B, C} are the parameters
fit separately for each PMT. 
These parameters are recalculated every year to account for possible change
in the PMT pulse shape.

\begin{figure}[!ht]
\centering
\includegraphics[width=0.50\textwidth]{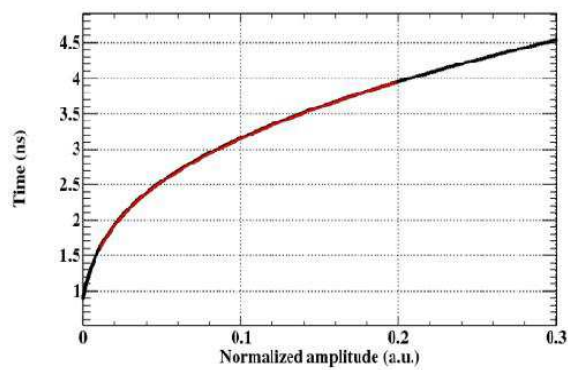}
\caption{Time delay versus amplitude relation with fit superimposed.}
\label{fig_time_amp}
\end{figure}

The correction in Eq.~\ref{twx} is applied on an event-by-event basis.
As shown in Fig.~\ref{fig_time_amp}, this function reproduces well
the experimental data.

The optimal LLT value balancing Time Walk immunity and noise rejection
was found by evaluating the timing resolution of each 
bar with dedicated tests (see Sect.~\ref{sec_reso}).

\begin{figure}[!ht]
\centering
\includegraphics[width=0.50\textwidth]{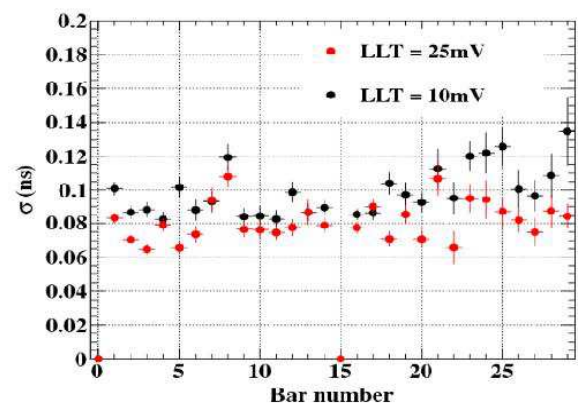}
\caption{Timing resolutions for different values of low-level threshold for
double bar events.}
\label{fig_confronto_low}
\end{figure}

Figure~\ref{fig_confronto_low} shows the comparison between the time
resolutions obtained with two different LLT values (10~mV
and 25~mV) on the double bar sample (see Sect.~\ref{sec:tres}).  
The difference between the time resolutions for the two thresholds
systematically favours 25~mV. 
From systematic studies of the double bar time resolution for LLT 
values ranging from 5~mV to 35~mV in 5~mV step, 
the optimal value of LLT was found to be 25~mV.

\subsubsection{Time offsets }
\label{timealignement}
Inter-bar time offsets due to the electronic chains are evaluated
using cosmic rays.
All PMTs are equipped with 10~m of low-loss signal cable to preserve the 
leading edge of the pulses \cite{hubersuhner}; the equal length 
for all PMTs minimises the possibility of time offsets drifting 
during the data-taking period
(several months) and, together with a
continuous monitoring of the relevant variables, ensures stable
operation. For each bar, the distribution of the time difference
between the inner and outer PMTs is acquired. Due to the uniformity and isotropy
of cosmic rays this distribution is expected to be flat and 
centred at zero; the mean values of these distributions 
are direct measurements of the relative offsets between PMTs on each
scintillating bar. The time offsets between different bars and
between the TC and the LXe detector are measured using positrons from Michel 
decays as well as different calibration approaches \cite{calibration_cw}.

\begin{figure}[!ht]
\centering
\includegraphics[width=0.50\textwidth]{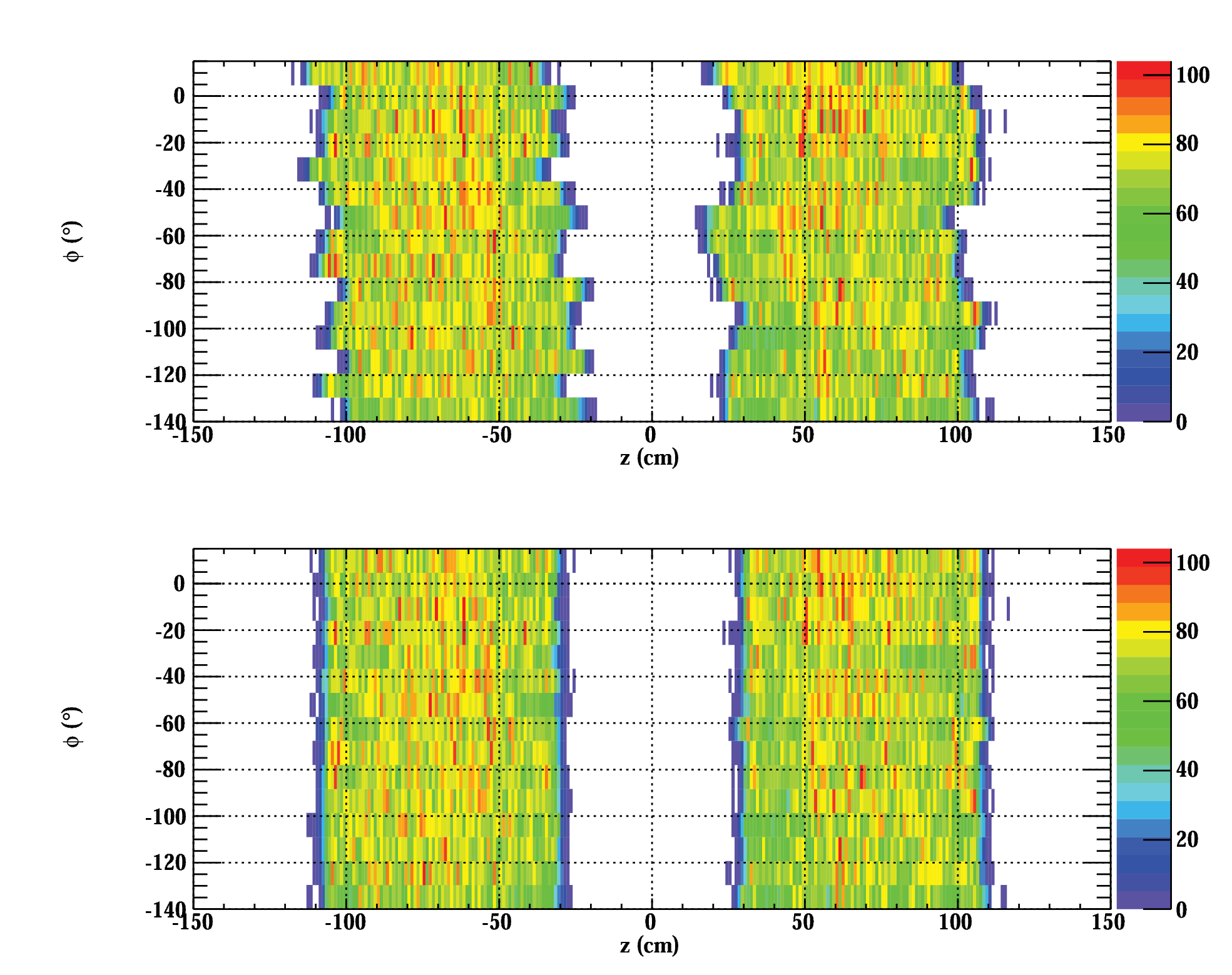}
\caption{Cosmic ray hit maps before and after bar time-offset subtractions.}
\label{fig_hitmap}
\end{figure}

Figure~\ref{fig_hitmap} shows the effect of accounting for time offsets
on a two-dimensional ($z$ versus $\phi$) cosmic ray hit map. 
Using the measured time offsets (lower panel), the $z$ reconstruction 
algorithm returns the correct position.


%

\subsection{Trigger settings}
\label{sect:online}
The TC plays a crucial role in trigger (see Sect.~\ref{sec:trigalgo}):
it detects the positron and  provides a preliminary estimation of its 
timing and direction.

The trigger position resolution is a key parameter for
the direction match with the $\gamma$-ray entering the LXe detector and 
is evaluated by comparing the trigger-level reconstruction of the impact point 
from Eq.\ref{z_charge} with the one obtained 
from Eq.\ref{z_time} in Sect.~\ref{sec:position}. It was $\sigma_{z} \sim 7.3$~cm
in 2009 and has improved to $\sigma_{z} \sim 5.0$~cm since 2010
after refining the algorithms.

\begin{figure}[!ht]
\centering
\includegraphics[width=.40\textwidth,angle=-90]{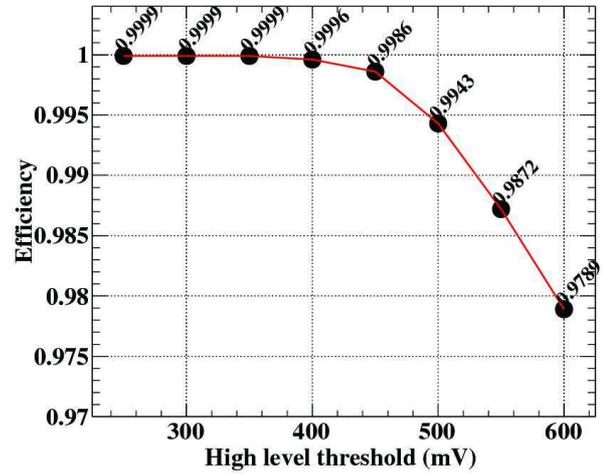}
\caption{DTD efficiency vs high threshold value. The efficiency
 plateau is reached for HT $\le 400$~mV.}
\label{fig_confronto_high}
\end{figure}

From the point of view of efficiency at the analysis level,
the TC trigger and DTD conditional efficiency studies are motivated by
the need for all triggered events to fire the DTD to
allow event reconstruction.
The relevant parameter is the threshold efficiency $\epsilon_{DTD}$
defined, for each bar, as the ratio of the number of events with
NIM signals to the total number of events triggered on the same bar,
The high-level threshold (HLT) is set slightly
lower than the trigger one, in order to acquire only events with
NIM pulse information. Figure~\ref{fig_confronto_high} presents the
efficiency $\epsilon_{DTD}$ versus the HLT. Averaging over all
bars yields $\epsilon_{DTD} \ge 99.9\%$ for 
$\mathrm{HLT} \le 400$~mV.

\subsection{Timing resolution of TC}\label{sec_reso}

If a positron impinges on a TC bar at time $T_{TC}$,
the times measured by the inner (${in}$) and outer (${out}$) PMTs read as:
\begin{eqnarray}\label{tcinouttime}
t_{in}&=&T_{TC}+b_{in}+TW_{in}+\frac{\frac{L}{2}+z}{v_{\mathrm{eff}}} \nonumber
\\
t_{out}&=&T_{TC}+b_{out}+TW_{out}+\frac{\frac{L}{2}-z}{v_{\mathrm{eff}}}
\end{eqnarray}
where $b_{in,out}$ are time offsets depending on the read-out chain,
$TW_{in,out}$ are contributions from Time
Walk effect, $v_{\mathrm{eff}}$ is the effective velocity of light in the bar
and $L$ is the bar length; the $z$ axis points along the main axis 
of the bar and its origin is in the middle of the bar.
From Eq.~\ref{tcinouttime} the impact time is :

\begin{equation} \label{tctime}
T_{TC}=\frac {t_{in}+t_{out}}{2} -  \frac{b_{in}+b_{out}}{2} - \frac{TW_{in}+TW_{out}}{2} - \frac{L}{2v_{\mathrm{eff}}}.
\end{equation}

A more detailed discussion on the determination of the factors $b$ is in Sect.\ref{timealignement}
and of the factors $TW$ in Sect.\ref{sec_timewalk}.

\begin{figure}[!ht]
\centering
\includegraphics[width=3.4in]{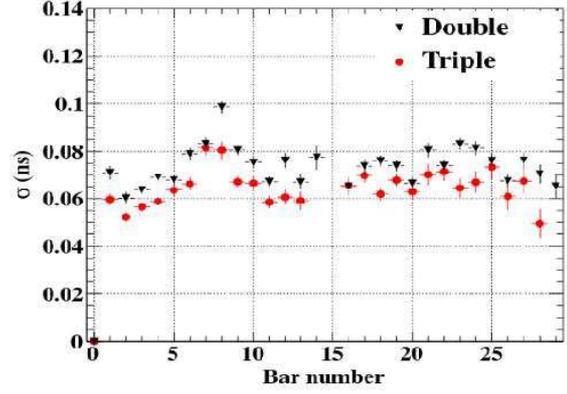}
\caption{Timing resolution on double (black markers) and triple (red
  markers) bar events, with LLT=$25$~mV. Due to the different
  rates for two-bar and three-bar events, the sample used here is
  different than the one in Fig.~\ref{fig_confronto_low}} 
\label{fig_reso}
\end{figure}

\label{sec:tres}
The timing resolution is determined from hits on multiple bars. On a
sample of two consecutive bars hit by a positron, the time difference
$\Delta T = T_{2} -T_{1}$ is studied. For each bar the impact time is
defined as in Eq.~\ref{tctime}. 
The time resolution can be written as
\begin{equation} 
\sigma_{t} =\sqrt{(\sigma^2_{1} + \sigma^2_{2}) + \sigma^2_{track}}
\end{equation}
where $\sigma_{1,2}$ are the single bar time resolutions and
$\sigma_{track}$ the contribution from the track length spread. Under
the hypothesis that the two bars have similar resolutions $\sigma_{1}
\simeq \sigma_{2} $ and neglecting the term due to 
track propagation, the single bar resolution is estimated from 
$\sigma_{t}/\sqrt{2}$. 
This method overestimates the timing resolution since the
track length spread term is not corrected for. Using triple bar
events and evaluating the quantity:
\begin{equation}
\Delta T = T_{2} - \frac{T_1+T_3}{2}
\end{equation}
the effect of the different path length between bars can be removed
at first order, resulting in a better estimate of timing
resolution. This is shown in Fig.~\ref{fig_reso} where the red and black
markers represent respectively
the resolutions obtained in triple and double bar samples.
However, the triple sample has significantly smaller statistics 
so the double bar sample is used as the reference tool for checking the detector
performance. 

A certain degradation of the timing performance between the Beam Test
($\sigma_t \sim 40-50$~ps) and the Physics Run configuration
($\sigma_t \sim 65$~ps) has been observed. A few factors
contribute to this effect: the need for a lower PMT gain in order to
withstand the high rate and match the dynamic range of the
DAQ/electronics chain; a slightly higher value for the low threshold
due to additional noise from the surrounding environment; intrinsic
uncertainties in both double and triple bar estimates; and 
contributions from DRS calibration.

\subsection{Positron timing resolution }\label{sec_postime}

\label{sec:postime}

In the proposal the timing resolution of the positron was assumed to be dominated
by the timing resolution of the TC. That was because of the optimistic assumption 
on tracking performances, as well as the neglect of the contributions of the 
calibration of the DRS boards and of TC counter offsets. Assuming a small 
contribution from the uncertainty on the track length, the design positron timing 
resolution was to be $\sigma_{t_e} = 50$~ps.

In Monte Carlo, which incorporates a fairly precise description of DCH and 
of the positron reconstruction algorithm, an additional contribution due to 
track length fluctuations $\sigma^{L,MC}_{t_e} = 50$~ps is present. Added to the 
TC intrinsic resolution  $\sigma^{TC,MC}_{t_e} = 40$~ps, it totals a positron 
time resolution $\sigma^{MC}_{t_e} = 64$~ps.

In the data, those contributions are evaluated to be  $\sigma^{TC}_{t_e} = 60$~ps 
and $\sigma^{L}_{t_e} = 75$~ps. Additionally, an estimated contribution from DRS 
calibration is  $\sigma^{DRS}_{t_e} = 25$~ps and from TC offset calibration 
$\sigma^{cal}_{t_e} = 40$~ps. The total positron resolution is 
 $\sigma_{t_e} = 107$~ps.

\subsection{Position resolution}\label{pos_tc}

\label{sec:position}
The positron impact point calculated from the time difference between the two pulses 
is obtained from Eq.~\ref{tcinouttime} as
\begin{equation} \label{z_pos}
\label{z_time}
z=\frac{v_{\mathrm{eff}}}{2}\times (t_{in} - t_{out} - (b_{in} - b_{out}) - 
  (TW_{in} - TW_{out}))
\end{equation}

The impact point can also be evaluated using the ratio of the 
charges delivered at the inner and outer PMTs \cite{knoll}: 
\begin{eqnarray} \label{qq}
Q_{in} &=& E\, G_{in} \,
e^{-\frac{\frac{L}{2}+z}{\Lambda_{\mathrm{eff}}}} 
\\
Q_{out}\hfill &=& E\, G_{out} \,
e^{-\frac{\frac{L}{2}-z}{\Lambda_{\mathrm{eff}}}} 
\label{eq:qfrac}
\end{eqnarray}
where $E$ is the energy released inside the bar, $G_{in,out}$ takes
into account several contributions (i.e. the scintillator yield, 
PMT quantum efficiency and gain), $\Lambda_{\mathrm{eff}}$ is the
effective attenuation length of the bar. Taking the ratio we obtain: 
\begin{equation} \label{qratio}
\frac{Q_{in}}{Q_{out}}=\frac{G_{in}}{G_{out}} 
\,e^{-\frac{2z}{\Lambda_{\mathrm{eff}}}} 
\end{equation}
which leads to:
\begin{equation} \label{z_charge}
z=\frac{\Lambda_{\mathrm{eff}}}{2}\left(\ln\frac{Q_{out}}{Q_{in}} -
  \ln\frac{G_{out}}{G_{in}}  \right) 
\end{equation}

Moreover, from Eq.~\ref{qq} the energy release in the bar can be
estimated without dependence on $z$: 

\begin{equation} \label{energy}
\sqrt{Q_{in}\,Q_{out}} = E\sqrt{G_{out}\,G_{in}}
\,e^{-\frac{L}{\Lambda_{\mathrm{eff}}}} 
\end{equation}

\begin{figure}[!ht]
\centering
\includegraphics[width=3.4in]{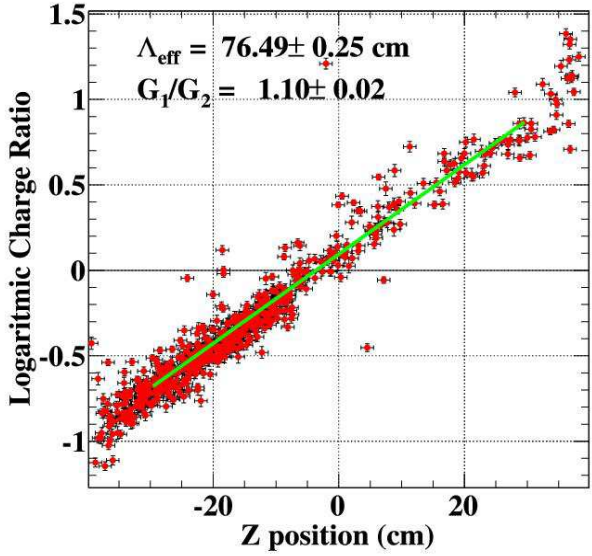}
\caption{Logarithm of the PMT charge ratio vs the $z$ coordinate
  measured using Eq.~\ref{z_pos} for bar number 1. 
The linear fit returns $\Lambda_{\mathrm{eff}}$ and the ratio of PMT
gains.} 
\label{lamqq}
\end{figure}

Note that the combination of Eq.~\ref{z_charge} and Eq.~\ref{z_pos}
provides a way to evaluate the ratio 
$\Lambda_{\mathrm{eff}}/v_{\mathrm{eff}}$ for each TC bar. Assuming
$v_{\mathrm{eff}} = 14.0\, \mathrm{cm/ns}$, the value of
$\Lambda_{\mathrm{eff}}$ is extracted from a linear fit as shown in
Fig.~\ref{lamqq}. The $\Lambda_{\mathrm{eff}}$ values extracted in 2011 for
all bars fall in the range  $40-90\,\mathrm{cm}$. This apparently large dispersion
of values is caused by variation in surface reflectivity due to
surface machining, hand-made polishing and wrapping, while 
the bulk attenuation is expected to be constant between the bars.

Two methods for impact point reconstruction are used in
different stages of the data acquisition chain. The on-line algorithm
for $z$ reconstruction in the  trigger, requiring fast response and 
moderate precision, relies on the charge ratio method, 
while the offline analysis, aiming at the best possible resolution, exploits
the PMT time difference. The need for offline calibrations to guarantee
the ultimate performance (both for $z$ and time resolutions)
is satisfied by using several tools, presented in the following.

\subsection{Efficiency}\label{tceff}
\label{sec:efficiency}

The TC efficiency for Michel and signal positrons emitted from the target can be estimated by Monte Carlo. 
In the acceptance region the efficiency is $\epsilon_{e,TC}^{MC,Michel} \sim 46\%$ for Michel 
($48\, \mathrm{MeV}<E_e$) positrons and $\epsilon_{e,TC}^{MC,Signal} \sim 59\%$ for signal positrons.
Such a low efficiency compared to a design value $>90\%$, is due to the DCH support frame that intercepts 
a large fraction of the positrons
exiting the DCH volume; these interacting positrons may lose a large fraction of their energy or change
their direction by a large angle due to multiple scattering. Therefore they do not reach the TC and 
follow a different trajectory from that predicted by the DCH and hence fail the DCH-TC matching cut.

Using the DCH-only trigger (see Sect.~\ref{sec:trigalgo}), the TC efficiency versus the reconstructed
positron energy was also estimated from data, with results compatible with the Monte Carlo simulation.

The total positron efficiency requires a track in the DCH, a hit in the TC and that the
track matches the TC hit. This requirements amount to an efficiency  $\epsilon_{e}^{MC,Michel} 
\sim 37\%$ for Michel positrons and $\epsilon_{e}^{MC,Signal} \sim 48\%$ for signal positrons,
as estimated by Monte Carlo. A direct measure from data is not possible.

Absolute detection efficiencies are anyway not used in the analysis, as explained in
Sect.~\ref{sec:dchefficiency}, therefore the values reported are just indicative.

%

\section{Liquid Xenon detector}
\label{sec:LXe}
\subsection{Physical requirements vs liquid xenon properties}
\label{sec:LXeproperty}
Liquid Xenon (LXe), with its high density (and short radiation
length) is an efficient detector medium for $\gamma$-rays.
Both ionisation charges and scintillation photons are produced by
radiation. In MEG, only the scintillation photons are used 
to simplify the detector construction 
and to utilise a prompt response of the detector.
The general properties of LXe are summarised in 
Table~\ref{table:LXeproperty}. 

\begin{table}[ht]
\begin{center}
\begin{tabular}{l r}
\hline\hline
Properties & Value \& Unit \\
\hline
Atomic/Mass number & 54 / 131.293 \\
Density at 161.4K & 2.978 g/cm$^3$ \\
Boiling/Melting point & 165.1/161.4 K \\
Radiation length & 2.77 cm \\
Moliere radius & 4.20 cm \\
\hline
Scintillation wavelength & 178 nm \\
$W_{ph}$ for electron / $\alpha$  & 21.6 eV / 17.9 eV \\
Decay time (recombination) & 45 ns \\
Decay time (Fast/Slow component) & 4.2 ns/22 ns \\
Absorption length & $>$100 cm \\
\hline\hline
\end{tabular}
\caption{Liquid xenon properties:$W_{ph}$ is the average energy for scintillation photon,
scintillation wavelength is the wavelength where the scintillation spectrum has its maximum.}
\label{table:LXeproperty}
\end{center}
\end{table}

The MEG $\gamma$-ray detector requires excellent 
position, time and energy resolutions to minimise
the number of accidental coincidences,
which are the dominant background process.
These requirements need high light yield.
Fast decay time helps to reduce pile-up events. 

On the other hand, the peak of the LXe scintillation emission spectrum 
is in Vacuum UltraViolet (VUV), $\lambda \sim$178~nm, which requires 
quartz windows in the PMTs, 
and the operational temperature is
$\sim$165 K, which requires cryogenic equipment.
Furthermore, the
scintillation light can be easily absorbed by impurities 
like H$_2$O, O$_2$, and N$_2$ etc., 
which must therefore be removed efficiently and continuously.

\subsection{Detector Design}
We developed detector prototypes to perform various
feasibility tests. The first used 2.3~$\ell$ of LXe with 32 PMTs,
and the second used 68.6~$\ell$ of LXe with 228 PMTs \cite{baldini_2005_nim}. 
The purpose of
the first prototype was a proof-of-principle detection 
of $\gamma$-rays at $\sim$2 MeV from radioactive sources 
in LXe with adequate resolutions. 
The second prototype was designed to confirm the good resolutions for
position, time and energy for $\gamma$-rays at $\sim$50 MeV and 
to study LXe detector operation including the purification system 
toward a full-size detector. 

\begin{figure}[htb]
\begin{center}
  \includegraphics[width=.45\textwidth,angle=0]{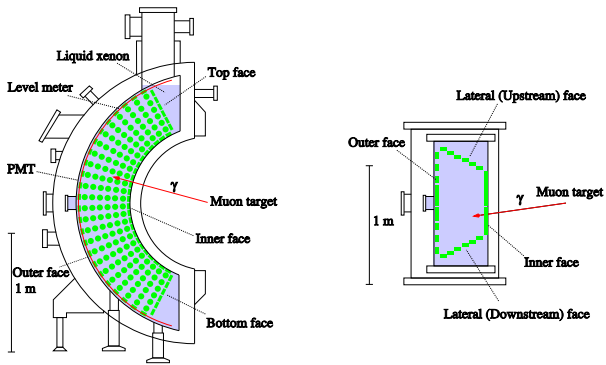}
 \caption[]{Schematic view of the LXe detector: from the side (left), 
from the top (right).}
 \label{XEC:xecdetector}
\end{center}
\end{figure}

Based on the prototype measurements, the final detector design 
employing 900~$\ell$ of LXe was completed in 2004. The construction 
began in 2005 and was completed in late 2007.
A schematic view of the
LXe detector is shown in Fig.~\ref{XEC:xecdetector}.

The C-shaped structure fits the outer radius of COBRA.
The active volume is read out by 
846 PMTs submerged directly in LXe to detect the scintillation light, 
placed on all six faces of the detector (inner, outer, upstream,
downstream, top, and bottom), with different PMT densities. 
The active volume of the detector is $\sim$ 800~$\ell$ and 
covers 11\% of the solid angle viewed from the centre of the 
stopping target. Its depth is 38.5 cm, corresponding to $\sim$ 14
X$_0$ and fully containing a shower induced by a 52.83 MeV
$\gamma$-ray.


 \subsubsection{Cryostat}
\begin{figure}[htb]
\begin{center}
  \includegraphics[width=.45\textwidth,angle=0]{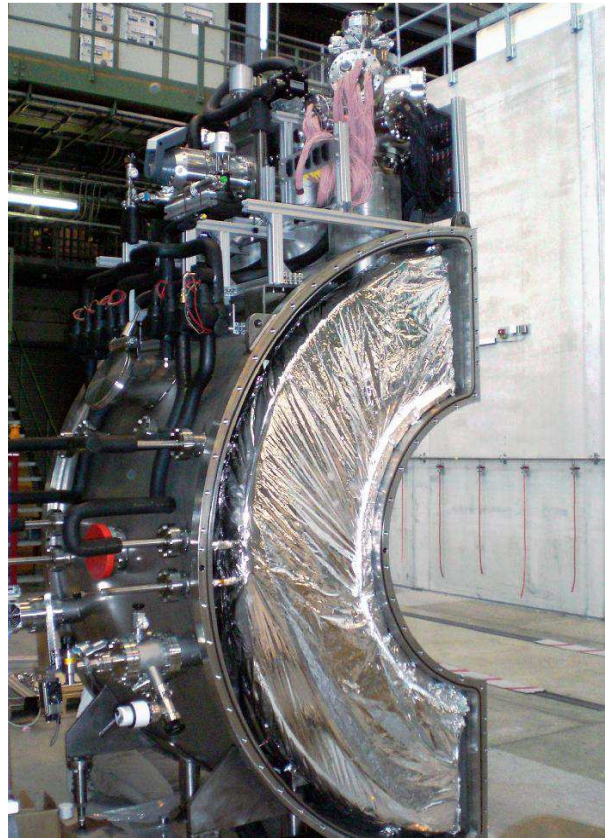}
 \caption[]{LXe detector under construction. A flange of the outer vacuum vessel was not closed yet when this picture was taken.}
 \label{XEC:picturexec}
\end{center}
\end{figure}

The LXe detector consists of the inner and the outer vessels. The inner vessel
holds 900~$\ell$ of LXe and the PMT support structure. 
The outer vacuum vessel is used as a thermal insulation layer.
To reduce the material traversed by incident $\gamma$-rays, the window
of the outer vessel consists of a thin stainless steel plate (0.4 mm thickness),
while that of the inner vessel is made of aluminium
honeycomb panels covered with carbon-fibre plates to withstand pressure up to
3 bar. The total thickness of the $\gamma$-ray entrance window
is 0.075 X$_0$.  A picture of the LXe detector is shown in Fig.~\ref{XEC:picturexec}

In order to monitor the detector condition,
27 temperature, two pressure, and two liquid-level sensors are
installed inside the detector. 
To maintain the LXe temperature and to enable the recondensation of
xenon, a custom-designed 200 W pulse-tube cryocooler 
\cite{haruyama_2004_aip} is installed on the top of the chimney.
It does not produce any electrical
noise or mechanical vibrations, thanks to the cooling principle.
In case more cooling 
power is needed, e.g. during liquefaction or purification, liquid
nitrogen lines are installed to cool down the gaseous volume 
and the  wall of the inner vessel.
From the other two chimneys, the cables of PMT high voltages, signals,
and sensors etc. are extracted via feedthroughs. 
A turbomolecular pump is connected to each vessel to evacuate the
detector, and a cryopump and a getter pump are installed in the
inner vessel to efficiently remove water vapour, which is the most dangerous
impurity.

 \subsubsection{PMTs}
\begin{figure}[htb]
\begin{center}
  \includegraphics[width=.45\textwidth,angle=0]{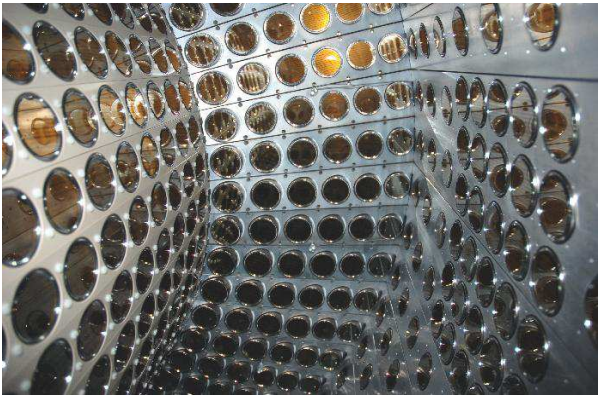}
 \caption[]{View of the LXe detector inside. A total of 846 PMTs are mounted on all surfaces. The face on 
 the left side is the incident one for $\gamma$-rays from the target.}
 \label{XEC:picturepmt}
\end{center}
\end{figure}

In total, 846 2" PMTs \cite{hamamatsu-lxe} are
internally mounted on all surfaces and submerged in LXe; the support structure is shown in Fig.~\ref{XEC:picturepmt}.
In order to allow the operation in LXe, the
PMTs are designed to withstand up to 3 bar. 
The coverage of the inner surface with active photo-cathodes is $\sim$ 35\%.
The PMT uses a metal-channel dynode structure to achieve a reasonable
gain inside a short package (the height is 3.2 cm) and a high
tolerance for magnetic fields as large as $\sim 10^{-2}$ T \cite{Mihara:2011zza}, directed either
along the transverse or the longitudinal direction.
The PMTs are operated at LXe temperature $\sim$~165 K, and are equipped 
with a quartz window that transmits VUV photons and a bialkali (K-Cs-Sb) photo-cathode
sensitive to VUV photons. Aluminium strips are added to the surface of the cathode
to reduce the sheet 
resistance, which increases at low temperature. 
Heat dissipation from the base circuit is minimised by optimising
the resistor chain (16 M$\Omega$) with Zener diode protection in the
last two dynodes, which keep the voltage constant under a high
counting rate. 
Typical PMTs show Quantum Efficiency (QE) $\sim 15\%$ 
and gain $\sim$ 1.8$\times$10$^6$ at 850 V.
Inside the cryostat, 3.0 m plus 1.6 m RG-196A/U coaxial cables are used for the PMTs signal, 
and 3.0 m RG-188A/U coaxial and 1.6 m wires for HV. 
These are connected to the feedthroughs. 
Outside the cryostat, signal cables are connected to the splitter boards (see Sect.~\ref{sec:splitter}),
and divided into trigger (see Sect.~\ref{trigger}) and DRS4 boards (see Sect.~\ref{DAQ:drs4}) which record
all the PMT waveforms.  


 \subsubsection{Purification System}
As discussed in Sect.~\ref{sec:LXeproperty}, scintillation light may be
absorbed by impurities in LXe, such as H$_2$O and O$_2$ at ppm level.  
We have developed two purification methods,
gas-phase \cite{baldini_2005_nim} and liquid-phase \cite{mihara_2006_cryogenics}, 
to remove those impurities.
Gas-phase purification removes impurities
in Gaseous Xenon (GXe) by means of a metal-heated getter.
A diaphragm pump is used to circulate GXe without introducing any additional impurities. 
Although the gas-phase purification successfully reduces such impurities,
its circulation speed is limited ($\sim$~0.6 $\ell$/h).
It turned out during the gas-phase purification study that H$_2$O was 
the dominant component for the absorption. 

Liquid purification was developed 
to improve the circulation speed and to remove mainly H$_2$O from LXe 
by using a cryogenic centrifugal fluid pump \cite{barber} and molecular sieves (MS13A). 
At the normal working point, the flow rate is $\sim$~70 $\ell$/h.
Molecular sieves can absorb more than 24~g of water, and the cartridge contains heaters
which enable a regeneration of the cartridge. 

 \subsubsection{Storage System}
In addition to the detector, two storage systems for LXe
and GXe have been developed so that 1000~$\ell$ of xenon can be
stored safely when the detector is not operated 
\cite{Iwamoto:2009}.
One is a 1000~$\ell$ Dewar with a pulse-tube cryocooler as well as liquid
nitrogen cooling lines and a thermal
insulation layer. The heat
income is estimated to be less than 20 W, and the cryogenic tank is
designed to tolerate a pressure up to 6 bar. As a
result, LXe can be stored safely without supplying
any cooling power for 100 h.
The Dewar is connected to the detector with flexible tubes
thermally insulated to allow a rapid LXe transfer.

\begin{figure}[htb]
\begin{center}
  \includegraphics[width=.5\textwidth,angle=0]{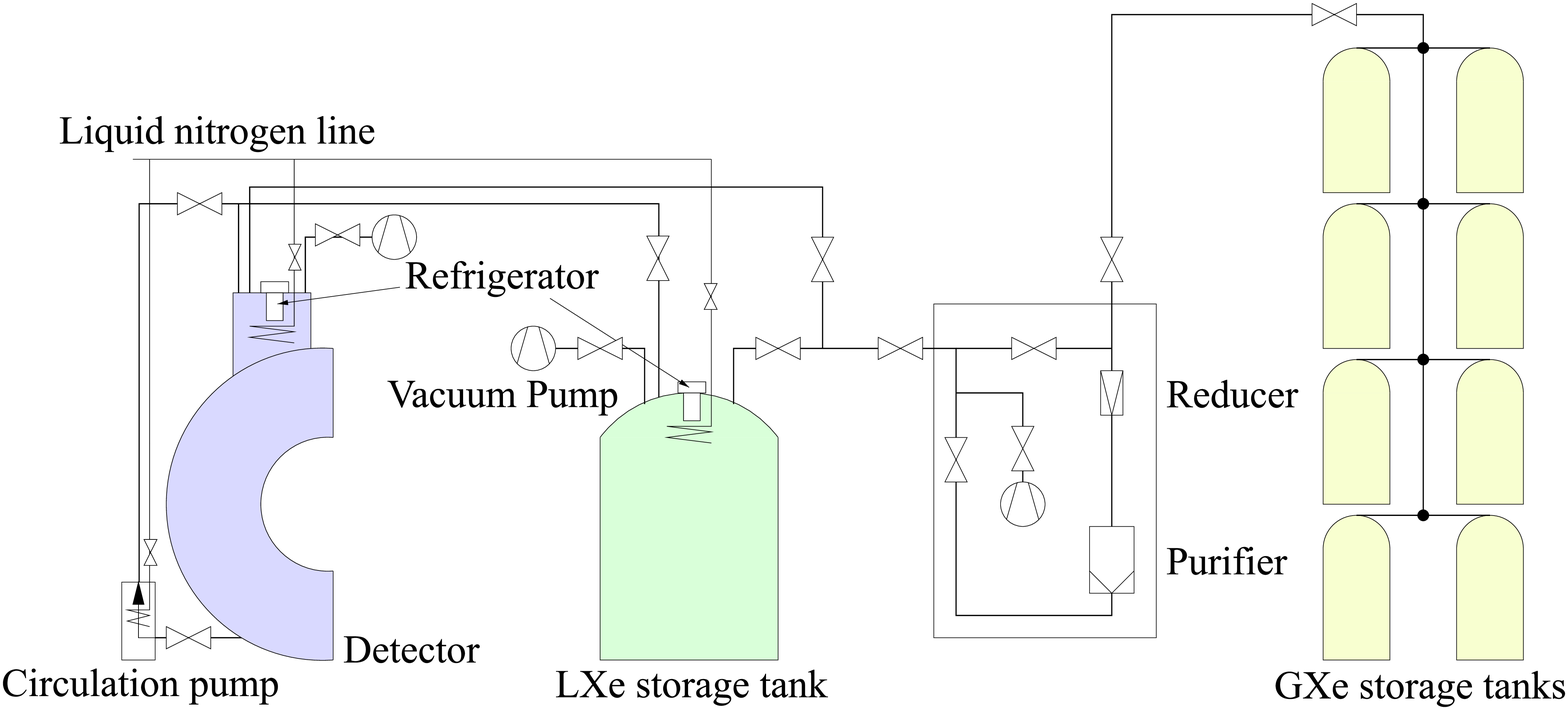}
 \caption[]{Schematic view of LXe storage and purification systems as
   well as of the detector.}
 \label{XEC:xecstorage}
\end{center}
\end{figure}

The other storage system consists of eight high-pressure gas tanks to
store GXe, each of which can contain up to 360 kg of xenon corresponding
to 120~$\ell$ of LXe. This can be used in long shutdown
periods. 
In Fig.~\ref{XEC:xecstorage}, a schematic view of the LXe detector
system including purification and storage systems is shown.
 
 \subsubsection{Detector Operation}
Before starting detector operation, the cryostat is
evacuated by means of turbomolecular and getter
pumps to minimise impurities inside the detector.
Meanwhile, GXe is liquefied
into the 1000~$\ell$ storage tank. LXe can be transferred from the
tank to the detector by means of the pressure difference between 
the pressurised tank and the detector. 
After the liquid transfer is complete, the pressure of the detector is
controlled to be 1.2--1.4 bar by means of the pulse-tube
cryocooler. 
If the cooling power is too much, heaters surrounding the
cryocooler are used to reduce the cooling power, and if the cooling power 
is not sufficient, liquid nitrogen is used to sustain the pressure.
In 2010, the
pulse tube cryocooler was sufficient to keep 900~$\ell$ of LXe in
the detector, and $\sim$30 W heaters are constantly used to produce
additional heat.
In case of a problem, or during a short maintenance period, we can 
transfer the LXe from the detector into the 1000~$\ell$ tank by means
of the centrifugal pump. 

\subsection{Reconstruction methods}
The reconstruction methods and algorithms are described.
In the analysis of the LXe detector, a special coordinate system {\it (u,v,w)} is used:
{\it u} coincides with {\it z} in the MEG coordinate system; {\it v} is directed along the 
negative $\phi$-direction at the radius of the inner face ($r\mathrm{_{in}}$=67.85 cm),
which is the direction along the inner face from bottom to top; $w = r-r\mathrm{_{in}}$,
measures the depth from the inner face.

 \subsubsection{Waveform Analysis}\label{sec:waveform}
\begin{figure}[htbp]
  \begin{center}
    \mbox{\begin{tabular}[t]{ll}
    \subfigure[]{\includegraphics[width=.95\linewidth] {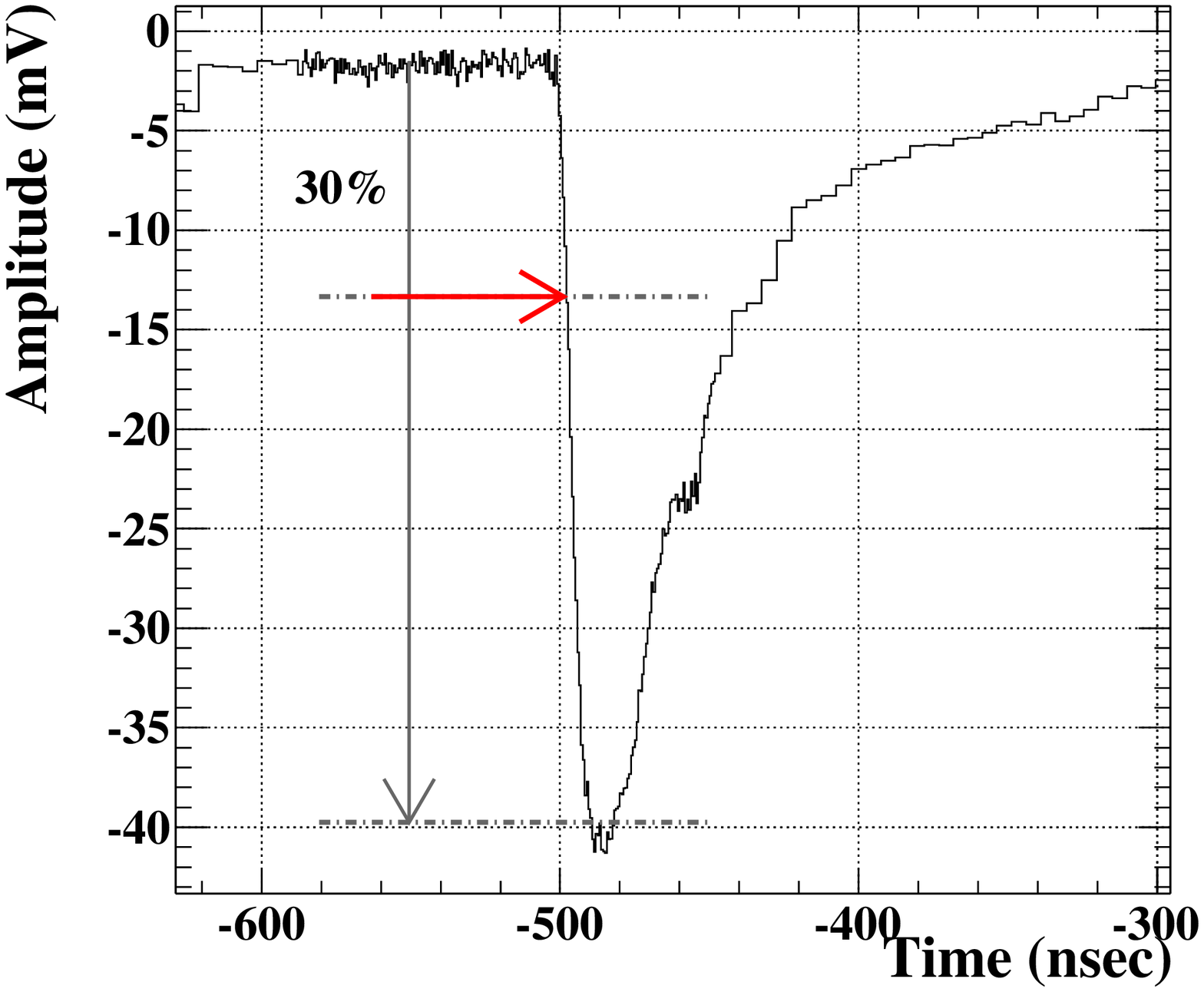}
    \label{XEC:rawwaveform}
    }  \\
    \subfigure[]{\includegraphics[width=.95\linewidth] {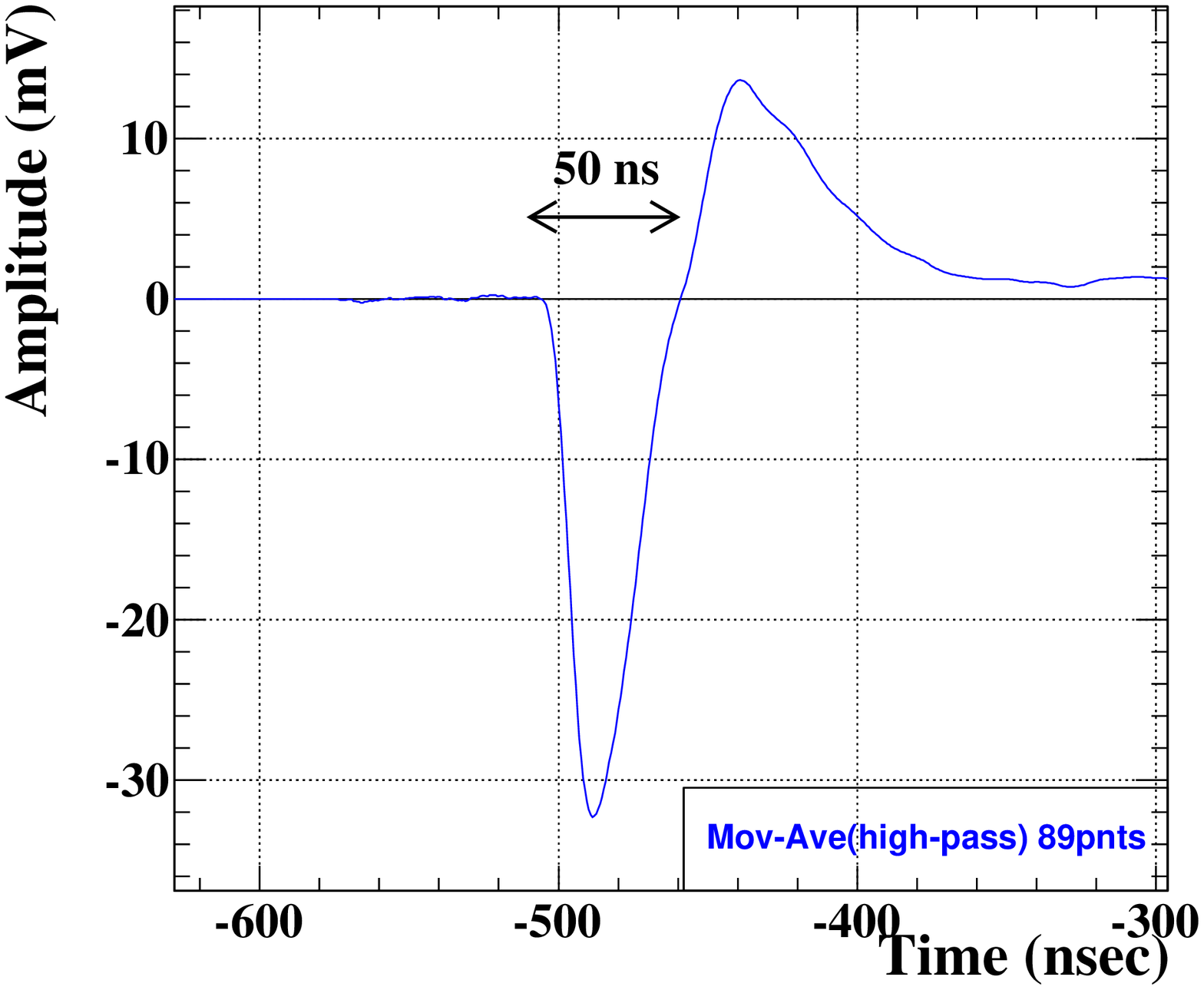}
    \label{XEC:filterwaveform}
    } \\
    \end{tabular}}
    \caption[]{a) An example of a PMT raw waveform from the waveform digitiser.
               b) An example of a waveform with high-pass filter.}
  \end{center}
\end{figure}

We extract the pulse time and charge from each waveform. 
A typical waveform is shown in Fig.~\ref{XEC:rawwaveform}.
The baseline is
estimated by averaging the points in the region before the pulse 
for each channel on an event-by-event basis. The digital constant 
fraction method is used to
determine the pulse time, defined as the time when the signal
reaches a predefined fraction (30\%) of the maximum pulse height. 
To maximise the signal-to-noise ratio for the determination of the
charge, a digital high-pass filter, based on a moving-average method, 
is applied as shown in Fig.~\ref{XEC:filterwaveform}. 
The charge is estimated by integrating the filtered PMT pulse
and is later converted into the number of photoelectrons. 
The number of average points is 90--100,
corresponding to about a 10 MHz cutoff frequency.  

%
%

\subsubsection{Position}
The $\gamma$-ray conversion point is reconstructed by using
the distribution of the light detected by the PMTs near the incident
position. The three-dimensional position is determined by fitting the
expected light distribution on the PMTs, calculated from the solid angles,
to the observed distribution.  
To minimise the effect of shower fluctuations, the fit is done
twice: the first with $\sim$45 PMTs and the second with $\sim$15 
PMTs around the first fit result.
The remaining bias on the reconstructed position along {\it u} and 
{\it w} coordinates is corrected using results from the Monte Carlo simulation.
The $\gamma$-ray direction is defined by the line segment between its
reconstructed conversion point in the LXe detector and the 
reconstructed positron vertex on the target.

 \subsubsection{Energy}
 \label{sec:lxeerec}
The reconstruction of the $\gamma$-ray energy E$_\gamma$ is based on the sum
of the number of photons detected by the PMTs corrected for their
geometrical coverage, which depends on the PMT locations:
the coverage on the outer face is 2.6 times less than that on the
inner face. This correction works well for almost all events except
for very shallow ones ($w_\gamma<3$~cm), which require special treatment
because they are very sensitive to the relative positions
of each PMT and the conversion point. In this case, 
the solid angle subtended by each photo-cathode 
at the conversion point can be a better variable than the coverage 
for estimating the collection efficiency. 

\begin{figure}[htb]
 \centerline{\hbox{
  \includegraphics[width=.45\textwidth,angle=0]
{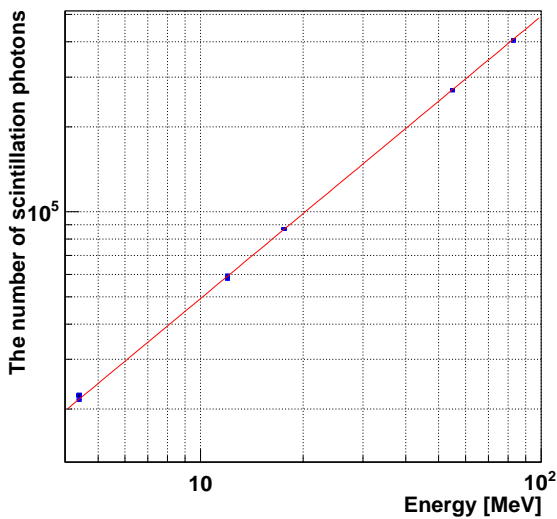}
 }}
 \caption[]{The number of detected scintillation photons versus the incident $\gamma$-ray energy:
 4.4, 11.6 and 17.6 MeV $\gamma$-rays are measured by C--W 
 calibration (see Sect.~\ref{sec:accelerator}), 
 while 55 and 83 MeV $\gamma$-rays by the CEX calibration (see Sect.~\ref{sec:cex}).
 Error bars assigned to data points are too small to be visible.
}
 \label{XEC:linearity}
\end{figure}

In Fig.~\ref{XEC:linearity}, the linearity of the relation between 
E$_\gamma$ and the number of detected scintillation photons is shown. 
The conversion factor from the number of photons to energy in order to define an absolute
energy scale is obtained from the 55 MeV $\gamma$-ray peak taken by the Charge EXchange
(CEX) calibration (see Sect.~\ref{sec:cex}).

 \subsubsection{Timing}
 \label{sec:lxetime}
Each PMT determines the time of arrival of the scintillation light from
the waveform analysis described in Sect.~\ref{sec:waveform}. 
To relate this time to the $\gamma$-ray conversion time, the 
propagation time of the scintillation light must be subtracted
as well as any hardware-induced time offset (e.g. cable length).
The scintillation light propagation time is subtracted by exploiting the distance 
between the reconstructed interaction point and each PMT 
assuming an effective light velocity $\sim$8 cm/ns.
After the subtraction of these effects, the $\gamma$-ray conversion time is 
obtained by combining the timings of those PMTs which contain more
than 50 photoelectrons and calculating the minimum value of $\chi^2$
defined as a sum of the squared difference between each
calculated and reconstructed time. Typically $\sim$150 PMTs are
used to reconstruct 50 MeV $\gamma$-rays. 
The PMTs with large contributions to $\chi^2$ are rejected
during this fitting procedure to remove pile-up effects.

Finally, the reconstructed time is corrected by the time-of-flight
between the positron reconstructed vertex  
on the target and the reconstructed conversion point in the
LXe detector.

\subsection{Performance}
The LXe detector performance is described in the following and 
summarised in Table~\ref{table:LXesummary}.

 \subsubsection{Position resolution}
\begin{figure}[htb]
 \centerline{\hbox{
  \includegraphics[width=.45\textwidth,angle=0]
{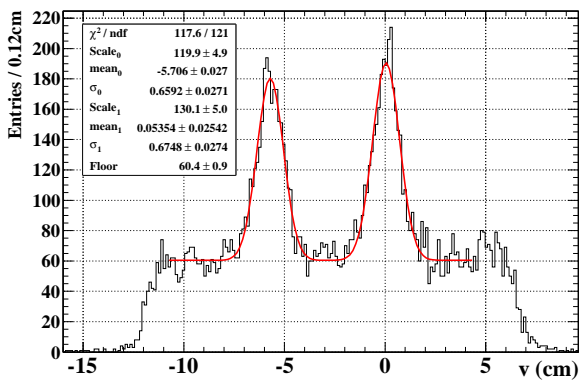}
 }}
 \caption[]{Reconstructed position distribution with a lead collimator in CEX runs. 
   There are two slits 1~cm wide in the 1.8~cm thick collimator.}
 \label{XEC:positionspectrum}
\end{figure}

The performance of the position reconstruction is evaluated by a Monte Carlo
simulation and is validated in dedicated CEX runs by
placing lead collimators at several positions in front of the LXe
detector. Figure~\ref{XEC:positionspectrum} shows the result of position
reconstruction with a 1.8~cm thick lead collimator installed in front
of the LXe detector during the CEX run. 
The lead collimator has two 1~cm wide slits along the {\it u} coordinate at 
$v=0$~cm and $v=6$~cm, and this figure shows the projection onto the $v$ 
coordinate. Events with the energy between 50~MeV $<E_\gamma<80$~MeV, and with the 
$\gamma$ conversion depth $w_\gamma>2$~cm are selected.
Since the thickness of the lead collimator is not sufficient to stop 55 MeV $\gamma$-rays,
the floor events which are penetrating the lead collimator are also observed.
A double Gaussian function plus a constant term is fitted
to extract the position resolution of 54.9MeV $\gamma$-rays, and the results are 6.6~mm, 6.7~mm in this example.
The average resolution at different positions is 6.9~mm.
This result contains the effect of the slit width itself and of the spread 
of $\pi^{0}$ decay points, and is to be compared with the average resolution (6.5~mm) 
from a Monte Carlo simulation of the same configuration.
The quadratic difference (1.8~mm, expected to come from the PMT QEs 
calibration uncertainty) between the data and the Monte Carlo simulation
is added into the position resolution map built from simulation.

Taking into account the difference between Monte Carlo simulation and the data, 
the average position resolutions are estimated to be 
$\sigma_{(u_\gamma,v_\gamma)}\sim$5~mm and $\sigma_{w_\gamma} \sim$6~mm, respectively, 
comparable with the design position resolutions. They are also close to
$\sigma^{MC}_{(u_\gamma,v_\gamma)}\sim 4$~mm and $\sigma^{MC}_{w_\gamma}\sim 6$~mm 
estimated by Monte Carlo simulation.

 \subsubsection{Energy resolution}
\label{sec:elxe}
The energy response of $\gamma$-rays at the signal energy is extracted from 
the CEX calibration. A small correction is applied to take into account the different background
conditions between the muon and the pion beams, and the opening angle between the two $\gamma$-rays.

\begin{figure}[htb]
 \centerline{\hbox{
  \includegraphics[width=.45\textwidth,angle=0]
{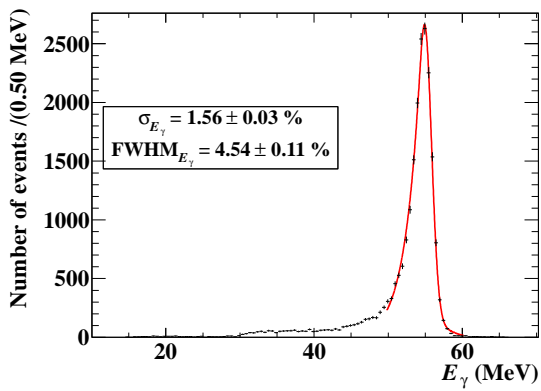}
 }}
 \caption[]{Energy response of the LXe detector to 54.9~MeV $\gamma$-rays for $w_\gamma>2$ cm in a
 restricted range of $(u_\gamma,v_\gamma)$. The fitting function is described in the text. 
 The resolution is $\sigma_{E_\gamma}=1.56$\% and FWHM$_{E_\gamma}$ = 4.54\%.}
 \label{XEC:energyspectrum}
\end{figure}

The response function of the detector for monochromatic $\gamma$-rays is 
asymmetric with a low-energy tail due to mainly two reasons. 
One is the interaction of $\gamma$-rays in the material in front of the LXe
active volume, and the other is the shower leakage from the front face.
Figure~\ref{XEC:energyspectrum} shows the LXe detector response to
54.9 MeV $\gamma$-rays. 
The distribution is fitted with an asymmetric function ${\it F}(x)$
convolved with the pedestal distribution ${\it h}(x)$ in the CEX
run. ${\it F}(x)$ is given by
\begin{displaymath}
 F(x) =
 \left\{
 \begin{array}[tb]{ll}
  A\exp({-\frac{(x-x_0)^2}{2\sigma^2_{E_\gamma}}}) &  (x>x_0+\tau),\\
  A\exp({\frac{\tau}{\sigma^2_{E_\gamma}}(\tau/2-(x-x_0))}) &(x\leq x_0+\tau),
 \end{array}
 \right.
\end{displaymath}
where $A$ is a scale parameter, $x_0$ is a peak position parameter, $\tau$ is
a transition parameter and $\sigma_{E_\gamma}$ is a resolution parameter 
indicating the distribution width on the high-energy side.
Since ${\it F}(x)$ shows the intrinsic resolution of the detector
without pedestal contribution, it can be used for any realistic
environment with a different pedestal distribution.

A 3-dimensional mapping of these parameters is incorporated into the
likelihood function for the final analysis since they depend on 
the position of the $\gamma$-ray conversion, mainly on its $w_\gamma$ coordinate. 
As an example, the average energy resolution is measured to be
$\sigma_{E_\gamma}=1.6$\% (3~cm $<w_\gamma$), 2.0\% (0.8~cm $<w_\gamma<$ 3~cm) and 2.7\% 
(0~cm $<w_\gamma<$ 0.8~cm) in 2011. Except for the acceptance edge along $v$ coordinate
($\sigma_{E_\gamma}\sim 2.5$\% with $|v|>68.2$ cm and $w_\gamma>$3~cm), 
the energy resolution depends weakly on the $u_\gamma$ and $v_\gamma$ coordinates.
This number is to be compared with the energy resolution 
of $\sigma^{MC}_{E_\gamma} = 1.2\%$ for (2~cm $<w_\gamma$ evaluated by Monte Carlo simulation.
The reason of this slightly worse resolution is not fully understood.
The behaviour of PMTs such as gain stability, angular dependence etc., or optical properties
of LXe such as convection might be possible sources.

The design resolution was $\sigma_{E_\gamma} = 1.7\%$ over all $w_\gamma$.

\subsubsection{Timing resolution}

To investigate the intrinsic time resolution of the LXe detector due to 
photoelectron statistics,
two PMT groups (even PMT IDs or odd PMT IDs) are defined, and the times of 
the same event are reconstructed by these two groups independently.
Then the intrinsic time resolution is estimated by the time difference between these 
two results, being dominated by photoelectron statistics, 
while the effects of electronics, the position reconstruction and the event-by-event 
shower spread cancel out resulting in $\sigma^{phe}_{t_\gamma} = 36$~ps at 55~MeV. 

\begin{figure}[htb]
 \centerline{\hbox{
  \includegraphics[width=.45\textwidth,angle=0]
{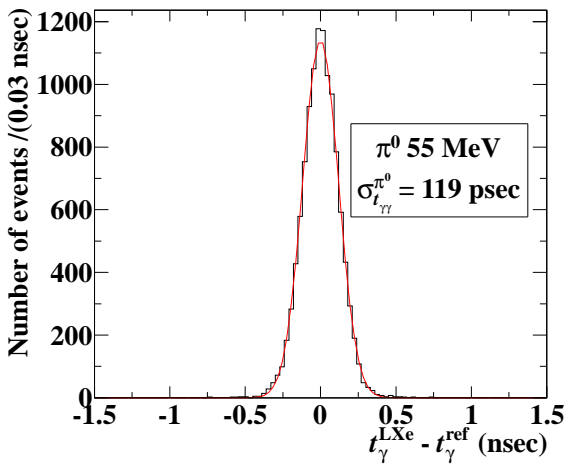}
 }}
 \caption[]{Time difference reconstructed by the LXe detector and a reference preshower counter
 for 54.9~MeV $\gamma$-rays. }
 \label{XEC:timingspectrum}
\end{figure}

\begin{figure}[htb]
 \centerline{\hbox{
  \includegraphics[width=.45\textwidth,angle=0]
{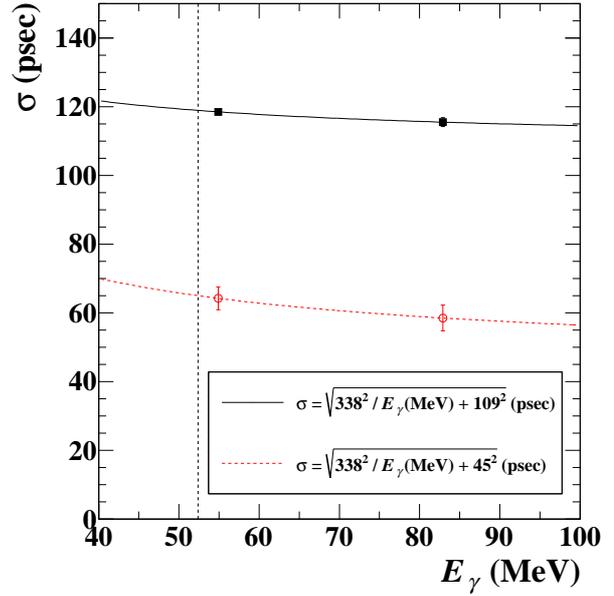}
 }}
 \caption[]{Energy-dependent timing resolution of the LXe detector (see text for details).}
 \label{XEC:timeres_energy_dep}
\end{figure}

The absolute timing resolution is evaluated from the time difference
between two $\gamma$-rays, emitted back-to-back from the $\pi^0$ decay during
CEX calibration (see Sect.~\ref{sec:cex}), reconstructed by the LXe detector 
and by a reference preshower counter. 
Figure~\ref{XEC:timingspectrum} shows the measured time difference distribution 
having a resolution of $\sigma^{\pi^0}_{t_{\gamma\gamma}} = 119$~ps, which includes not 
only the LXe detector timing 
resolution but also the contributions due to the uncertainty of the $\pi^0$ decay 
position and to the reference counter. 
The former is evaluated to be $\sigma^{\pi_0}_{t_\gamma} = 58$~ps by 
the $\pi^{-}$ beam spread ($\sim 8\times 8$~mm$^{2}$ beam spot size).
The reference counter has two plastic scintillator plates and both sides are 
read out by 2" fine mesh PMTs. The timing resolution of each plate is estimated 
by the time difference between the two PMTs, and the timing resolution of the 
counter is estimated by these resolutions by taking into account the correlation 
of the plates.
Finally, the resolution of the reference counter is evaluated to be 
$\sigma^{ref}_{t_\gamma} = 81$~ps.
The absolute timing resolution of the LXe detector is estimated by subtracting 
these contributions, resulting in $\sigma_{t_\gamma} = 65$~ps at 55~MeV.
Figure~\ref{XEC:timeres_energy_dep} shows the energy-dependent timing resolution 
of the LXe detector.
Black squares show the measured timing resolution $\sigma^{\pi^0}_{t_{\gamma\gamma}}$, 
while the red circles show the LXe detector timing resolution $\sigma_{t_\gamma}$.
The black and the red smooth 
curves are the fit results, and their functions are shown in the figure. 
A vertical dotted line shows the $\gamma$-ray signal energy (52.83~MeV). 
The timing resolution improves at higher energy, which indicates that 
the photoelectron statistics still contributes significantly.
This small energy dependence is taken into account in extracting the timing 
resolution for signal $\gamma$-rays to obtain $\sigma^{LXe}_{t_\gamma} =67$~ps. 
This number is in good agreement with the timing resolution of 
$\sigma^{LXe,MC}_{t_\gamma} =69$ ps evaluated by Monte Carlo simulation.

The breakdown of the time resolution is as follows: 
\begin{equation}
\sigma^{LXe}_{t_\gamma} = \sigma^{phe}_{t_\gamma} \oplus \sigma^{ele}_{t_\gamma} \oplus 
\sigma^{TOF}_{t_\gamma} \oplus \sigma^{sho}_{t_\gamma}
\end{equation}

where $\sigma^{phe}$ is defined above, 
$\sigma^{ele}_{t_\gamma}= 24$~ps from electronics contribution, 
$\sigma^{TOF}_{t_\gamma}= 20$~ps from the $\gamma$-ray time of flight uncertainty
(which corresponds to depth reconstruction uncertainty),
and $\sigma^{sho}_{t_\gamma}= 46$~ps from position reconstruction uncertainty
and the shower fluctuation.

The final resolution on the $\gamma$-ray timing $\sigma_{t_\gamma}$ is obtained combining 
$\sigma^{LXe}_{t_\gamma}$ with the additional $\sigma^{tar}_{t_\gamma}$ spread due to the uncertainty in
the muon decay vertex on the target as measured by extrapolating the positron
at the target plane (see Sect.~\ref{sec:vertex}). This spread is no more than 
$\sigma^{tar}_{t_\gamma} \sim 5$ ps giving a negligible contribution.
The results are $\sigma_{t_\gamma} = 67$ ps and $\sigma^{MC}_{t_\gamma} = 69$ ps.

These results are to be compared with the design resolution $\sigma^{LXe}_{t_\gamma} \sim 43$ ps,
that was calculated taking into account approximately only the contribution from 
position reconstruction uncertainty. Taking into account also the uncertainty 
in the muon decay vertex the design resolution is again $\sigma_{t_\gamma} \sim 43$ ps.

\subsubsection{Detector uniformity and energy scale}

A spatial non-uniformity of the detector, due mainly to its geometry, 
is measured by means of the C--W accelerator 
(see Sect.~\ref{sec:accelerator}) and CEX (see Sect.~\ref{sec:cex})
and corrected to equalise the detector response.

The $E_\gamma$ scale is constantly monitored by looking at the
reconstructed 17.6~MeV energy peak from photons produced by C--W protons on Li 
(see Sect.~\ref{sec:accelerator}).
The $E_\gamma$ scale and its resolution are cross-checked by fitting the  $\gamma$-ray
background spectrum 
to a theoretical spectrum 
obtained with a Monte Carlo simulation including radiative muon decay, positron annihilation-in-flight and
$\gamma$-ray pile-up, folded with the detector resolution (see Sect.~\ref{sec:LXebackground}). 
The energy scale and the
energy resolution are free parameters in the fit, and we can compare the fit results
with the calibrated energy scale and the measured energy resolution in CEX.
The systematic uncertainty on the energy scale 
on the basis of these measurements is $\sim$0.3\%, in which the dominant source is 
from the detector uniformity correction ($\sim$0.2\%).

\subsubsection{Rejection of pile-up events}
 \label{sec:pileupandcr}
It is important to recognise and unfold pile-up events in a high muon 
intensity environment in order to avoid a loss of signal efficiency. 
At $3\times 10^7\,\mu^+$/s beam rate,
around 15\%\ of triggered events suffer from pile-up.
The $\gamma$-ray pile-up signature is identified by three main methods, 
the shape of the waveform sum of all PMTs,
the charge distribution of the inner and outer face PMTs,
and the $\chi^{2}/NDF$ distribution in the time reconstruction. 

The first method identifies multiple $\gamma$-ray events with different timing
by searching for peaks in the summed waveform of all PMTs. 
If a pile-up $\gamma$-ray is detected by this method, 
its contribution is subtracted by using a template waveform, 
and the waveform of the main $\gamma$-ray is used for energy estimation.

The second method identifies multiple $\gamma$-ray events in different positions 
by searching the PMT charge distribution on the inner and 
outer faces for spatially separated peaks. If the event has two peaks, the 
main peak is retained after removing the secondary $\gamma$-ray. 
A pile-up removal method was developed, to accomplish this efficiently.
A position-dependent table containing the average charge of each PMT 
in presence of 17.6~MeV $\gamma$-rays is prepared beforehand. 
Once a pile-up event is identified, 
the energy of the event is estimated by fitting the PMT charges to the table 
without using PMTs around the secondary $\gamma$-ray. 
Then, the PMT charges around the secondary $\gamma$-ray are replaced with 
the charges estimated by the fit. Finally, the energy is 
reconstructed with the replaced PMT charges as
described in Sect.~\ref{sec:lxeerec}.

In the third method multiple $\gamma$-ray events with different timings
are identified using the algorithm of the first method.
If an event has multiple $\gamma$-rays, 
the $\chi^{2}/NDF$ in the time reconstruction becomes large. 
Although the PMTs with large contributions to the $\chi^{2}$ are discarded when the $\gamma$-ray 
timing is reconstructed (see Sect.~\ref{sec:lxetime}), all the PMTs observing 
more than 50 photoelectrons are used to identify pile-up events effectively. 
If the pile-up is not identified in the summed waveform or in the light distribution,
and is only identified in the timing distribution, the event is discarded.
Similarly, if the pile-up is only found in the light distribution 
but the subtracted energy is negative or larger than 10\%\ of
the total energy, the event is discarded as well. 
This cut has a 98\% efficiency, estimated by the time sideband data
around the signal region.

\subsubsection{Rejection of cosmic ray events}

\begin{figure}[htb]
 \centerline{\hbox{
  \includegraphics[width=.45\textwidth,height=.35\textwidth]
   {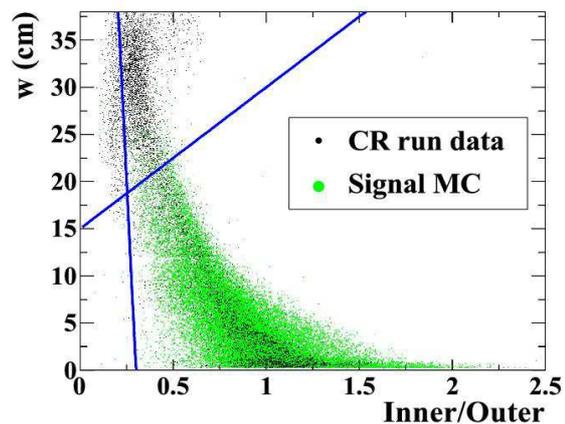}
 }}
 \caption[]{Two-dimensional plot of the charge ratio collected on the inner/outer faces 
 vs the $\gamma$-ray interaction depth. Black points show the data from 
 dedicated cosmic ray runs, and green ones the signal Monte Carlo events. 
 The two lines (blue in the online version) show the selection criteria for rejection of cosmic ray events. }
 \label{XEC:crcut}
\end{figure}

Cosmic ray events are rejected using topological cuts because these events 
mostly enter the detector from the outer face. Therefore the ratio of the charges collected 
on the inner and the outer faces of the detector 
is smaller in cosmic ray events than in muon-beam events, which enter the detector from the inner face.
The reconstructed depths of cosmic ray events are significantly larger 
than those of signal $\gamma$-rays for the same reason. 
Fig.~\ref{XEC:crcut} shows the two-dimensional scatter plot of the charge ratio 
vs depth. 

In this two-dimensional parameter space, the selection criteria are defined to maximise 
cosmic ray rejection efficiency while keeping a signal efficiency of 99\%. The cut discards 
56\% of cosmic ray events.

The combined analysis efficiency of this and the pile-up cuts is 97\%.

\subsubsection{$\gamma$-ray background spectra}
\label{sec:LXebackground}

\begin{figure}[htb]
 \centerline{\hbox{
  \includegraphics[width=0.45\textwidth,angle=0]
  {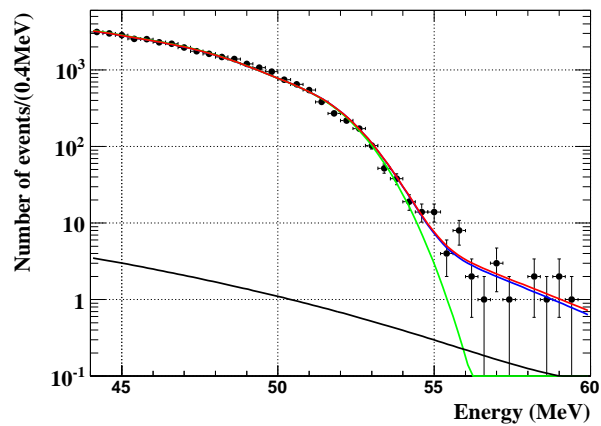}
 }}
 \caption[]{$\gamma$-ray background spectrum during physics run in 2010 
(see text for details). }
 \label{XEC:gammabackground}
\end{figure}

In Fig.~\ref{XEC:gammabackground} the $\gamma$-ray background data are shown, while
the green line shows the fitted spectrum from the
Radiative Muon Decay (RMD) plus positron Annihilation-In-Flight (AIF) spectrum,
the blue line the $\gamma$-ray RMD + AIF spectrum taking into account the effect of pile-up,
the black line the cosmic ray spectrum,  
and the red line the fit result taking into account all the background spectra.
In the 51--55~MeV region with $w_\gamma>2$ cm, the RMD + AIF spectrum accounts for 94--96\% 
of the total background spectrum,
cosmic ray spectrum for less than 1\%, 
and the remaining components from pile-up, energy response with higher energy tail, etc. for 3--5\%. 

\subsubsection{Detection efficiency}
The fiducial volume of the LXe detector is defined as $|u|<$ 25~cm, $|v|<$ 71~cm 
and 0~cm $<w_\gamma<$ 38.5~cm. 
The $\gamma$-ray detection efficiency is estimated using the Monte Carlo
simulation, and is confirmed by the measurement in CEX runs and by the
$\gamma$-ray background rate in physics runs. 
The low-energy tail in Fig.~\ref{XEC:energyspectrum} comes from
interactions with the material in front of the active volume and shower
escape from the inner face. 
On the path of a $\gamma$-ray entering the LXe active volume, 
there are the drift chamber frame, its support structure, 
the COBRA magnet, an entrance window of the LXe cryostat, the PMTs 
and their support structure, etc.
The largest source of inefficiency is due to the COBRA magnet, 14.2\%.
Within the analysis window 48.0 MeV $<E_{\gamma}< 58.0$~MeV, 
the average efficiency in Monte Carlo simulation is found to be $\epsilon_\gamma =65\%$, 
taking into account the positron event distribution.
This efficiency was measured using NaI single trigger data in the CEX 
run. By tagging an 83~MeV $\gamma$-ray from $\pi^0$ decay with the NaI
detector, the detection efficiency of the detector for
55~MeV $\gamma$-rays is found to be  $\epsilon_\gamma = 64-67\%$,
consistent with the Monte Carlo estimation. 

Taking into account the analysis efficiency of 97\% as discussed in Sect.~\ref{sec:pileupandcr}), 
the combined $\gamma$-ray analysis and detection efficiency is
$\epsilon_\gamma = 0.65\times0.97 = (63\pm3)$\%.
This value is to be compared 
with an expected value $\epsilon_\gamma \sim 60\%$.

\begin{table}[ht]
\begin{center}
\begin{tabular}{l r}
\hline\hline
 Energy resolution [$w_\gamma>3$ cm] & 1.6\%  \\
 Energy resolution [$0.8<w_\gamma<3$ cm] & 2.0\%  \\
 Energy resolution [$0 <w_\gamma<0.8$ cm] & 2.7\% \\
Timing resolution & 67 ps \\
Position resolution & $5(u_\gamma,v_\gamma)$, $6(w_\gamma)$ mm\\
Detection efficiency & 63\% \\
\hline\hline
\end{tabular}
\caption{LXe detector performance summary.}
\label{table:LXesummary}
\end{center}
\end{table}

%
%
\section{Calibrations}
\label{sec:calsec}


For each kinematic variable it is necessary to know the resolution, supplemented by the absolute 
scale (for the energies) or the position of the zero (for the relative time and direction) 
and the degree of stability of these parameters over the experiment lifetime.
We refer to the knowledge of a single quantity at a given time as ``calibration'', 
and call the ensemble of these periodically repeated calibrations ``monitoring''. 

Continuous monitoring of the apparatus is essential for two reasons: firstly for
early detection of misbehaviour of the apparatus and secondly the knowledge of the 
parameters has a direct influence on the systematic uncertainties on  physical variables, 
acceptances and thresholds.

The issues of calibration and monitoring were incorporated into the design of MEG at an 
early stage, so that several auxiliary devices were designed, assembled and integrated 
for this purpose~\cite{papa_2010}. 

The measurement of positron and $\gamma$-ray kinematic variables can be obtained from one 
detector (the $\gamma$-ray energy and positron momentum) or 
from the combination of the information from several 
detectors ({\em e.g.} the relative time and direction). 
Hence we are prepared to deal with calibration of each detector separately, or of 
multiple detectors simultaneously.


Firstly the devices involved in the calibration of the LXe detector are presented:
LEDs and point-like $\alpha$-sources inside the LXe detector, a movable AmBe source and 
a 9.0~MeV $\gamma$-ray generator made of a nickel/polyethylene sandwich coupled to a 
pulsed neutron generator.

Then the calibration of the spectrometer by means of Mott scattering of positrons
is described.

Finally the devices that allow the simultaneous calibration of multiple detectors:
the 1~MeV Cockcroft--Walton (C--W) accelerator and
the liquid hydrogen and NaI/BGO setup for the Charge EXchange (CEX) reaction
are discussed.

A complete calibration cycle is composed of LED, $\alpha$-source, lithium, boron, 
neutron generator runs. It takes about three hours to complete, and it is 
usually performed two/three times per week.

\begin{figure}
\begin{center}
\includegraphics[width=0.70\columnwidth]{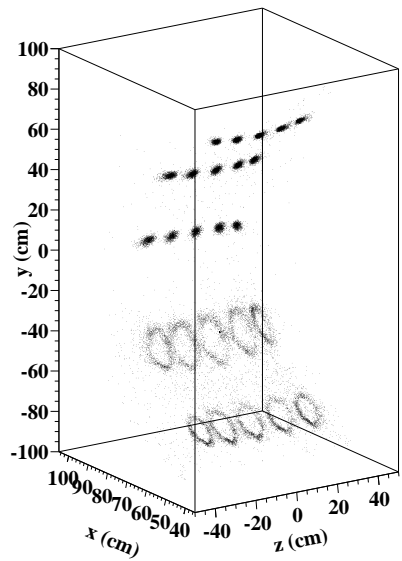}
\caption{\label{fig:alpha2008}Reconstructed position of the 25 $\alpha$-sources inside the LXe detector.
In this particular run the detector was partially filled with liquid xenon. The different reconstructed
shape of $\alpha$-sources in gas and in liquid xenon is visible (see text for details).
}
\end{center}
\end{figure}

\subsection{LXe detector calibration}

Unlike the spectrometer, where the magnitude of the positron momentum is determined uniquely by the knowledge of the 
geometry of the DCH and of the magnetic field, the value of the $\gamma$-ray energy must be 
extracted from the number of photons detected by the PMTs surrounding the LXe volume, once the proper proportionality 
factor is known. 
This factor contains the LXe light yield, the photo-cathodic coverage, the 
PMT gains and the Quantum Efficiencies (QE). All these quantities may depend on time: some 
more than others (e.g. the purity of LXe may change, some PMTs may be switched off for some runs).

\begin{table*}[th]
\caption{Typical calibrations performed to determine the 
LXe detector performance (energy scale, linearity, etc.) with their energy
range and frequency.}
\begin{center}
{\begin{tabular}{cccc}
\hline
\multicolumn{2}{c}{\bf Process}  & {\bf Energy(MeV)} & {\bf Frequency} \\
\hline
Charge exchange & $\pi^- p \to \pi^0 n$  & $54.9,82.9$ & yearly \\
& $ \pi^0 \to \gamma \gamma$ & & \\[1.2mm]
Charge exchange & $\pi^- p \to n \gamma$  & $129.0$ & yearly \\
Radiative $\mu^+$ decay  & \radiative & 52.83 endpoint & weekly \\[1.2mm]
Proton accelerator & $^7 {\mathrm Li} (p, \gamma_{17.6(14.8)}) ^8 {\mathrm Be}$ & 14.8, 17.6 & weekly \\[1mm]
& $^{11} {\mathrm B} (p,  \gamma_{4.4}\gamma_{11.6}) ^{12} {\mathrm C}$ & 4.4, 11.6 & weekly \\[1.2mm]
Nuclear reaction & $^{58} {\mathrm Ni}(n,\gamma_{9.0}) ^{59}{\mathrm Ni}$ & 9.0 & daily \\[1.2mm]
AmBe source & $^9\mathrm{Be}(\alpha_{^{241}\mathrm{Am}},n)^{12}\mathrm{C}_*$ & 4.4 & daily \\
& $^{12}\mathrm{C}_*\to ^{12}\mathrm{C} \gamma_{4.4}$ & & \\[1.2mm]
\hline
\end{tabular}}
\label{tab:calibrations} 
\end{center}
\end{table*}

For this reason a number of calibration lines are available to check the energy 
scale over the full energy range. Low-energy 
calibrations by means of radioactive sources are easier and performed more 
frequently. Although they are of limited use to set the absolute energy scale in 
the signal region, they are helpful in finding gross variations of LXe purity. 
At the opposite end, $54.9~$MeV $\gamma$-rays from $\pi^0$ decays make possible
to directly measure the detector response, uniformity and resolution close to 
the signal energy. A drawback of this calibration method is the need to change the beam polarity 
and momentum (from $\mu^+$ to $\pi^-$) and the usage of a liquid hydrogen target.

Table~\ref{tab:calibrations} presents a list of these lines, which span a broad energy range:
\begin{enumerate}
\item In the low-energy region $4.4$~MeV $\gamma$-rays from an AmBe source and 
$5.5$~MeV $\alpha$-particles from $^{241}$Am sources deposited on thin wires are used to 
monitor the PMT QEs and the LXe optical properties on a daily basis. 
In addition, 9.0~MeV $\gamma$-rays from capture by nickel of thermalized neutrons produced by a neutron
generator are also available 
(see Sect.~\ref{alphaambe}--\ref{neutron}).
\item In the intermediate-energy region a C--W accelerator is used, 
two/three times per week, to accelerate protons, in the energy range 400--900~keV, 
onto a Li$_2$B$_4$O$_7$ target. 
$\gamma$-rays of 17.6~MeV energy from $^7 {\mathrm Li} (p, \gamma_{17.6}) ^8 {\mathrm Be}$ monitor the LXe detector energy scale 
and resolution, while time-coincident 4.4~MeV and 11.6~MeV $\gamma$-rays 
from $^{11} {\mathrm B} (p, \gamma_{4.4}\gamma_{11.6}) ^{12} {\mathrm C}$ are used to intercalibrate the relative timing of the LXe detector
with the TC (see Sect.~\ref{sec:accelerator}).
\item In the high-energy region
measurements of $\gamma$-rays from $\pi^0$ decays produced by the $\pi^-$ 
CEX reaction in a liquid hydrogen target are performed once/twice a year
(see Sect.~\ref{sec:cex}).
\end{enumerate}

\subsubsection{LEDs and gain evaluation}
\label{sec:led}
PMT gains are estimated by using blue LEDs immersed in the LXe at
different positions. In total, 44 LEDs are installed in the detector.
To minimise the position dependence of the detected photons,
11 LEDs are flashed simultaneously. In dedicated gain measurement runs,
performed every second day on average, 
LEDs are flashed at several different
intensities, and the PMT gains are evaluated from the photoelectron
statistics following the method described in \cite{baldini_2005_nim}.
There are two kinds of known PMT gain instabilities, 
a long-term gain decrease, and a rate-dependent gain shift. 
The former is typically 0.1\%/day during physics 
runs and 0.4\%/day during CEX calibration runs (see Sect.~\ref{sec:cex}).

The latter is observed when starting to use the $\mu$ beam, the
typical shift being $\sim +2$\%.
It was found that the LED light intensity was stable enough to check the
long-term stability of each PMT gain.
To monitor and correct long-term instabilities, constantly flashing LED data have
been taken during physics runs since 2009 at a rate of $\sim$0.2~Hz: the LED peak 
position for each PMT, and its variation with the time, is used to bridge the PMTs 
gain variation between two consecutive dedicated gain measurement runs.

\subsubsection{Point-like $\alpha$-sources}
\label{alphaambe}
A calibration technique based on a lattice of $^{241}$Am point-like 
$\alpha$-sources was developed and applied, for the first time, 
in a prototype LXe detector~\cite{Baldini:2006}. 
The sources are prepared by fixing small portions of a $^{241}$Am 
foil to a gold-plated tungsten wire by a thermo-compression method. 
The resulting wire diameter, after the source mounting, is $< 150~\mu$m.

In the LXe detector we mounted five wires, each one hosting five point-like 
sources of $\sim 1$~kBq/source, for a total activity of $\approx 25$~kBq. 
The wires are positioned in a staggered fashion to optimise 
the range of angles and distances from which they are viewed from the PMTs (see
Fig.~\ref{fig:alpha2008}).

In GXe the sources are reconstructed as 3-dimensional spots of $\sim 1$~cm diameter 
(the range of $\alpha$-sources in GXe is $\sim 8$~mm). In LXe, due to 
the much smaller range ($\sim 40~\mu$m, comparable to the wire 
diameter) part of the scintillation light impinges on the wire itself, 
which projects a shadow on the PMTs opposite to
the $\alpha$-source emission direction.
This produces reconstructed positions of  $\alpha$-sources in the form of 
rings surrounding each wire. 
It is still possible to identify all $25$~sources and furthermore one can 
select the preferential $\alpha$-source emission direction ({\em e.g.} 
towards the front face) by applying a topological cut on the rings.

Dedicated $\alpha$-source calibration runs are collected on a daily basis. 
Thanks to their different pulse shapes in LXe, 
the trigger is 
able to classify each event as either low-energy $\gamma$-ray or
$\alpha$-source also during normal data taking (beam-{\sc on}).

The constant monitoring of the combined peak from all $\alpha$-sources is used to control 
the light yield and transparency of LXe.
A comparison of the observed light distribution on each PMT with the one 
obtained by a detailed Monte Carlo simulation results in a measurement of 
the LXe attenuation length~\cite{baldini_2005_nim}. This has turned out to be much larger than the dimension
of our detector during all physics runs ($\lambda_{\rm Abs} > 300$~cm).
  
\subsubsection{Quantum Efficiency evaluation}

The relative QE of each PMT is also measured by using the $\alpha$-sources.
LEDs are not suitable for this task since the LED wavelength
is different from that of LXe scintillation light, and the
PMT response depends on the wavelength.

QE\footnote{The measured QE is actually the product of the photocathode
quantum efficiency times the collection efficiency of the first dynode.
The distribution of the measured QE of all PMTs in in the ($10 - 20$)\% range.}
is first evaluated from the observed $\alpha$-peak charge compared 
with that from simulation \cite{Baldini:2006} and then
corrected by using the $\gamma$-rays from Li peak events
(see Sect.~\ref{sec:accelerator})
to minimise systematic uncertainties related to the
$\alpha$-source positions and angles between sources and PMTs.
The measured time dependence of the QE of each PMT has also been corrected for.

\subsubsection{Americium/Beryllium source}
\begin{figure}
\centering
\includegraphics[width=0.99\linewidth]{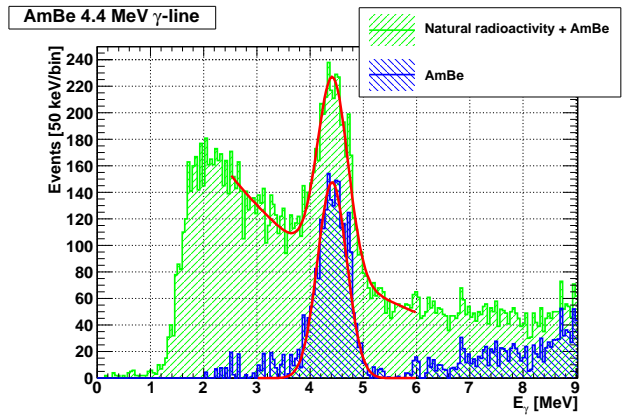}
\caption{The AmBe $\gamma$-ray spectrum measured by the LXe detector. A Gaussian and exponential fit is superimposed on the green spectrum. A Gaussian fit is used after subtracting the exponential natural radioactivity background.}
\label{fig:ambe_spectrum}
\end{figure}

Since LXe has a different response for $\alpha-$particles and $\gamma$-rays,
A low-energy calibration point (see Fig.~\ref{fig:ambe_spectrum}) is provided by 
$4.4$~MeV $\gamma$-rays from an $^{241}$Am/Be source. 
A Gaussian fit (blue spectrum) is used to extract the peak position and the energy resolution.
A 50~kBq AmBe source \cite{ambeez} stored 
in a lead repository can be moved, by means of a compressed air circuit, 
to the front of the LXe detector. 
This allows a rapid check of the LXe light yield and of the correct 
time constants of the scintillation waveforms of 
$\alpha-$particles and $\gamma$-rays in the same energy range. 
This calibration is possible only during beam-{\sc off} periods due to the 
high rate of low-energy $\gamma$-rays present during normal data taking. 
In the latter condition the method described in the Sect.~\ref{neutron} is used.
%
%
%

\subsubsection{Neutron Generator}
\label{neutron}
\begin{figure}
\centering
\includegraphics[width=0.70\linewidth,angle=-90]{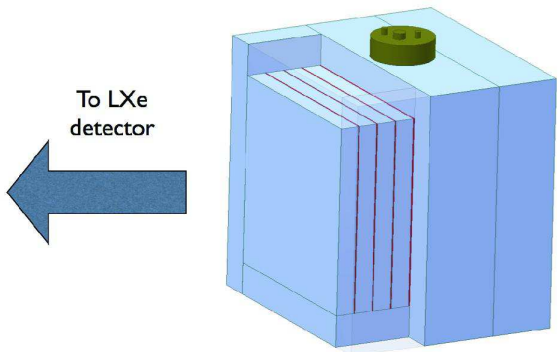}
\caption{The nickel slabs (red), moderator (blue) and neutron generator (green).}
\label{fig:Nigeom}
\end{figure}

A 9.0~MeV $\gamma$-ray line from capture of thermal neutrons in nickel 
$^{58}$Ni$(n,\gamma_{9.0})^{59}$Ni by using a Neutron Generator (NG), allows 
a rapid and frequent calibration and 
monitoring of the LXe detector in normal (beam-{\sc on}) conditions.
The LXe detector response can be studied when 
electromagnetic radiation associated with the beam (radiative muon decay 
and annihilation in flight) illuminates the detector. It is therefore 
appropriate for checking the stability of the LXe detector behaviour in case 
of variable $\mu$-beam intensity. 
Neutrons of
2.5 MeV kinetic energy are produced by a pulsed D-D neutron generator, based on the 
$d(d,^3$He)$n$ nuclear reaction, with a tunable frequency (1-100~Hz) and intensity 
(up to $2.5\times10^4\, n/$pulse)~\cite{Thermofisher}.
The neutron generator is placed at the centre of a polyethylene block placed 
in front of the LXe detector. The calorimeter-facing side consists of alternating 
slabs of nickel and polyethylene. 
A large fraction of the neutrons are thermalized in polyethylene and then captured 
by nickel nuclei. The sandwich structure is optimised to allow a large fraction of 
the $9.0$~MeV $\gamma$-rays from the capture of thermal neutrons to reach the LXe detector 
(see Fig.~\ref{fig:Nigeom}).
This structure and the materials are able to accommodate several requirements: 
to minimise the unthermalized neutron flux, to avoid background from fast 
neutron interactions in the detector, and to minimise the thermal neutron flux out of the 
moderator-nickel box in order to avoid material activations. 

Nickel was selected instead of other materials because of its relatively 
large probability ($\sim 34\% $\cite{Nidatabase}) of emitting a single 
capture $\gamma$-ray at 9.0~MeV.

The pulsed mode of the NG is crucial to reach a good signal-to-noise 
ratio when the muon beam is operated.
Thermal neutron capture events follow the NG emission by a typical 
average delay of 50--100 $\mu$s.
Therefore a special ``neutron generator'' trigger is used, in which a 100~$\mu$s 
gate is opened $15~\mu$s after the NG pulse. In this way the uncorrelated 
$\gamma$-rays from the beam are reduced, and prompt fast neutron interactions
are suppressed. The neutron generator intensity is tuned in such a way as to have 
only one photon per pulse on average in the LXe detector.

\begin{figure}
\centering
\includegraphics[width=0.99\linewidth]{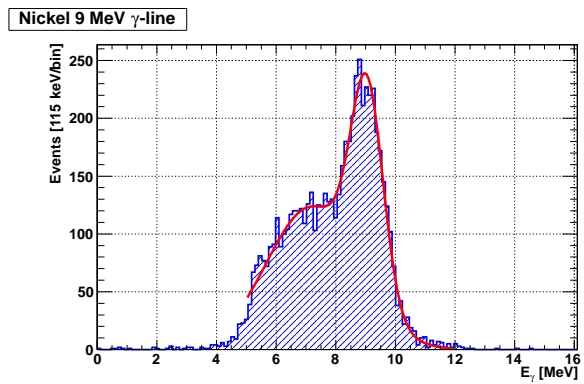}
\caption{The 9.0 MeV $\gamma$-ray line from neutron capture in $^{58}$Ni as reconstructed in LXe detector.}
\label{fig:Nispectrum}
\end{figure}

\begin{figure}
\centering
\includegraphics[width=0.99\linewidth]{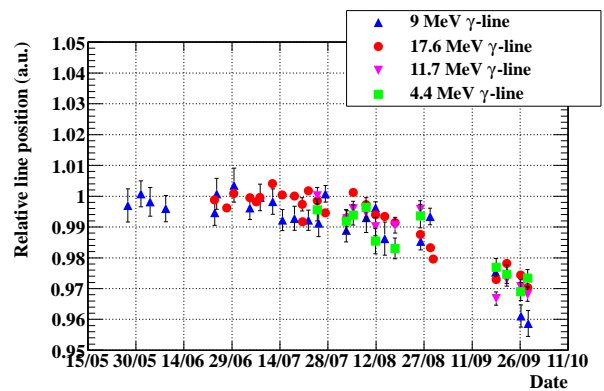}
\caption{The relative position of the NG 9.0 MeV $\gamma$-ray line compared with other $\gamma$-ray line relative positions, as a function of time.}
\label{fig:NiandCW}
\end{figure}

A typical $\gamma$-ray spectrum is shown in Fig.~\ref{fig:Nispectrum}. A double Gaussian fit is used to extract the peak position and energy resolution. 
Besides the 9.0~MeV line, the $\gamma-$ray continuum from other Ni level is visible above threshold.
Data are acquired 
with the muon beam {\sc on} and {\sc off}. No energy offset is observed between these data sets, 
if the gain calibration with beam on/off is applied, proving that the LXe detector 
is well calibrated. 
Figure~\ref{fig:NiandCW} shows the consistency of this method with the other ones.
\subsection{Monochromatic positron beam}
\label{sec:mott}
The abundant positron component of the MEG beam (eight times more intense than the 
$\mu^+$ component, but normally separated and rejected) is used for selecting a 
monochromatic positron beam. The positrons hit the MEG target and are scattered 
into the COBRA spectrometer by coherent Mott scattering on carbon.

This method has several potentialities which were extensively studied with MC simulations:
 
\begin{itemize}
\item A measurement of the momentum and angular resolutions of the spectrometer 
with trajectories (double-turn tracks) similar to the ones of the  signal.
\item A measurement of the muon polarisation based on the comparison 
between the angular distribution of positrons from muon decays at rest and that of the Mott scattered positrons.
\item A measurement of the spectrometer acceptance.
\item An independent check of the spectrometer alignment.
\end{itemize}

\begin{figure}
\centering
\includegraphics[width=0.99\linewidth]{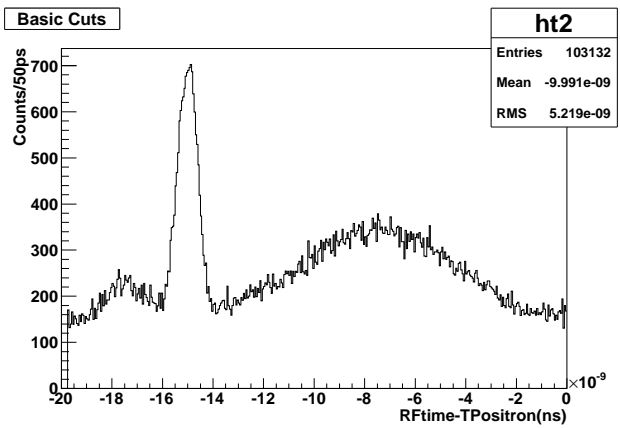}
\caption{Positron timing distribution relative to the accelerator 50 MHz RF-signal (see text).}
\label{fig:RF}
\end{figure}

A preliminary experimental study of the positron beam properties and of Mott scattering events 
was performed. The average momentum of Mott events is centred close to the incident beam positron 
momentum of 53 MeV/c with a measured one-sigma spread of 450 keV/c, this being largely due 
to the minimum achievable momentum-slit opening.

\begin{figure}
\centering
\includegraphics[width=0.99\linewidth]{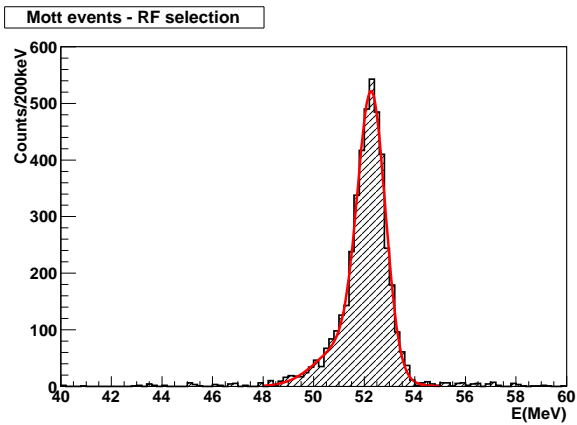}
\caption{The momentum distribution of RF-selected Mott events.}
\label{fig:Mottbasic+RF}
\end{figure}

The incident positron beam, associated with prompt production in the proton target, is bunched and 
synchronous with the accelerator 50 MHz RF-signal. This signal can be used to measure the 
time-of-flight of particles from the production target to the spectrometer and so distinguish beam 
positrons from other backgrounds, as shown in Fig.~\ref{fig:RF}.
Here a sharp peak due to beam positrons is seen at a relative time of $-15$ ns within the modulo 
20 ns timing window (50 MHz). The broad peak on the right is associated with Michel positrons 
originating from cloud-muons interactions in the MEG target. These muons are produced from pion 
decay-in-flight in the surrounds of the proton production target. The smaller peak on the left 
is consistent with the timing expected from pions produced in the production target. Finally 
the constant background is associated with Michel positrons originating from muon decays after 
the Wien filter.
With appropriate timing cuts imposed, the selected Mott positron line is shown in Fig.~\ref{fig:Mottbasic+RF}
with superimposed a fit with two Gaussian,
yielding a momentum resolution of 450 keV/c (one sigma) on the high-energy side.

\subsection{Intercalibration of detectors}

\subsubsection{The Cockcroft--Walton accelerator}
\label{sec:accelerator}

A 1 MeV proton C--W accelerator~\cite{cwhve}
is coupled to MEG for calibrating and monitoring the performance of the LXe detector, 
the relative inter-bar timing of the TC and the relative timing between the TC and 
LXe detector~\cite{calibration_cw,papa_2007}. 
Its properties are listed in Table~\ref{tab:comp}.

\begin{figure}
\begin{center}
\includegraphics[width=0.9\columnwidth]{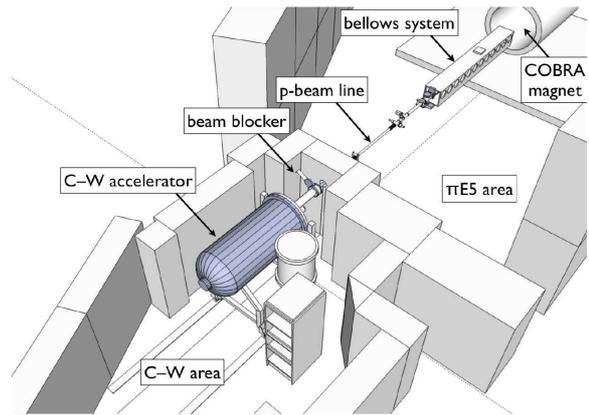}
\caption{\label{fig:areaCW}Schematic layout of the area where the Cockcroft--Walton 
accelerator is placed, with respect to the $\pi E 5$ area.}
\end{center}
\end{figure}

The C--W accelerator is placed in a separate and independently radiation-surveyed
area. in which it can be serviced and tested (see Fig.~\ref{fig:areaCW}).

Calibration measurements are carried out with $\gamma$-rays from two different 
nuclear reactions: $^7 {\mathrm Li} (p, \gamma_{17.6}) ^8 {\mathrm Be}$ produces monochromatic 
$17.6$~MeV $\gamma$-rays and a broad resonance centred at $14.8$~MeV;
$^{11} {\mathrm B} (p, \gamma_{4.4}\gamma_{11.6}) ^{12} {\mathrm C}$ produces a pair of simultaneous $\gamma$-rays of $4.4$~MeV and 
$11.6$~MeV. The two reactions have a threshold at a proton energy of $440$~keV and 
$163$~keV, with a peak cross section of $\sim 6 \times 10^{-3}$~barn
and  $\sim 2 \times 10^{-4}$~barn, respectively.
To increase the number of generated $\gamma$-rays, especially for the reaction on boron,
protons of energy higher than the threshold are used.
These protons are captured during the slow-down process in the target, leading to
an increase in the event rate without 
increasing the $p$-beam intensity.

The proton beam energy is therefore set to $500$~keV for $^7 {\mathrm Li} (p,\gamma_{17.6}) ^8 {\mathrm Be}$ and to $900$~keV 
for $^{11} {\mathrm B} (p, \gamma_{4.4}\gamma_{11.6}) ^{12} {\mathrm C}$. A lithium tethraborate (Li$_2$B$_4$O$_7$) target 
($p$-target to distinguish from the $\mu$-target described in Sect.~\ref{target}) 
is used to generate $\gamma$-rays from both reactions. 

\begin{figure}
\begin{center}
\includegraphics[width=.95\columnwidth]{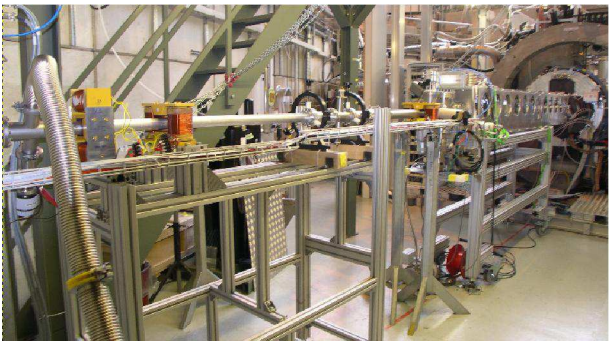}
\caption{The Cockcroft--Walton accelerator beam line }
\label{fig:beamline}
\end{center}
\end{figure}

During calibration runs, the target, contained in a vacuum pipe connecting 
the C--W accelerator to the MEG area, is positioned at the centre of COBRA 
(see Fig.~\ref{fig:areaCW} and ~\ref{fig:beamline}). The $p$-target is 
oriented at $45^\circ$ relative to the proton beam direction, to reduce 
the amount of material on the path of the $\gamma$-rays directed to the LXe detector.

\begin{table}[htbp]
  \caption{\small Characteristics of the MEG Cockcroft--Walton accelerator}
  \label{tab:comp}
  \begin{center}
    \begin{tabular}{lr}
      \hline
       {\bf Proton beam properties}  & {\bf MEG C--W} \\ 
      \hline
      Energy [keV]   & 300-1000   \\
      Energy spread (FWHM)[keV] & $<\, 0.5$ \\
      Angular divergence (FWHM) [mrad $\times$ mrad] & $<\, 3 \times 3$  \\
      Spot size at 3 meter (FWHM) [cm $\times$ cm ] & $<\, 3 \times 3$ \\
      Energy setting reproducibility [ \%  ] & 0.1 \\
      Energy stability (FWHM) [ \%  ]  & 0.1  \\
      Range of the average current [$\mu$A]  & 1--10  \\
      Current stability [ \% ]  & 3 \\
      Current reproducibility [ \% ]  & 10  \\
      Duty cycle [ \% ]  & 100 \\
      \hline
    \end{tabular}
  \end{center}
\end{table}
%

During normal data taking the $p$-target is positioned downstream outside the COBRA spectrometer.
When starting a calibration, the $\mu$-target is removed from the beam line by means 
of a compressed 
helium system and the $p$-target is inserted to the centre of COBRA by means of an 
extendable bellows system of $\sim 2$~m stroke. The insertion (or extraction) is computer 
controlled and takes ten minutes. 
At the end of the test the inverse operation 
is performed, and the $\mu$-target reinserted. The reproducibility of its positioning has
been visually inspected and surveyed to be better than our spatial resolutions.

\begin{figure}
\begin{center}
\includegraphics[width=0.95\columnwidth]{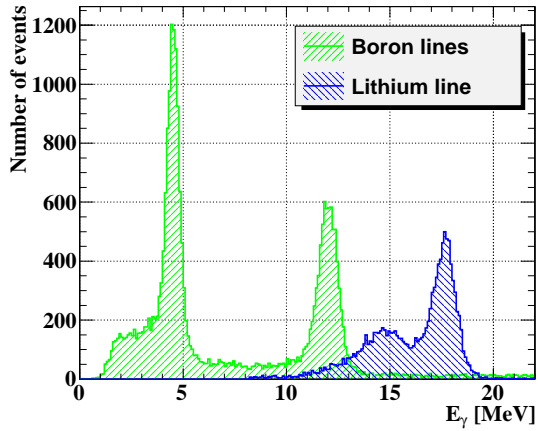}
\caption{\label{fig:LiB}Measured calibration lines from the reactions 
$^{11} {\mathrm B} ( p, \gamma_{4.4} \gamma_{11.6}) ^{12} {\mathrm C}$ (green) and
$^{7} {\mathrm Li} ( p, \gamma_{17.6} ) ^{8}{\mathrm Be}$ (blue)}
\end{center}
\end{figure}

Steering magnets and monitors are available along the proton beam line 
(see Fig.~\ref{fig:beamline}) for centring the beam on the $p$-target
and for measuring the proton beam properties.
The data from the lithium reaction are recorded by a low-threshold trigger, 
while a LXe-TC coincidence trigger 
is used to record the two boron $\gamma$-rays (see Fig,~\ref{fig:LiB}).

By means of the C--W calibration lines the energy scale of the experiment is constantly monitored,
as are possible drifts in the relative timing between the LXe detector and the 
TC bars. This allows knowing of the energy scale in the LXe detector at a few-per-mil
level, and a time alignment better than 20~ps.

\subsubsection{$\pi^-$ beam and charge exchange set-up}
\label{sec:cex}

To calibrate the LXe detector at an energy close to that of the signal we use 
$\gamma$-rays from neutral pion decay ($\pi^0 \to \gamma \gamma$). A neutral pion is 
produced in the CEX reaction of negative pions on protons at rest $\pi^-p \to \pi^0 n$.
The resulting $\pi^0$ has a momentum of $\sim 28$~MeV/c in the laboratory frame and decays 
immediately to two $\gamma$-rays. 
The photons are emitted back-to-back in the $\pi^0$ rest
frame with an energy of 
\[ E^\ast_\gamma = \frac{m_{\pi^0}}{2} \simeq 67.5~{\rm MeV}.
\]
In the laboratory frame, the photon energies are
\begin{equation}
E_{\gamma_{1,2}} = \gamma \frac{m_{\pi^0}}{2} \left( 1 \pm \beta \cos \theta^\ast 
\right)
\label{eq:gammaenergy}
\end{equation}
where $\beta$ is the $\pi^0$ velocity and $ \theta^\ast$ the center-of-mass
angle between the photon and the $\pi^0$ direction. 
 
Differentiating Eq.~(\ref{eq:gammaenergy}), the
energy spectrum of the two photons in the laboratory frame 
\begin{equation}
\frac{{\rm d}N}{{\rm d}E_\gamma} = \frac{{\rm d}N}{{\rm d}\cos \theta^\ast} \times
\frac{{\rm d}\cos \theta^\ast}{{\rm d}E_\gamma} 
\end{equation}
is flat (because dN/d$\cos \theta^\ast$ is constant)
between the energies
\begin{equation}
E_{\gamma, \rm Min} = \frac{m_{\pi^0}}{2} \sqrt{ \frac{1-\beta}{1+\beta} } \qquad
E_{\gamma, \rm Max} = \frac{m_{\pi^0}}{2} \sqrt{ \frac{1+\beta}{1-\beta} }.
\end{equation}
A strong correlation exists between the laboratory relative angle between photons 
and energies. In particular the extremal energies are obtained for 
photons which are emitted at relative angle 180$^\circ$ in the laboratory, that is along
the $\pi^0$ flying direction.

This strong correlation between energy and opening angle can be used 
to precisely define the energy of one photon by tagging the direction of the 
opposite one.
By selecting back-to-back $\gamma$-rays in the laboratory, by means of an auxiliary 
detector opposite to the LXe detector (an array of NaI or BGO in our case), the energies 
are at the edge of the spectrum, and the higher the collinearity, the more monochromatic the 
two $\gamma$-rays. For instance by requiring an opening angle larger than
175$^\circ$ (170$^\circ$) the energy spread $\sigma_{E_\gamma}/{E_\gamma}$ is better than 0.2\% (1\%). 
A few percent of the time one of the two $\gamma$-rays converts into a $e^+ e^-$ pair, 
either internally (Dalitz decay) or in the target material.

Additionally the $\pi^-$ radiative capture on protons produces a 129~MeV $\gamma$-ray 
coincident with a $9.0$~MeV neutron.

A negative pion beam of 70.5 MeV/c is chosen, partly due to the maximum current available 
for the BTS and partly because at this momentum a maximum TOF separation between pions and 
all other particles is achieved at the centre of COBRA. The measured intensity is ~ 1.7 MHz at 
a proton current of 2~mA (lower intensities selectable via slit setting) with a round 
spot-size of 8 mm (sigma), fully illuminating the target cell.

\begin{figure}
\centering
\includegraphics[width=0.95\columnwidth]{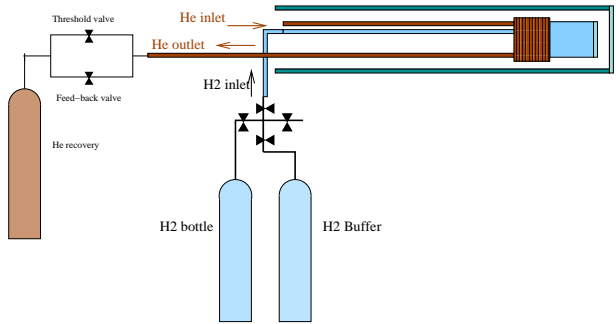}
\caption{Scheme of the liquid-hydrogen target with dedicated transfer lines. 
The 100~litres buffer is filled by means of a standard hydrogen bottle up to 2.5~bar: this 
quantity is sufficient to have a liquid volume of 150~cc in the cell.
When the cell is cooled down the hydrogen flows from the buffer to the target cell.}
\label{fig:lh2}
\end{figure}

A Liquid-Hydrogen (LH$_2$) target is used for the CEX reaction. It is a cylindrical cell 
of $50$~mm diameter, 75~mm length placed at the end of a 2~m long support. Both the LH$_2$ 
target (cold) window and the vacuum (warm) window are made of 135~$\mu$m Mylar in order 
to minimise $\pi^-$ multiple scattering and absorption outside the hydrogen-filled volume.
The LH$_2$ target with dedicated transfer lines is shown in Fig.~\ref{fig:lh2}. 
During normal operation the cell is filled with $150$~cc LH$_2$ at
20~K, kept cold by a continuous flow of liquid helium. The usage of a mechanical refrigerator 
is not possible due to the high magnetic field. 

The operation of the target is relatively safe, since all hydrogen is contained in a 
$\sim 100$~$\ell$ reservoir (the ``buffer'') at $2.5$~bar at room temperature. When liquid helium cools 
the target cell, hydrogen is liquefied at the centre of COBRA.
It takes three days to set up the LH$_2$ target 
and a few hours to liquefy the hydrogen. A 250~$\ell$ liquid-helium 
Dewar allows operation of the target for a little more than 48 hours.

To select $\gamma$-rays with a back-to-back topology a sodium iodide (NaI) detector
(2008--2009) or a Bismuth Germanate (BGO) detector (2010--2012) is 
placed opposite to the LXe detector. 

\begin{figure}
\centering
\includegraphics[width=0.65\columnwidth]{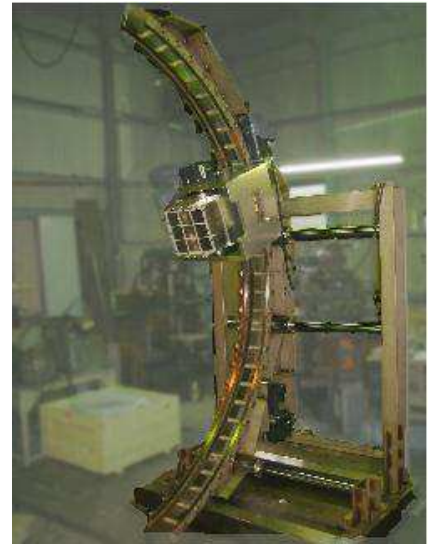}
\caption{The NaI mover.}
\label{fig:naimover}
\end{figure}

This ``opposite side''  detector is placed on a movable stage (see Fig.~\ref{fig:naimover}) that is 
remotely controlled to allow the mapping of the entire front face of the LXe detector. 
When displaced from the $z=0$ plane, the NaI detector is always tilted 
to face the centre of the LH$_2$ target.

The NaI detector consists of nine crystals of 
NaI(Tl) scintillator and two layers of lead/plastic scintillator for timing measurements 
(the lead/plastic scintillator can be removed during energy measurements). 
Each crystal has a size of $62.5 \times 62.5 \times 305$~mm$^3$ (11.8~X$_0$) and 
is read out by a $10 \times 10$~mm$^2$ APD \cite{hamamatsu-apdnai}. The lead/scintillator 
detector is made of two plastic scintillators measuring $60 \times 60 \times 7$~mm$^3$ 
preceded by a lead converter 5~mm thick, and is placed in front of the central NaI 
crystal when needed.

\begin{figure}
\centering
\includegraphics[width=0.99\linewidth]{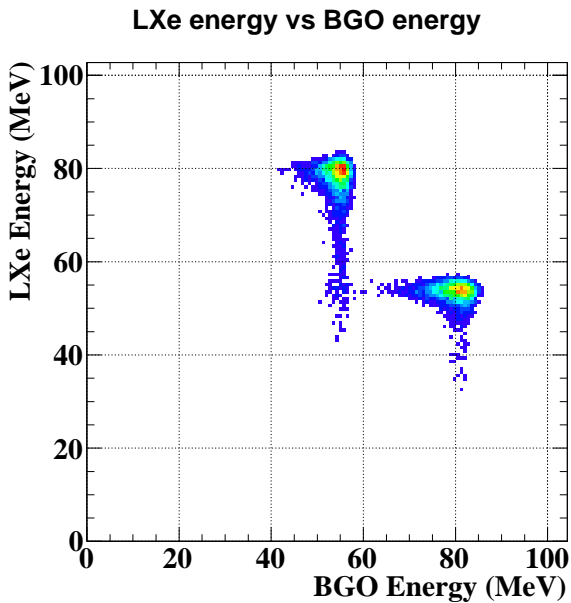}
\caption{Energy recorded by the LXe detector vs the energy recorded by the BGO detector during a CEX calibration run. The 
extremal energies of the calibration $\gamma$-rays at 54.9~MeV and 82.9~MeV anti-correlated in the two detectors are visible.}
\label{fig:pi02d}
\end{figure}

The NaI detector was replaced for the 2010 CEX run by a 
BGO detector composed of $4 \times 4$ crystals, because of the better energy and position
resolution of the latter. This translated into a better monochromaticity 
in the back-to-back topology and into a better efficiency.
A typical correlation spectrum of the energy seen by the LXe detector vs the energy
seen by the ``opposite side'' detector in a back-to-back topology is shown in 
Fig.~\ref{fig:pi02d}.

\subsubsection{Radiative muon decay (RMD)}
\begin{figure}
\begin{center}
\includegraphics[width=0.9\columnwidth]{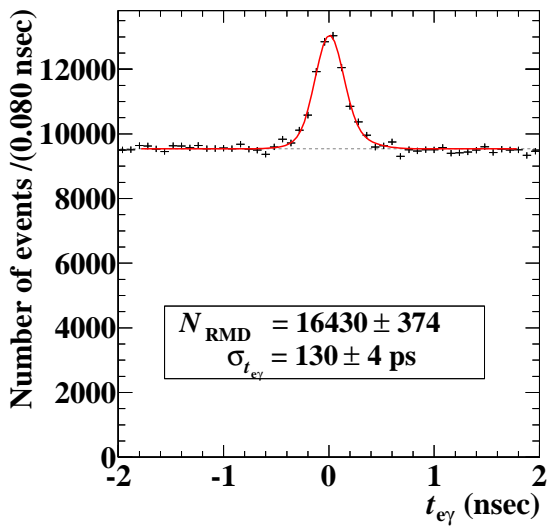}
\caption{\label{fig:RDPeak}
Distribution of $t_{{\rm e} \gamma}$ for MEG standard trigger. The peak is from RMDs, 
the flat component is from accidental coincidences.}
\end{center}
\end{figure}

Observation of the RMD process \radiative\ demonstrates that the 
apparatus can detect coincident $e$-$\gamma$ events, and 
measures directly the global time resolution and the relative offset. 
At the first stages of the experiment we took runs at 
low beam intensity to reduce the accidental background; later, the time resolution 
was good enough to also clearly see the RMD peak at the nominal beam intensity 
of $\sim 3 \times 10^7\,\mu^+$/s.
Figure~\ref{fig:RDPeak} shows the distribution of the $e$-$\gamma$ relative time 
$t_{e \gamma}$ for RMD decays using the standard MEG trigger.

In order to obtain the resolution on $t_{{\rm e} \gamma}$ for signals, 
the resolution in Fig.\ref{fig:RDPeak} must be corrected for the 
$\gamma$-ray energy dependence shown in Fig.\ref{XEC:timeres_energy_dep} and 
for the positron energy dependence using Monte Carlo resulting in 
$\sigma_{t_{{\rm e} \gamma}} = 122$~ns.

%
%
\section{Trigger}
\label{trigger}

\subsection{Requirements}
An experiment to search for ultra-rare events in a huge beam-related background needs a quick and efficient event selection, which demands the combined use of high-resolution detection techniques
and fast front-end, digitising electronics and trigger. The trigger system plays an essential role in 
processing the detector signals to find the signature of $\meg$ events in a high-background environment. 

The most stringent limitation to the trigger latency originates from the matching with the DRS (see
Sect.~\ref{DAQ:drs4}), whose
digitisation needs to be stopped at latest 450 ns after the event occurrence. If the trigger were later, 
the charge stored in part of the DRS cells may be overwritten and a few samples of the waveform 
(in particular those recording the leading edge of PMT signals) may be lost.
The speed objective must be achieved while keeping the signal efficiency close to unity. On the other hand, the event selection needs to provide a significant background rejection. The trigger rate,
as explained in Sect.~\ref{sec:effi}, should be kept below 10 Hz so as not to overload the DAQ system, which would result in a significant increase in the dead time. 
\subsection{The trigger scheme}
In order to accomplish these goals, we developed a system consisting of VME boards arranged in a 
tree-structure (see Sect.~\ref{sec:tree}), based on on-board Field-Programmable Gate Arrays 
(FPGA) to process the detector signals and reconstruct the observables needed for the event 
selection \cite{trigger2013}.
The signals are digitised by means of 100 MHz flash-ADCs so as to obtain an estimate of the signal 
amplitude (at the level of a few-per-mil accuracy) and timing (with a few ns resolution).
The advantage of choosing an FPGA-based digital approach is manifold:
\begin{itemize}
\item operation of reconstruction algorithms to match with the above speed and efficiency 
requirements;
\item availability of RAM memories for storing the digital data stream (for both
signal input and algorithm output) into cyclic buffers, which makes the trigger system an 
independent digitiser system as well, to be used as a back-up for the DRS;
\item versatility of the trigger scheme, as it is possible to implement selection criteria, other than  
for $\meg$ decays, related to calibration and other event types; the acquisition of pre-scaled triggers 
associated with those events during normal data taking enables us to monitor the stability of both 
detector response and trigger efficiency as a function of time.
\end{itemize}
\subsection{Online algorithms}
\label{sec:trigalgo}
The set of observables to be reconstructed at trigger level includes:
\begin{itemize}
\item the $\gamma$-ray energy;
\item the relative $e$-$\gamma$ direction;
\item the relative $e$-$\gamma$ timing.
\end{itemize}
The stringent limit due to latency prevents use of any information from the DCH; the electron 
drift time toward the anode wires may be as long as 200 ns,
too long to match the latency requirement. Therefore a reconstruction of the positron momentum 
cannot be obtained at the trigger level\footnote{
Although excluded from the $\meg$ trigger, DCH anode signals are collected indeed by the system
to provide stand-alone triggers associated with tracks in the DCH. For each chamber, anode signals 
are grouped in two sets (as coming out of inner and outer wires) and fanned-in linearly before being fed
to Type1 boards. The track condition is met if four out of five consecutive chambers are hit.}. 
Even if there is no explicit requirement on the $e^+$ in the trigger algorithm, the requirement of a TC hit
is equivalent asking for a positron with momentum $\gtrsim 45$~MeV.

The $\gamma$-ray energy is the most important observable to be reconstructed. 
The steep decrease of the spectrum at the end-point (equal to the signal 
energy for \meg) emphasises its role in suppressing the background. 
For this reason the calibration factors for LXe PMT signals 
(such as PMT gains and QEs ) are continuously monitored and periodically 
updated in a dedicated database. The energy deposited in LXe is estimated
by the linear sum of PMT pulse amplitudes, which is accomplished by 
cascaded adder stages (with the proper bit resolution) up to the master trigger
board. The sum of about 250 channels (one for each inner PMT, 
one to four PMTs for other LXe faces) is sensitive to coherent noise 
in the electronics. To cope with that, adder stages are preceded by the 
subtraction of the signal pedestal, updated online by the average of 
out-of-pulse samples. 

The interaction time of the $\gamma$-ray in LXe is extracted by a fit of the 
leading edge of PMT pulses, with a resolution of the order of 2 ns. 
The same procedure allows us to estimate the time of the positron
hit on the TC with similar resolution. The relative time, a useful tool 
to suppress our accidental background, can thus be obtained from 
their difference; fluctuations due to the time-of-flight of each particle 
to their respective detectors are within the resolutions.

The amplitudes of the inner-face PMT pulses are also sent to comparator 
stages to extract the index of the PMT collecting the most photoelectrons. 
This provides a robust estimator of the interaction vertex of the $\gamma$-ray in LXe, 
the Moli\`ere radius being similar to the PMT size (see Table~\ref{table:LXeproperty}). 
This vertex and the target centre provides an estimate of the $\gamma$-ray direction. 
On the spectrometer side, lacking any DCH information, one has to rely on the 
coordinates of the TC online reconstructed hit. Under the assumption of the momentum being that 
of the signal and the direction opposite to that of the $\gamma$-ray, 
by means of Monte Carlo simulated events, the PMT index is associated 
with a region of the TC indexed by a bar number and a $z$-segment.
If the online $z$-coordinate of the TC hit (see Sect.~\ref{sect:online}) 
falls in this region, the relative 
$e$-$\gamma$ direction is compatible with the back-to-back condition. 
\subsection{Hardware implementation}
\label{sec:tree}
\begin{figure}[htb]
\begin{center}
	\includegraphics[width=0.33\textwidth, angle = 270]{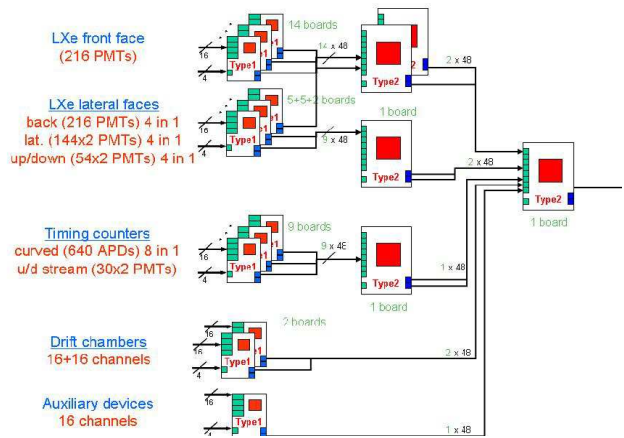}
\end{center}
\caption{Hierarchy of the trigger system: (from left to right) Type1 boards; Type2 intermediate boards; Type2 master board.}
\label{fig:layout}
\end{figure}

The trigger system is organised in a three-layer hierarchical structure, as shown in 
Fig.~\ref{fig:layout}, which consists of:
\begin{description}
\item[1)] VME 6U boards (so-called ``Type 1''), each to receive and sample
16 input analog detector signals at 100 MHz by the 10-bit 
flash-ADCs \cite{fadc} to be processed by means of digital algorithms implemented on a 
Virtex XC2VP20 FPGA chip \cite{fpga};
\item[2)] VME 9U boards (named  ``Type 2''), each equipped with a Virtex XC2VP20 FPGA chip \cite{fpga}, to collect digital pieces of information from
Type1 boards associated with different detectors and combine them to obtain estimates of the 
kinematic observables of interest (namely energy, direction and time of flight) of decay particles.
\end{description} 
The input stage of Type1 boards consists of plug-in Front-End (FE) mini-cards to 
receive the signals from detectors (in AC coupling) via the active splitters 
(see Sect.~\ref{sec:splitter}). A low-pass filter with cutoff at 33 MHz 
(to match Nyquist's criterion) is obtained by means of an RC integrator 
through the feed-back branch of an AD8138 differential driver. 
The common-mode voltage of the output signal can be further adjusted by 
using a programmable AD5300 DAC to match the dynamic range of the FADCs 
(which is set to 0--2~V in the case of LXe inner-face PMTs and 0--1~V for all other channels). 

The FPGA digital output is arranged in a 48-bit word, encoded with charge, time and other pieces
of information (which depend on the detector) relevant to event reconstruction. The output bus is transmitted to the upper-layer boards by means of a DS90CR483 Low Voltage Differential Signal (LVDS) serialiser clocked at 100 MHz, the same frequency as the FADC sampling and FPGA algorithm execution.  

\begin{figure}[ht]
	\centering
		\includegraphics[width=0.5\textwidth]{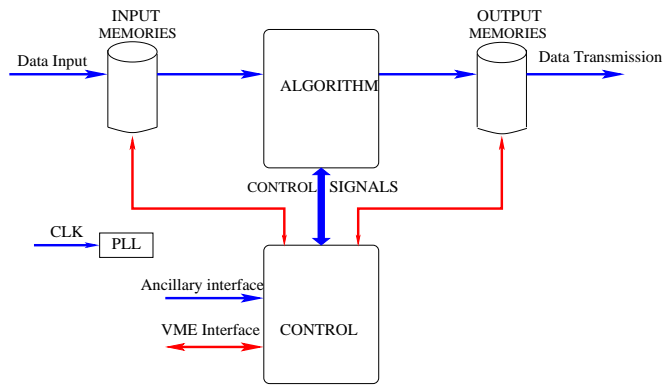}
		\caption{FPGA firmware structure.}
		\label{fig:FPGA}
\end{figure}

The FPGA inputs and outputs are both recorded into cyclic memory buffers, as shown in 
Fig.~\ref{fig:FPGA}. The comparison of the output memory content with the result of a simulator running on the same input data enabled debugging of the execution of reconstruction algorithms.

At the intermediate layer, Type2 boards collect and combine the information transmitted 
by nine Type1 boards via the companion DS90CR484 LVDS deserialisers. Again algorithm 
output arranged in 48-bit buses (up to two per board) is transmitted to the trigger 
master board via LVDS serialisers and there received by deserialisers. Finally, 
the task of the master is to assemble all event characteristics and generate a Stop 
signal whenever a trigger condition is met.
Moreover, this board is the one to receive Busy signals raised by each digitiser 
during data transfer and to issue a Start command as soon as that Busy condition 
has been released by all digitisers. 
 
The synchronous operation of the system (FADC digitisation, algorithm execution 
and data flow) is guaranteed by a set of ``Ancillary'' boards distributing clock 
signal and control (Start and Stop) signals to all digitising electronics. 
A master ancillary board hosts a SARONIX SEL3935 oscillator to deliver the 19.44 MHz 
master clock which, although being not essential for trigger synchronisation, 
is quite important for ensuring accurate detector timing. On the trigger side, each 
board utilises a PLL (based on a CY7B994V Roboclock \cite{cypress}) to multiply 
the input clock frequency by a factor 5 so as to generate 16 replicas of the 
100 MHz clock signal throughout a board. The relative phase between these can 
be programmed to compensate for the relative timing jitter between chips 
(and, in the case of the FPGA, even between lines in a bus) and ensure proper
synchronisation over the whole trigger tree.

Trigger boards are hosted in three VME crates and accessed by front-end 
code running on as many online PCs. Each board is equipped with a CoolRunner 
XC2C384 chip \cite{fpga} as an interface between all the chips and the VME bus.
Input and output data are read out through a standard, double-edge block transfer 
VME master-slave access to ensure a throughput rate close to 80 MB/s. 
Single word (both read and write) VME commands are used to access configuration 
registers to set behavioural parameters (such as thresholds, look-up table entries, 
calibration factors, clock skews, etc.) as well as to get the status of control 
signals (namely the Busy) from all the digitisers. 
\subsection{Operation and performance}
\label{sec:effi}
\begin{figure}[htb]
\begin{center}
	\includegraphics[width=0.45\textwidth]{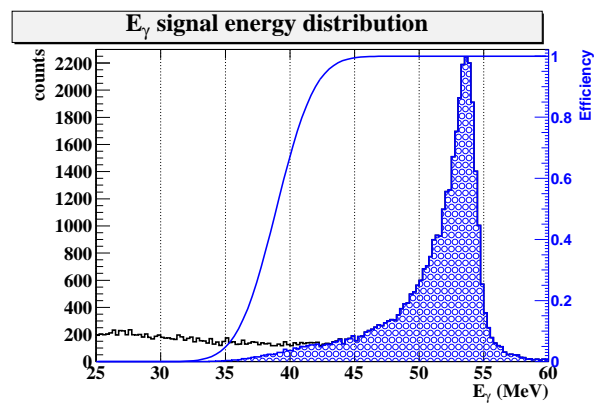}
\end{center}
\caption{Spectrum of $E_\gamma$ measured offline by the LXe detector, as obtained 
with the threshold at 45~MeV on the online measured $\gamma$-ray energy, superimposed 
to a lower-threshold one. The corresponding 
efficiency curve, resulting from the ratio of the two spectra, is also shown.}
\label{fig:eneffi}
\end{figure}

The $\gamma$-ray response function is obtained with CEX calibration close to 
the signal energy and continuously monitored (although at lower
energies) by using C--W proton-induced $\gamma$-ray lines; the resolution is 
estimated to be $3\%$ at $52.83$~MeV. The equivalent energy threshold for 
\meg\ events is computed by comparing the energy spectrum with the one obtained at 
lower thresholds (from 30 MeV to 40 MeV) thanks to a pre-scaled trigger operated in 
parallel with the main one. The ratio of the two spectra is shown in 
Fig.~\ref{fig:eneffi} and follows a step behaviour smeared with a Gaussian function
with an energy resolution in agreement with that estimated by the fit of the 55 MeV 
line from CEX data. The threshold on the $\gamma$-ray energy measured online is set 
to 45 MeV, which guarantees almost full 
efficiency ($\gtrsim 99.5\%$) for signal events; on the other side, the single rate 
drops from a few hundred kHz (the bulk of which is related to $\gamma$-rays from
radiative muon decays) to 1.5 kHz.

To assess the relative $e$-$\gamma$ timing, a 20 ns-wide window is fixed to look for coincidences 
between the two particles. Relative offsets are calibrated by reconstructing events due to simultaneous 
$\gamma$-rays (one detected in LXe, the other in the TC) from B$(p,\gamma\gamma)$C (see Sect.~\ref{sec:calsec}).
The resolution is estimated {\it a posteriori} by fitting the edges of the window steps with an integral Gaussian function and amounts to 3 ns, in agreement with the single-detector timing estimate. The coincidence window is wide enough to achieve full efficiency.

The $e$-$\gamma$ Direction Match (DM) enables a further reduction of the accidental 
background trigger rate, which is proportional to the square of the opening angle \cite{kuno_2001}, 
by one order of magnitude at least, as expected by detector and trigger simulation, 
as shown in Table~\ref{table_rate}.
\begin{table}
\begin{center}
\begin{tabular}{ l c c }
	\hline
	Selection & Measured (s$^{-1}$) & Expected (s$^{-1}$) \\
	\hline
	\hline
	$E_\gamma > 2$ MeV & $3\times 10^5$ & $2\times 10^5$ \\
      \hline 
	$E_\gamma > 45$ MeV & $1.5\times 10^3$ & $1.3\times 10^3$ \\
	\hline 
	$\Delta t_{{\rm e} \gamma} < 20$~ns & 150 & 120 \\ 
	\hline
	$\Delta \theta_{{\rm e} \gamma} \sim \pi$ & 12 & 10\\
	\hline
\end{tabular}
\end{center}
\caption{Trigger rates as a function of progressively applied selection criteria ($\gamma$-ray energy, 
relative $e$-$\gamma$ timing and direction) at $3\times 10^7\,\mu^+$/s stop rate on target. 
Measured values agree within 20\% with expectations based on detector simulation and 
reconstruction algorithm emulation.}
\label{table_rate}
\end{table}

In Table~\ref{table_trg} a subset of the available selections both for physics and calibration 
are listed following their priority order.
For calibration triggers refer to Sect.~\ref{sec:calsec} for details.

\begin{figure}[htb]
\begin{center}
	\includegraphics[width=0.4\textwidth]{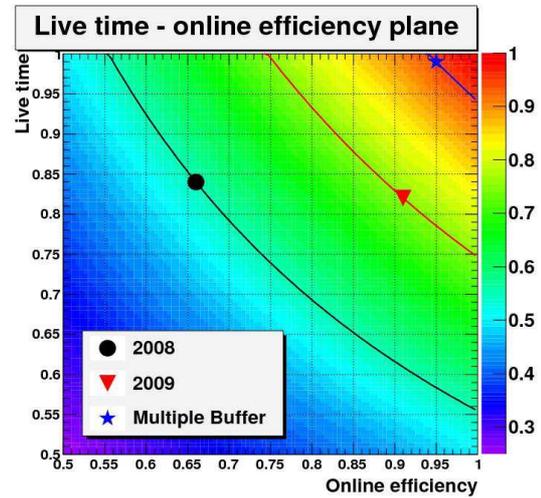}
\end{center}
\caption{DAQ efficiency as a function of trigger efficiency and DAQ live time. Superimposed markers
correspond to optimal values reached during 2008, 2009/2010 and 2011/2012 runs (marked as ``Multiple Buffer'').}
\label{fig:DAQtrig}
\end{figure}

The final goal is to optimise the overall DAQ efficiency, defined as the product of the trigger 
efficiency times the DAQ live time. Maximising the former is clearly advisable, but may be 
wasted if accompanied by a significant increase in the trigger rate and, as a result, in the dead time. 
The best trade off between the two effects is shown in Fig.~\ref{fig:DAQtrig}; due to limitations 
inherent in the DAQ (the digitiser read-out lasting $\sim 24$~ms/event), the TC acceptance region is 
modified until the DAQ efficiency reaches the maximum allowed value. 
Therefore the only possibility left to overcome such limitations was to reduce the dead time by 
implementing the multi-buffer read-out scheme, as described in Sect.~\ref{sec:multibuffer}. After this major 
improvement, 
we could loosen the DM-match selection criteria and the trigger efficiency increased to 97\%.
\begin{center}
\begin{footnotesize}
\begin{table}
\begin{tabular}{lccc}
	Name & E$_{\gamma}$ & $|\Delta \rm{t_{e\gamma}}|$ & DM\\
	\hline
	\hline
	MEG & 45~MeV & 10~ns & narrow\\ 
	MEG Low E$_{\gamma}$ & 40~MeV & 10~ns & narrow \\ 
	MEG Wide DM & 45~MeV & 10~ns & wide \\
	MEG Wide $|\Delta \rm{t_{e\gamma}}|$ & 45~MeV & 20~ns & narrow \\
	Radiative Decay & 45~MeV & 10~ns & no \\	
	\hline
	\hline
\end{tabular}
\begin{tabular}{lc}
	Name & Selection criteria \\
	\hline
	\hline
	LXe alone & Threshold on E$_{\gamma}$ \\
	DCH alone & DCH hit multiplicity \\ 
	TC alone & TC hit multiplicity \\ 
	$\pi^0$ decay & Coincidence LXe \& BGO \\
	$\alpha$ & Pulse shape discrimination \\
	LED & Discrimination on driver \\
	Neutron Generator & Discrimination on driver \\
	Pedestal & Random trigger \\
	\hline
	\hline
\end{tabular}
\caption{List of the most important triggers available for physics and calibration runs
 (see text for details).
    }
\label{table_trg}
\end{table}
\end{footnotesize}
\end{center}

%
%
\section{Front-end electronics and Data Acquisition system}
\label{sec:daq}
\subsection{Requirements}

The Data AcQuisition system (DAQ) is dedicated to recording the signals from all
detectors of the experiment. The requirements for this high-precision and 
high-rate experiment are challenging. Energy has to be measured with an accuracy
better than a few per mil, even for small signals. The timing
accuracy of the electronics has to be better than 40 ps, and the pile-up
detection has to distinguish events separated in time down to 10 ns.

The energy from the LXe detector is determined by adding several hundred channels.
Even a small level of common noise on these channels, such as 50 Hz line
noise, would result in a huge offset of the summed signal. It has therefore been
decided to digitise the waveforms of all detectors with high resolution and high
speed. This allows an efficient pile-up detection and an event-by-event
baseline measurement to eliminate any common low-frequency noise. In order to
obtain the required timing resolution, the signals must be digitised with a
sampling speed above one Giga-Samples Per Second (GSPS).

\subsection{Front-end electronics }

While FADCs with sampling speeds beyond one GSPS are commercially
available, they are impractical to use because of their costs, space and power
requirements considering the 3000 channels of the
experiment. An alternative is the use of Switched Capacitor Arrays (SCA).
They sample input signals in analog storage cells at high speed. After a
trigger, the cells can be read out at much lower speed in the MHz range, 
and digitised by a commercial ADC with typically 12-bit resolution. Since no
suitable SCA was available when the experiment was designed, the 
design of a new SCA chip fulfilling all requirements was started
\cite{Ritt2010486}. The latest version of this chip called Domino Ring
Sampler 4 (DRS4) has been used successfully in the experiment since 2009.

The sampling frequency was chosen as 1.6 GHz for the TC and LXe detectors,
which perform high-precision timing measurements, and 0.8 GHz
for the DCH, which has less stringent timing requirements.

\subsubsection{The DRS4 Switched Capacitor Array chip}
\label{DAQ:drs4}

\begin{figure}[htb]
 \centerline{\hbox{
  \includegraphics[width=.45\textwidth,angle=0] {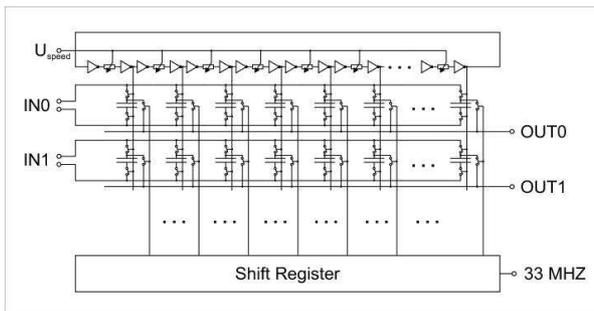}
 }}
 \caption[]{Simplified schematics of the inverter chain of the DRS4 chip.}
 \label{DAQ:invchain}
\end{figure}

To obtain the high sampling speed of a SCA chip, the write switches of the 
analog sampling cells are operated from the inverter chain shown in 
Fig.~\ref{DAQ:invchain}, and called the domino wave circuit. This 
eliminates the need to use a GHz clock, which would be very hard to distribute 
over thousands of channels.

\begin{figure}[htb]
 \centerline{\hbox{
  \includegraphics[width=.45\textwidth,angle=0] {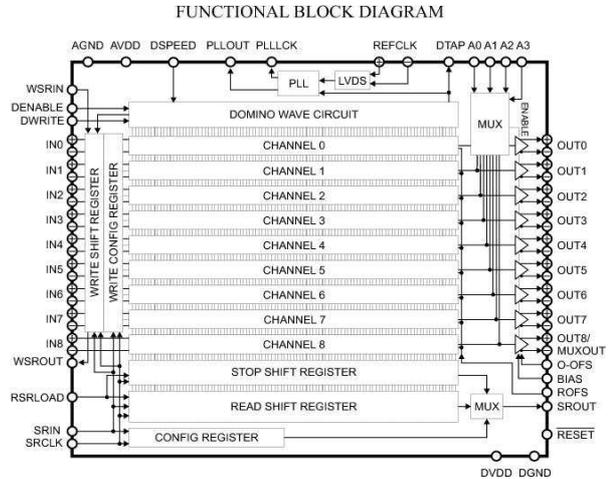}
 }}
 \caption[]{Functional block diagram of the DRS4 chip.}
 \label{DAQ:drs4_diagram}
\end{figure}

The variable resistors between the inverters, together with the parasitic 
input capacitance of the next inverter, a delay element, which is used to vary
the sampling speed with a control voltage $U_{speed}$. The analog inputs and
outputs of the DRS4 chip are differential, which reduces the crosstalk between
channels. The read-out shift register can be operated up to 33 MHz. However, we use
only half this frequency, since the chip read-out dead
time is not an issue and the lower frequency improves the analog performance of
the chip. Figure~\ref{DAQ:drs4_diagram} shows the functional block diagram of the DRS4 chip.

\label{DAQ:PLL}
The analog control voltage to operate the inverter chain is obtained from a 
Phase-Locked Loop (PLL), which fixes the sampling speed with high precision to 
an external reference clock over a wide temperature range. The chip contains 
eight data channels and a ninth timing channel used to sample the external 
reference clock directly. Since the same domino wave circuit operates all nine 
channels, any timing jitter in that circuit can be measured in the timing 
channel and corrected offline to improve the timing accuracy. 
The analog output of all nine channels is multiplexed into a single external 
12-bit ADC, which significantly reduces the cost and space requirements.

\subsubsection{Splitters}
\label{sec:splitter}

\hyphenation{dia-me-ter re-co-gni-tion}

\begin{figure}[htb]
 \centerline{\hbox{
  \includegraphics[width=.45\textwidth,angle=0] {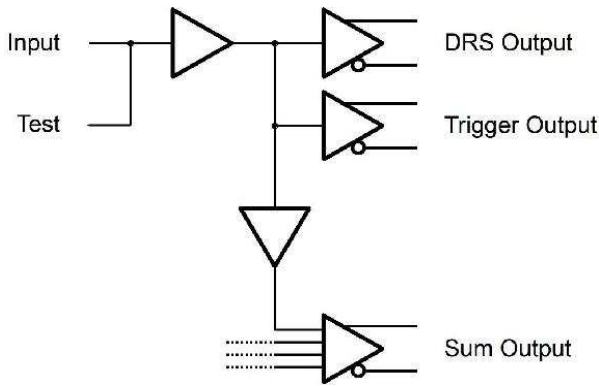}
 }}
 \caption[]{Block diagram of a splitter channel for the LXe detector.}
 \label{DAQ:splitter}
\end{figure}

A system of active splitters receives the signals from the LXe detector and 
from the TC through the passive splitters. Each splitter board receives 16 
signals through coaxial cables 
and makes two copies for the DRS and trigger. Differential signals are used to 
improve the noise immunity. Output signals are sent to digitization systems 
through flat cables (high-density twisted-pairs cable, 0.635 mm pitch, for the DRS 
and low-density twisted-pairs cable, 1.270 mm pitch, for the trigger). The gain 
for each channel is one and the bandwidth is about 1 GHz for the DRS output and 100 MHz for the trigger output. 
Careful choice of the components and their placement ensure a small integral non-linearity.
A dedicated test measurement resulted in a combined non-linearity of the
splitter system and the DAQ boards of $\pm 0.3$\%.
The crosstalk between each channel for the full bandwidth section is \textless 0.8\%. Furthermore, the splitter 
board provides an analog sum $4 \div 1$ of the LXe detector's lateral face signals, 
which is used by the trigger system. 
A block diagram of the splitter section is shown in Fig.~\ref{DAQ:splitter}.

\subsubsection{DAQ boards}

The DAQ boards are based on the VME PMC Carrier (VPC) (PCI Mezzanine Card (PMC)) 
board developed at PSI \cite{keil_2005}. 
This general-purpose VME board contains two Virtex-II Pro FPGAs \cite{fpga}, one of which 
is used to implement the VME64x protocol, while the second contains 
application-specific code. The board contains two PMC slots, which carry two 
DRS4 mezzanine boards with a total of 32 channels. The high channel density 
makes it possible to fit 640 channels into a single VME crate. The experiment
uses three VME crates to host 1728 channels for the DCH read-out, and
two crates to host 846 channels for the LXe detector read-out and 120 channels for
the TC read-out. 

\subsubsection{Trigger bus}

The synchronisation between the DAQ boards and the trigger is done via the
trigger bus, which is a cable running between the master trigger unit and the
DAQ boards. The first 12 lines send signals from the trigger to the boards. They
contain the main trigger, a synchronisation signal, five bits carrying the
trigger type and five bits with the least significant bits (LSB) of the event
number. The trigger type is one of the 32 types defined in the main trigger (see
Sect.~\ref{trigger} and Table \ref{table_trg}). The boards and the front-end program can decide to
handle different events flexibly: for example keeping the full waveforms for
physics events, while analysing online the waveforms and storing only integrals
and timings for calibration events.

The last eight lines of the trigger bus send
signals from the DAQ boards to the trigger. The main signal is the ``busy" line,
blocking the trigger until the board read-out is finished. Since all lines of
the trigger bus run between FPGAs, their meaning can be changed by modifying the
firmware only. This has proved to be very flexible and was successfully used
several times to implement new functionality such as a delayed trigger.

\subsubsection{DAQ synchronisation}
\label{DAQ:Synchronization}

While the trigger signal can be successfully used to start the read-out of all
DAQ boards, the timing it carries is rather poor. The bus structure of the
signal distribution causes this signal to arrive at slightly different times at
each board. In addition the signal is routed through the FPGAs to trigger the
DRS4 chips. This causes a jitter of the signal of $\sim 100$ ps, which exceeds
the experiment timing requirements. To overcome this problem, a dedicated 
high-precision clock distribution was implemented. A low-jitter 19.44 MHz master
clock is generated in the trigger and fanned out on a clock tree with 200 Low
Voltage Differential Signalling (LVDS) lines, each of them running to an
individual mezzanine board. The clock is fed into a clock conditioner chip
\cite{clockcond} on each mezzanine board, which removes most of the jitter
picked up by the cable, leaving a residual clock jitter below 4 ps. One output
of the chip is directly digitised by a dedicated timing channel in each DRS4
chip and used as a timing reference for each event.

The performance of the system was tested with a 50 ns square pulse with rise
and fall times of 2.5 ns. It was passively split and the two copies were sent to
two arbitrary channels of the digitising electronics. Their relative timing was
measured by waveform analysis. In 2009 the timing accuracy of the DRS4 chip was found to
be $\sim$~110--170 ps. This poor timing had two causes. The first was the choice
of the external components for the loop filter of the PLL (see
Sect.~\ref{DAQ:PLL}). A re-optimisation of these components improved the timing
significantly. The second cause was the continuous digitisation of the DRS4 chip
output by the ADC. That generated digital noise which affected the timing
jitter of the DRS4 chip. A firmware modification was then applied to stop the
ADC when the DRS4 chip is actively sampling data, and starting it only during
read-out. Minimising these two effects, the timing accuracy became $\sim$ 40--50
ps, comparable to or smaller than the detector timing resolution so as not to 
degrade the experiment's performance. All mezzanine boards were modified in June
2010, so all data since then were taken with good timing resolution.
  
\subsection{DAQ System }

\begin{figure*}[htb]
\begin{center}
\includegraphics[width=2\columnwidth] {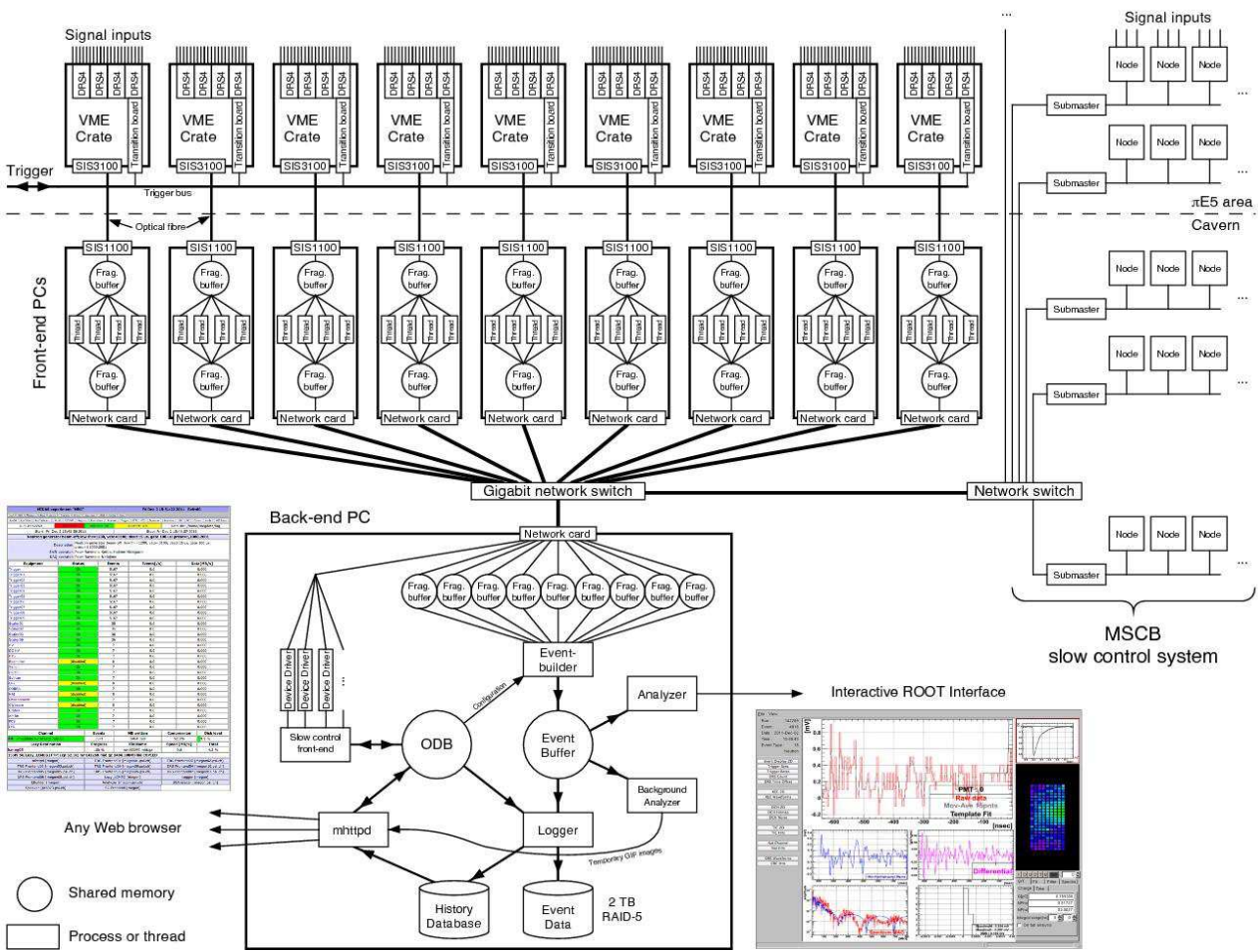}
\caption{Simplified schematic of the DAQ system. The individual
components are described in the text. The upper part resides 
next to the detector, the lower one just outside the experimental area.
The electronic logbook, the system
messages and the alarm system are omitted for clarity.}
\label{DAQ:daq_layout}
\end{center}
\end{figure*}

The data acquisition system (see Fig.\ref{DAQ:daq_layout}) is dedicated to reading out the front-end
electronics, combining the information into events and storing them on disk. It
consists of a cluster of PCs connected to the front-end electronics and an 
event-building PC connected through a Gigabit Ethernet switch. They run the DAQ
software, which handles all aspects of event-based data acquisition and slow
control.

\subsubsection{Online cluster}

The VME boards of the front-end electronics and trigger system are located in
nine VME crates. Since they produce a significant amount of data, each crate is
read out by a separate PC via the Struck SIS3100/1100 VME controller over a 15 m
optical cable \cite{struck}. The PCs contain two dual-core Xeon CPUs running at
2.8 GHz under Scientific Linux 5.1. The multi-core functionality is important
since the PCs that read out the DRS chips have a high computing load for the
online calibration and zero suppression of the raw waveform data. If the
front-end program run on a single core, the maximum event rate would be only 12
Hz. Running separate threads for the VME readout, calibration and 
network transfer of calibrated events increases the rate to 38 Hz. The PCs have
redundant dual power supplies and hot-pluggable mirrored hard disks, so that the
failure of a power supply or a disk would still allow data taking to continue.
The front-end PCs are connected to the central back-end PC via a Gigabit
Ethernet switch. This PC uses the same technology, but it has more disks. A
Redundant Array of Independent Disks (RAID 5) system with a net capacity of 2 TB
can buffer about 2--3 days of data taking. A dedicated process copies the data
from these disks constantly to the central PSI computing centre, where a
dedicated computer cluster analyses the data (offline cluster). The offline
cluster contains 16 nodes, each with four CPU cores, and a disk system with a
total of 280 TB, and is used both for data analysis and for simulation.

\subsubsection{MIDAS software}
\label{DAQ:MIDAS}

The data acquisition software used for the MEG experiment is based on the 
Maximum Integrated Data Acquisition System (MIDAS) \cite{midas}. MIDAS 
consists of a library written in C and C++, several tools and a general 
program framework to assist the development of experiment-specific front-end 
programs as well as analysis programs for online and offline use. The MIDAS 
library implements the following functionality:

\begin{itemize}

\item {\bf System Layer}. All Operating System (OS) specific functionalities are
covered by a set of functions using conditional compilation. In this way the
upper level of all MIDAS programs can use these general functions and compile on
different OSes. Over the last twenty years, many OSes have been supported. Some
of them have been phased out such as MS-DOS and VMS, others such as OSX (Darwin)
have been added. Since the system layer is relatively small (about 7\% of the
MIDAS code), new systems can be added relatively quickly.

\item {\bf Online Database (ODB)}. The online database is a hierarchical
database which stores all relevant information in a central location.
For performance reasons it is implemented completely in shared memory, so that
all experiment processes can exchange data very rapidly at an access rate 
$\sim 1$~MHz. The MEG ODB is about 5 MB in size and shared by about 20 processes. The
ODB contains configuration parameters for the trigger system, DAQ VME boards and
event building, as well as the current values for the slow control data and
event statistics. A special feature of the ODB are ``hot-links". A process can
register to be notified through a call-back routine when a specific value in the
ODB changes.

\item {\bf Buffer Manager (BM)}. The buffer manager is a set of library 
functions, over which processes can exchange event-based data. Buffers are 
segments of shared memory to which processes can attach as data producers or 
consumers. Data is passed very efficiently in a first-in first-out (FIFO) 
manner either locally or over the network using a dedicated TCP/IP protocol. 
On the Gigabit Ethernet network of the MEG experiment a net transfer rate of 
98 MB/s has been measured. Each consumer can either register to receive only 
a subset of all events (such as for monitoring) or request all events (such as 
the logger writing data to disk). In the latter case the consumer 
automatically slows down the data producer in case the data cannot be digested 
as fast as produced (commonly known as ``back-pressure").

\item {\bf System Messages (SM)}. Using the BM with a dedicated buffer, each 
process sends or receives text messages to inform the users about errors or 
status changes such as run start/stops. All SM are centrally recorded during 
the duration of the experiment.

\item {\bf Slow Control System (SC)}. MIDAS contains a framework to access ``slow
control" devices (see Sect.~\ref{slowcontrol}). The framework takes care of
transferring information between a device and the ODB (read-out) and between the
ODB and the device (control) by means of a device driver implementing two
functions for reading and writing.

\item {\bf History System (HS)}. While the ODB contains only the current value 
of all slow control values, it is desirable to see changes of the variables 
over time. The history system implements a database for storing and retrieving 
all slow control variables over the duration of the experiment. An 
optimised binary database is capable of writing up to $10^6$ variables 
per second to disk, and retrieving the time variation of any variable over 
days or months in a couple of seconds.

\item {\bf Event Builder (EB)}. The event builder is a dedicated process which
combines event fragments from nine different front-end computers into a single
event. Software and hardware counters are used to ensure that only fragments
belonging together are combined. In case of a fragment mismatch the event
builder stops the current run and issues an error message.

\item {\bf Logger}. The logger is an experiment-independent program 
which stores the events produced by the EB on disk. In addition, it writes the 
slow control data from the ODB into the HS utilising the above described 
hot-link scheme. The logger contains basic run control functionality such as 
stopping a run after a predefined number of events and automatically starting 
the next run. The logger writes data to a local disk array. A second 
logging process copies the data through a FTP interface to a central computing 
cluster, from where they are transferred to a tape archive.

\item {\bf Alarm System}. The alarm system is a part of the MIDAS library 
which runs in each process attached to the experiment. It checks individual 
values in the ODB or for the presence of a process. If a value exceeds a 
predefined limit or if a process dies, the alarm system can produce a system 
message, stop the run and execute a shell script, which in turn can send emails 
or text messages to mobile phones. 

\item {\bf Online Analyzer}. An analysis program based on the ROME framework
\cite{rome,romeurl} runs in the online environment sampling and analysing a
subset of the data stream. It combines a full physics reconstruction with a
single event display. Using various panels, sub-detector hit patterns as well as
short-term histograms can be inspected by the shift crew to identify potential
problems. A second copy of the program runs on the offline cluster to analyse
all events in near-to-real time. Histograms containing all events are available
10--20 minutes after the data has been taken, and allow monitoring of effects
which become visible only with adequate statistics. More information about the
analysis software can be found in \cite{software2010,softwaretns}.

\item {\bf Web Interface}. A dedicated web server (mhttpd) is used to control
the whole experiment through web browsers. It contains an interface to the
ODB, so the experimenter can view or change any ODB variable. In addition, the
interface implements panels for run control (start/stop a run plus run status
display), system messages, the alarm system, the history system and a process
interface to start and stop individual programs. While the mhttpd program is
experiment independent, ``custom pages" can be defined to monitor and control 
a specific part of the experiment (see Sect.~\ref{Slowcontrol:customsec}). The web
interface is also available from off-site locations for remote monitoring.

\item {\bf Run control}. While the standard MIDAS system has basic run control
functionality, more sophisticated run sequencing is needed. For example, for
the LED calibration of the LXe detector one needs a sequence of several runs, each with a
different LED flashing amplitude. To fulfil these needs, a dedicated utility
called ``runsubmit" was written. This tool reads an Extensible Markup
Language (XML) script and executes a sequence of tasks. It can start and stop
runs after a certain number of events, and change values in the ODB. This allows
the tool to change the trigger settings and prescaling factors, configure the
data read-out mode and even change slow control values such as high voltages or
beam line settings. Using the sequencing, several runs can be taken
automatically with different settings. 

\item {\bf Electronic Logbook (ELOG)}. Instead of using paper logbooks, the
collaboration uses an electronic logbook based on the ELOG program \cite{elog}.
This program, originally an integral part of MIDAS, is now available as a
stand-alone package. The usage of an electronic logbook has proved
extremely helpful for communication between the on-site shift crew and off-site
collaboration members, as well as for the retrieval of old logbook entries
tagged with certain keywords. An additional feature is a shift checklist,
filled out once by each shift, requiring the inspection of critical elements in
the experimental area not covered by automatic sensors. This checklist is a web form
integrated into the ELOG program and filled out using a tablet computer in the
experimental area. This allows a truly paperless operation of the experiment.

\end{itemize}

The MIDAS software runs very stably; in MEG the DAQ system is only stopped 
for configuration changes. If left in production mode, it runs for weeks, 
making most effective use of the beam time.

\subsection{Performance }

The performance of the DAQ system can be defined as the fraction of recorded
events for a given trigger rate. This includes dead time caused by the read-out,
lost events during data transport, events not recorded between runs, and events
lost during the downtime of the DAQ system due to re-configurations or crashes.
Muon beam downtime and calibration runs do not contribute to DAQ inefficiency.

\subsubsection{Dead time during event read-out}
\label{sec:multibuffer}

The DRS is stopped on each trigger and read out before being re-activated. One
waveform contains 1024 samples, each stored in a 16-bit value, digitised at 16.5
MHz. Each DRS chip uses four channels for detector waveforms and one for the
global reference clock. Two DRS chips are read out through a multiplexer by a
single ADC, which requires $61\ \mathrm{ns} \times 1024 \times 5 \times 2 =
625\,\mu s$. Each VME board contains two mezzanine boards each having four DRS
chips. This amounts to 1024 (bins) $\times$ 5 (channels/chip) $\times$ 4
(chips/mezzanine) $\times$ 2 (Bytes/sample) = 40 kB per event and mezzanine
board. A full VME crate contains 20 boards with 40 mezzanine boards, which
totals to 1.56 MB per crate and event. The SIS3100/1100 VME interface has a
setup time of $125\ \mu$s per DMA transfer and then a line speed of 83 MB/s.
The transfer of the 40 data blocks therefore takes $125\ \mu\,\mathrm{s} \times
40 + 1.56\, \mathrm{MB} / 83\, \mathrm{MB/s} = 23.8$ ms. The data read-out rate
from the trigger system is smaller and therefore does not pose a bottleneck.
Before 2011, the front-end electronics was only re-enabled after all data were
read out through the VME bus, thus causing a dead time of $23.8\
\mathrm{ms} + 0.625\ \mathrm{ms} = 24.4\ \mathrm{ms}$ per event. This limited
the maximum event rate to 41 Hz and caused a live time of 83.2\% at our average
event rate of 6 Hz, given the Poissonian time structure of event triggers. At
the end of 2010, a multi-buffer read-out scheme was deployed in the firmware of
the DRS and trigger DAQ boards. This scheme uses a FIFO buffer between the
front-end read-out and the VME transfer, capable of holding up to three events
for the DRS part and up to four events for the trigger part. This allows us to
re-enable the front-end electronics after the read-out time of $625\,\mu s$ and
do the data transfer in the background. If another trigger happens during the
VME transfer of the data, it is written into the FIFO buffer without interfering
with the data currently being transferred. The probability that three or more
events occur during the 23.8 ms of the VME transfer is $5\times 10^{-4}$ at a
trigger rate of 6 Hz and can be neglected. The dead time is therefore determined
by the probability to have another event during the DRS read-out time of
$623\,\mu \mathrm{s}$, which causes a dead-time of 0.4\%\ and leads to a live 
time of 99.6\%.

\subsubsection{Other sources of dead time}

The data read out through VME are transferred through Gigabit Ethernet to 
the back-end PC. Since the TCP/IP protocol is used, no packets are 
lost. Since a network switch with enough buffers and backplane bandwidth has 
been chosen, the probability of data collisions and necessary network 
re-transfers is very small. Constant data rates of 98 MB/s over hours at the 
input port of the back-end computer have been observed, which are more than 
enough for the typical overall data rate of 20 MB/s. At the back-end PC, the
event fragments are combined into full events. The average event size of a typical
run is 2.4 MB, consisting of ADC waveforms from the trigger system (12\%), DRS waveforms
for the LXe detector (30\%), the TC (8\%) and the DCH (45\%). The remaining 5\% covers
special waveform channels for timing synchronization and the slow control
data such as high voltage currents, temperatures and so on. The full event is then
compressed using the bzip2 algorithm, which results in a final event size of 0.9 MB
(62\% compression ratio). A special parallel version of this algorithm is utilized
that makes use of all eight CPU cores, so that the full 10 Hz event rate can be
compressed in real time.

A typical MEG run takes about six minutes. This time was chosen to produce raw
data files $\sim 4.8$ GB large. The time it takes to stop and start a new run is
six seconds, leading to an additional dead time of 1.7\%. Although this time is
small, there are plans to decrease it further by using multi-threaded control
programs for starting and stopping runs.

%
%
\section{Slow Control}
\label{slowcontrol}

Slow Control in this context means the monitoring and control of the physical
quantities in the experiment, that vary ``slowly" over time, such as
temperatures, pressures or high voltages. It is typically sufficient to read these
values once every few seconds and record them together with the data 
from the various detectors. This allows us to find during the offline
analysis any correlation between changes in detector performance
and variations in ambient quantities such as the temperature. Other quantities
need adjustments during the data taking: the beam line needs to be tuned
differently for calibrations and for data taking and the high voltages for 
DCH and PMTs need to be changed from time to time in order to
compensate for drifts in gain and efficiency.

Most experiments use many different systems for the slow control subsystems.
This approach has the problem of combining the data from different subsystems into
the central DAQ system. While this is viable in large collaborations, it is
difficult in smaller experiments such as MEG to maintain many different
subsystems. Thus the collaboration decided since the beginning to design a
common slow control system for the whole MEG experiment. The challenge of such a
design is to cover all requirements of the experiment with a single approach. A
field bus was clearly necessary to connect spatially separated sensors, and
local intelligence was required for control loops such as automatic high-voltage
ramping and trip recovery. The system had to be flexible and easily extendable,
very reliable and still operational with one part under maintenance.  

\subsection{MSCB System}

The new system was named Midas Slow Control Bus (MSCB), indicating the close
integration into the MIDAS DAQ system. It consists of a communication layer
through an RS-485 field bus and through Ethernet, and uses 8-bit microcontrollers
as communication end points and for local control loops. A crate-based 
high-voltage system was designed based on this concept. A general-purpose
control box SCS-2000 was built which can be equipped with various daughter
cards for up to 64 channels of various analog or digital inputs/outputs.

\subsubsection{Communication}

\begin{figure}[htb]
 \centerline{\hbox{
  \includegraphics[width=.45\textwidth,angle=0] {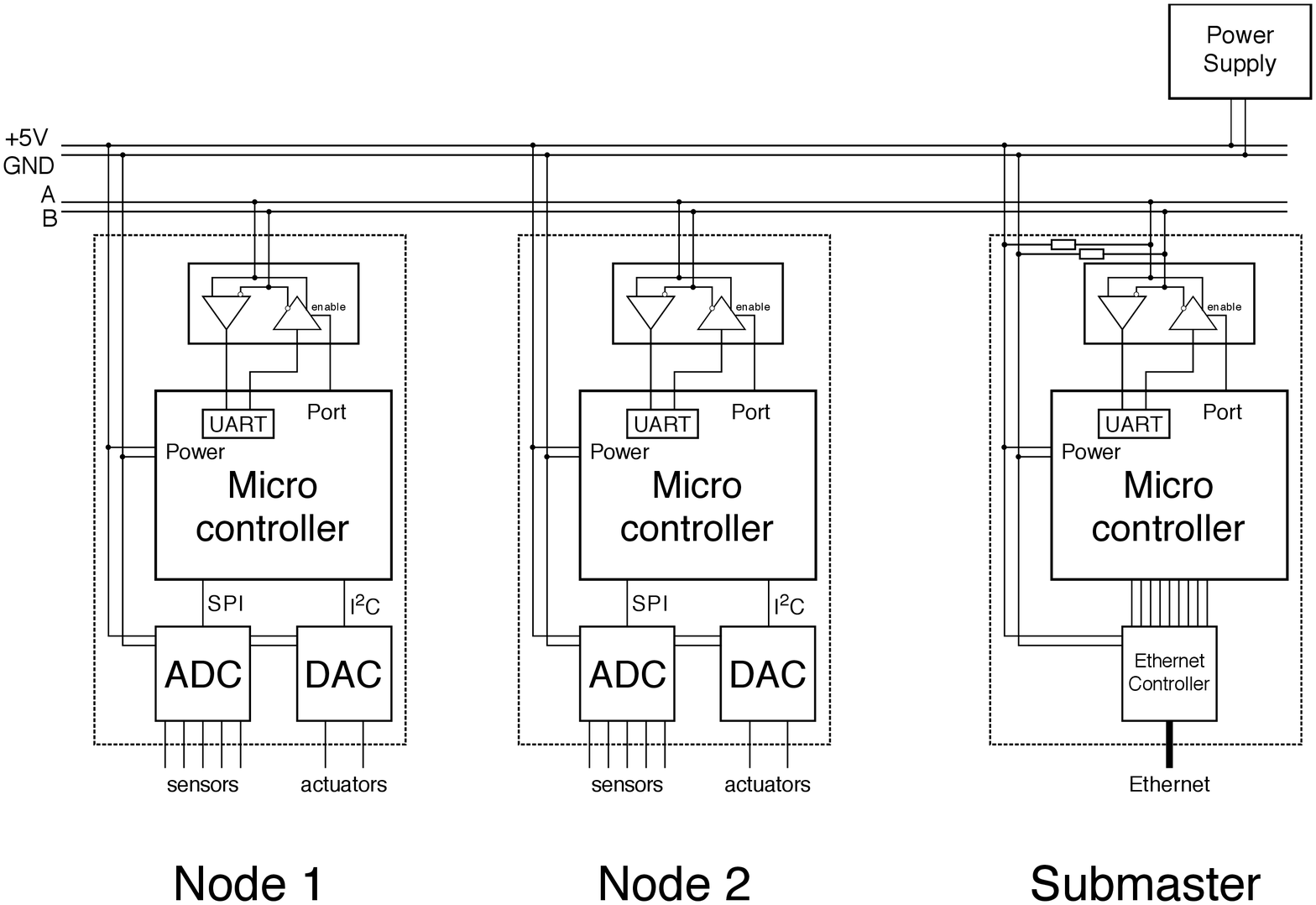}
 }}
 \caption[]{Simplified communication schematics of the MSCB system}
 \label{Slowcontrol:mscb}
\end{figure}

Physical quantities in MEG are measured at different locations. To collect them
in a centralised system, an electronic connection between the sensors and the
DAQ is necessary. Since analog connections are prone to electronic noise pickup,
it is desirable to digitise the quantities close to the sensors and transfer the
information digitally. For the MSCB system we choose a field bus based on the
RS-485 standard. This bus allows a multi-drop topology, where many devices are
connected in parallel to the same physical bus, which uses a 5V TTL differential
signal. The used baud rate is 115.2 kBaud/s, which is low enough so that bus
lengths below 100 meters do not require termination. Measurement nodes were
designed to use a 8-bit microcontroller which implements the MSCB protocol
and connects to the serial bus via a RS-485 transceiver. The low-level bus
protocol follows exactly the RS-232 standard (one start bit, one stop bit, no
parity), so that the internal microcontroller UART (Universal Asynchronous
Receiver/Transmitter) is used. On the other side the microcontroller
communicates to various ADCs and DACs through different protocols such as SPI
(Serial Peripheral Interface bus) or I2C (Inter-Integrated Circuit), as shown in
Fig.~\ref{Slowcontrol:mscb}.

The used RS-485 bus transceivers (MAX1483) have a reduced bus load, so that up
to 256 nodes can be connected to a single physical bus. To connect to the data
acquisition PC, a so-called ``submaster" was built to act as a master on
that bus segment. On one side it connects to the MSCB bus, on the other side it
encapsulates all MSCB telegrams into network packets and communicates with any
PC through standard 10Base-T Ethernet. The Ethernet submaster concept has proved
to be extremely versatile. MSCB nodes belonging together logically are connected
to one MSCB bus segment and interfaced via a dedicated Ethernet submaster. The
MEG experiment uses a total of 15 submasters, which are connected to the various
subsystems and to the central experiment network switch. If one segment is down
for maintenance, the other segments can still be operational, which was very
useful during commissioning phases. The submasters can also be accessed remotely
for debugging, development and safety, which can happen in parallel to the
normal DAQ operation. In case the central DAQ system is down, a remote backup PC
can temporarily take over the slow control monitoring, without the need to
change any physical cable or reboot any system.

\subsubsection{Communication protocol}

The protocol for the MSCB system was designed for minimal overhead. A
sequence of just a few bytes is enough to read or write a slow control value.
Therefore it is possible to monitor many hundred nodes per second in spite of
the relatively slow baud rate of 115.2 kBaud on the RS-485 bus. Each packet
starts with a one-byte command, followed by optional arguments and a 8-bit CRC
code. Usual commands are reading and writing of values, obtaining information
about variables, rebooting a node or updating the node firmware through the
RS-485 bus. 

Since many nodes can be connected to the same bus, an addressing scheme has to
be used to communicate with individual nodes. Each node contains a unique
16-bit address stored in the EEPROM of the node microcontroller. To address a
certain node, an address packet containing the node address has to be sent over
the bus. If a node recognises its own address, it switches to the ``addressed
mode" and interprets all consecutive commands until another node becomes
addressed. 

\begin{figure}[htb]
 \centerline{\hbox{
  \includegraphics[width=.45\textwidth,angle=0] {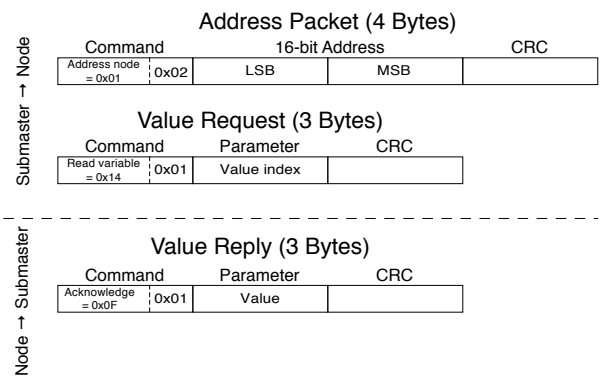}
 }}
 \caption[]{Typical communication protocol example between a submaster and a node.}
 \label{Slowcontrol:protocol}
\end{figure}

Figure~\ref{Slowcontrol:protocol} shows an example of the communication between a
submaster and a node. A first packet addresses the node, while a second packet
requests a certain slow control value from that node. The reply to this command
contains a command code indicating an acknowledge, followed by the actual value
and the CRC code. 

The MSCB protocol is encapsulated by the Ethernet submaster into UDP packets. A
software library on the PC side communicates with the submaster using these UDP
packets. The library is written in C and available for common operating systems
such as Linux, Windows and OSX. In addition, virtual instruments (VI) have been
written to export the functionality to the LabVIEW system from National
Instruments. All MSCB commands require an acknowledge. The commands writing a
value get a one-byte acknowledge from the node after applying the new value,
while the read command returns the actual value read. This scheme protects not
only lost or corrupted packets on the RS-485 bus, but also lost UDP packets. If
an acknowledge is not returned in a certain time window, the packet is re-sent
several times with an increasing timeout. This protocol has been optimised for
local area network connections, where timeouts of the order of a few hundred
milliseconds are needed to achieve maximal throughput. Tests have shown that
this is more effective than a TCP connection. A simple password authentication
scheme ensures that none of the several experiments at PSI using the MSCB system
talk to each other accidentally. The connectionless protocol ensures
that the system will recover automatically after a network failure or a reboot
of a node or the submaster.

\begin{figure}[htb]
 \centerline{\hbox{
  \includegraphics[width=.45\textwidth,angle=0] {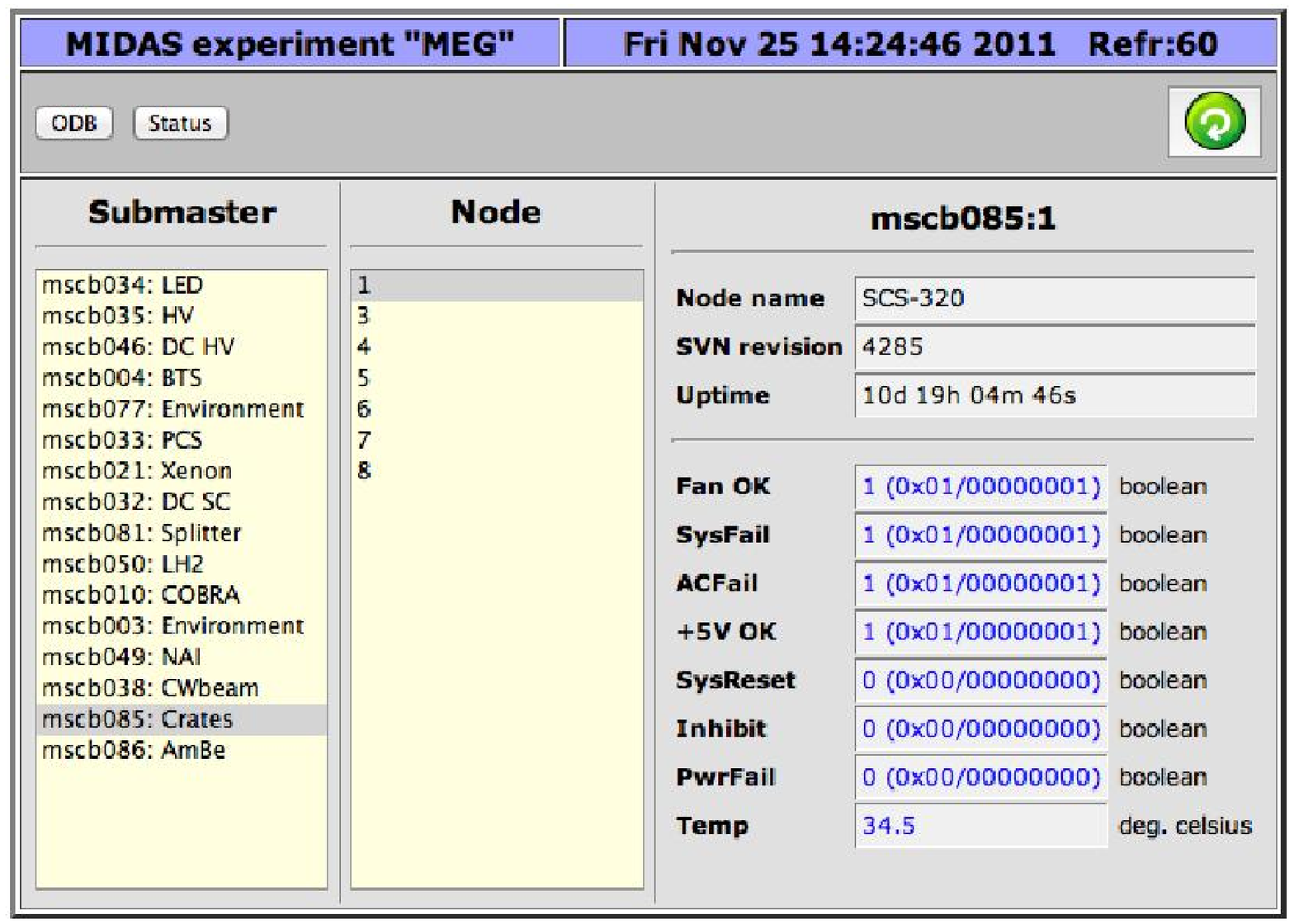}
 }}
 \caption[]{Web interface page to the MSCB system. The left column shows all available submasters, the second column all nodes connected to the selected submaster, and the right column all network variables of the selected node.}
 \label{Slowcontrol:mhttpd_mscb}
\end{figure}

The MSCB system acquires data and controls values by exchanging certain values
called network variables. Each MSCB node can implement up to 256 network
variables. Each variable has a name, a type (such as byte or 4-byte floating
point) and a physical unit. This variable definition resides on each node and
can be queried using the MSCB protocol. 
Since all necessary information can be obtained directly from the MSCB nodes, 
it is possible to use general-purpose programs to connect to the MSCB system 
and display all network variables of a certain submaster without using a central configuration database.
Adding new nodes or removing nodes is trivial. The MIDAS web interface
(mhttpd) described in Sect.~\ref{DAQ:MIDAS} contains a dedicated MSCB page which
can be used as an user interface to all nodes of the MEG experiment, as in 
Fig.~\ref{Slowcontrol:mhttpd_mscb}.

\subsubsection{Local intelligence}

Each MSCB node and submaster requires a microcontroller to implement the MSCB
protocol and to realise local intelligence needed e.g. in control loops. The
requirements for those microcontrollers are low power consumption, small size,
integrated analog electronics such as ADCs and DACs and user reprogrammability.
This lead to the C8081F121 microcontroller 
\cite{silabs_url} used in most MSCB nodes, which contains 128kB of EEPROM, 8kB
of RAM, a 12-bit ADC and two 12-bit DACs and runs at a clock speed of 100 MHz.

The microcontroller is programmed in C using the Keil PK51 development kit
\cite{keil_pk51}. The software is divided into a framework implementing the MSCB
protocol and general housekeeping functions and the application-specific part.
Both parts communicate via network variables described above. The available
program space of 128kB is more than enough to accommodate the C runtime library
including floating point functions, the MSCB framework consisting of about 3500
lines of C code, and still leaves space for several thousand lines for
application-specific functionality such as regulation loops. The DCH 
pressure control system for example uses elaborate code including sensor
linearisation, alarm functions and Proportional-Integral-Differential (PID)
loops to keep the gas pressure stable to better than 10$^{-5}$ bar, all
implemented in about two thousand lines of C code.

One special feature of the MSCB system is that it can be reprogrammed over the
network. The microcontrollers can write to their own code space from the user
application. A small boot loader has been realised and is located in the upper memory
of the code space, implementing a subset of the MSCB protocol. Using this boot
loader, the main program space can be re-programmed. Firmware updates of all
MSCB devices are therefore very easy. Since the protocol is encapsulated fully
in the Ethernet communication layer, the update can even be sent through the
Ethernet submaster from remote locations. A typical update contains 50 kB of
code and requires ten seconds. After the update, the node reboots in about 0.5
seconds, reducing downtime to a minimum. 

\subsubsection{High-voltage system}
\label{High_voltage_system}

\begin{figure}[htb]
 \centerline{\hbox{
  \includegraphics[width=.45\textwidth,angle=0] {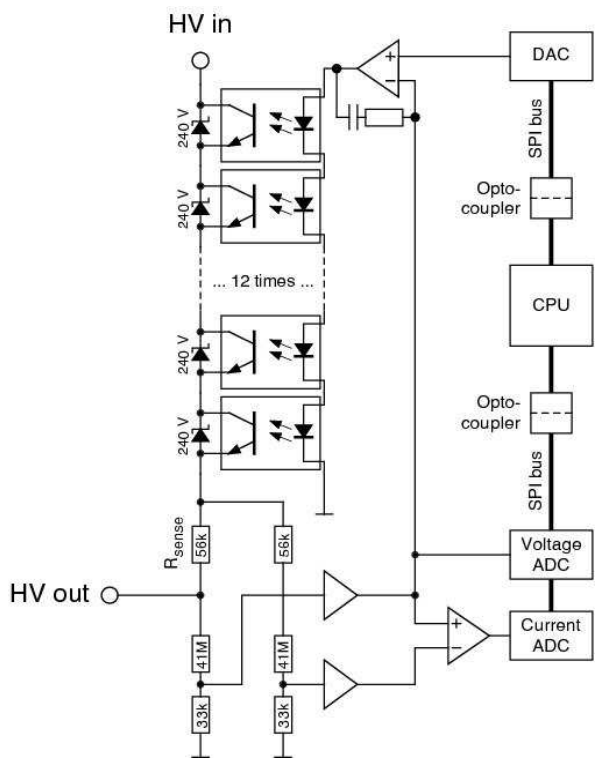}
 }}
 \caption[]{Simplified schematics of the high-voltage regulator. The 240 V Zener diodes protect the optocouplers, so that no more than 240 V drop across the transistors of the optocouplers.}
 \label{Slowcontrol:hv}
\end{figure}

The MEG experiment requires high-voltage supplies for the PMTs of the LXe 
and the TC, the APDs of the TC and the
DCH. The MEG experiment requires an accuracy and long-term stability
better than 0.1 V, which was hard to achieve with commercial systems at the time
of the design of the experiment. Therefore a high-voltage supply based on the
MSCB system was designed. Since the experiment has many channels which need
about the same high voltage, it was easier to make a voltage regulator which is
fed from a common external source. All 848 channels for the LXe detector 
are for example powered from a single high-current commercial high-voltage
source, and the voltage is adjusted for each channel separately by a MSCB
voltage regulator. The regulator uses a string of optocouplers, each capable of
sustaining 240 V. A total of twelve optocouplers allows voltage regulation from
zero to 2800 V. The LEDs of the optocouplers are controlled by a regulation
loop which compares the output voltage with the demand value obtained from a DAC,
as shown in Fig.~\ref{Slowcontrol:hv}.

The output voltage is measured by means of an ADC and a voltage divider consisting of
two resistors with 41 MOhm and 33 kOhm. The current is determined by measuring the
voltage drop across a resistor $R_{sense}$ with a differential
amplifier. Using 24-bit sigma-delta ADCs (AD7718 from Analog Devices), very
good resolutions can be achieved. The ADCs and the DAC are controlled by the
microcontroller through the SPI bus via optocouplers for noise immunity. The
accuracy depends only on the resistors of the voltage divider and the
current sense resistor, which can be calibrated using an external
multimeter. The calibration is then stored in the nonvolatile memory of the
microcontroller. Accuracies of 30 mV in voltage and a few nA in current have
been achieved at 2000 V output voltage. 

\subsubsection{RS-232 and GPIB interface nodes}

Although the MSCB system handles many sensors and actuators directly, some
devices, e.g. COBRA, have their own control systems. These devices
typically have a RS-232 or GPIB interface. To integrate these devices into the
MEG slow control system, MSCB interface nodes were designed for the RS-232
and the GPIP busses, containing only the microcontroller with the RS485
interface chip (ADM2486 from Analog Devices), and a second interface chip for
the RS-232 bus (MAX232 from Maxim) or the GPIB bus (directly connected to the
CPU). A 5V DC-DC converter galvanically decouples the bus power from the local
power. This allows the nodes to be powered from the bus without the danger of
introducing ground loops between different devices through their earthing system.

The interface nodes have two modes of operation. The so-called transparent mode
encapsulates the device communication into MSCB data packages. The PC side
contains routines which allow direct writing or reading from the RS-232 or GPIB
bus. The advantage of this mode is that new devices can be connected and tested
without modification of the MSCB firmware since this system works only as the
transport layer. The disadvantage is that each device needs a dedicated driver
on the PC side which understand the device-specific ASCII protocol.

The second mode encapsulates the device specific protocol in the firmware of the
interface node. If connected to a power supply for example, the voltage and
current of the power supply are polled periodically and put into MSCB network
variables. These variables can then be read out without the knowledge of the
device-specific protocol to query the voltage and current of the power supply.
Parameters in attached devices can be changed in a similar way. Writing to the
associated network variable triggers a callback routine in the microcontroller
firmware which generates a write command to the device through the RS-232 or GPIB
bus. Using this mode, all devices attached to the MSCB system can be talked to
in the same way (namely network variables) independent of the specific device.
This mode has therefore been chosen for all devices in the MEG experiment.

\subsubsection{SCS-2000 control box}

The MSCB-based SCS-2000 control box was designed for analog and digital inputs and outputs.
The box contains an Eurocard-size main board with a microcontroller, 
a battery-buffered real-time clock and a 1 MB static RAM. User
interaction happens through a LCD display and four push buttons. The SCS-2000
unit can either be inserted in a 3 HE crate or be housed in a stand-alone box
with an external 24V power supply. A single box can accommodate eight daughter
cards, each with eight input/output channels of various kinds. Analog input
cards typically host a 24-bit sigma-delta ADC with eight input channels. Other
cards were developed for analog and digital input and output, including
a capacitance meter for the liquid xenon level gauge or
a 10V LED pulser for the LXe detector calibration.

A generic firmware scans all plugged-in daughter cards and displays their
parameters on the LCD display. Using the four buttons on the front panel, the
user can scroll through all values and change values of those cards which
support digital or analog output. This is very useful for controlling devices
locally with an attached SCS-2000 without the need of using a laptop or other
computers. The parameters are visible as network variables and are accessible
through the MSCB bus. 

\subsection{DAQ Integration}

Most slow control parameters in the MEG experiment are acquired locally by
SCS-2000 units or other MSCB nodes, but they need to be integrated into the full
DAQ system. Measured values have to be written to disk together with the event
data. Furthermore, the temporal changes of parameters have to be accessible by
the shift operators in order to ensure the stability of the experiment. This is
achieved by a dedicated slow control front-end program, together with the MIDAS 
online database and history system, which are described in Sect.~\ref{DAQ:MIDAS}.

\subsubsection{MIDAS slow control front-end}

A dedicated process runs on the back-end computer which connects to all slow
control submasters through the Ethernet network. All controlled devices are
partitioned into separate entities called equipments, such as high-voltage
supplies, beam line magnets or ambient temperature sensors. Each equipment is
read out periodically with a dedicated thread inside the front-end process for
optimal read-out frequency. The readout threads copy all parameters into the
ODB, where they become immediately available to other processes or to the shift
operator via the mhttpd web interface. Control parameters are placed into the
ODB as well, and linked to their devices via hot-links (see 
Sect.~\ref{DAQ:MIDAS}). If the user or a process changes a control parameter in the
ODB, a callback routine in the slow control front-end is called, which in turn
writes the new value to the associated slow control device. By dumping the ODB
into a snapshot file, the current status of the whole experiment can be captured
at a specific time, and be reconstructed later from that dump.

In addition, the slow control front-end generates events from the slow control
parameters and injects them periodically via the buffer manager into the event
stream written to disk for later analysis. 

\subsubsection{MIDAS Custom pages}
\label{Slowcontrol:customsec}

\begin{figure}[htb]
 \centerline{\hbox{
  \includegraphics[width=.45\textwidth,angle=0] {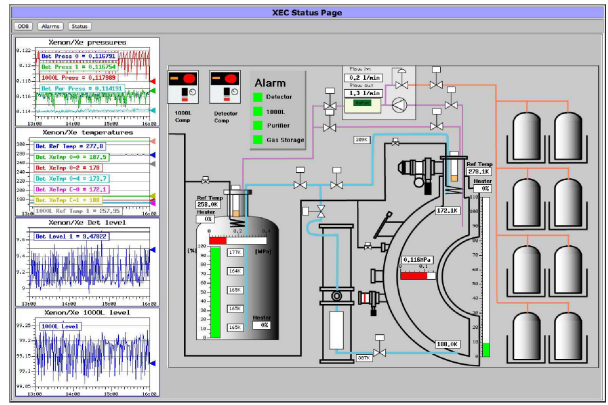}
 }}
 \caption[]{Custom web page of the LXe detector cryogenic system. The plots on the left are taken from the history system. }
 \label{Slowcontrol:custom}
\end{figure}

The whole mhttpd web interface is non-specific, meaning that the same pages are
used for different experiments. This is useful for monitoring the general status,
starting and stopping runs and changing ODB variables. There is however the need for
experiment-specific pages such as the one shown in 
Fig.~\ref{Slowcontrol:custom}. Such pages condense information from one or several
subsystems into a single display and provide the shift operator with a quick overview of
a part of the experiment. To realise these so-called ``MIDAS custom pages", three
techniques are implemented into the mhttpd web server:

\begin{itemize}

\item The HTML language is extended by an $<$ODB [path] [flags]$>$ tag. When the
mhttpd process parses the user-written HTML file for the custom page, the ODB
tag is replaced by the actual value of the parameter stored in the ODB indicated
by its path. Optional flags allow various number formats and the possibility to
change the ODB value though the web interface.

\item Dynamic GIF images can be created by mhttpd. They consist typically of a
static background image combined with overlays of labels, bar graphs or coloured
areas based on simple script statements. The image is created in memory using
the GD library \cite{GD_url} and directly streamed to the browser for maximum
efficiency. The right side of Fig.~\ref{Slowcontrol:custom} is based on this
technique.

\item An Asynchronous JavaScript and XML (AJAX) library is available which
contains functions for asynchronous communication with mhttpd. These functions
can be used to change parts of a web page using the Document Object Model (DOM)
in the background without the need to reload the complete page. Functions are
implemented for reading and writing ODB values, to start/stop a run and to
generate alarm and system messages. 

\end{itemize}

The combination of the three methods has proved to be extremely powerful in
designing web pages which almost look and behave like a dedicated program, but
have the advantage that only a web browser is required to use them.

The usage of custom pages became a very efficient tool to control and monitor
the complete experiment with only a web browser. This was a prerequisite for
running remote shifts, which were introduced in 2011 in the MEG experiment.

%
%

\section{Simulation}
\label{sec:simul}
This section provides an overview of the packages used in the Monte Carlo (MC) 
simulation of the MEG experiment.

The MC simulation is based on the physics event simulation ({\bf megeve}), the 
detector simulation ({\bf gem}), the read-out simulation and the event mixing 
({\bf megbartender}).
These packages make use directly or indirectly (via {\bf megdb2cards}) of the 
MySQL \cite{mysql} based Data Base (DB) (more details are in \cite{software2010,softwaretns}).

\subsection{The simulation architecture}

The simulation architecture was designed with a number of constraints from the time schedule, 
manpower and technical skills available at the start of the project. 
Therefore the physics and detector simulation packages were written in FORTRAN 77, while
{\bf megbartender} was written in C++.

The FORTRAN 77 packages were organised following a modern programming paradigm, 
i.e. to use an Object-Oriented (OO) approach organised in a custom-designed framework called {\bf rem}.
The basic unit manipulated by the framework is a Module, corresponding to an OO class.

Modules can be either sub-detector simulation 
sections or service tools such as graphics or input/output. 
The {\bf megeve} and {\bf gem} packages are steered through cards read by the FFREAD package. 

\subsection{Physics event simulation}

In a Module, {\bf megeve}, several event types of relevance for the 
experiment can be generated. A non-exhaustive list including exclusive and inclusive
decays is (particles in brackets are not generated)

\begin{basedescript} {\desclabelstyle{\pushlabel}\desclabelwidth{6em}
}
\item[Signal]\hfill  \meg 
\item[Signal $e^+$]\hfill   $\mu^+\to {\rm e}^+(\gamma)$ 
\item[Signal $\gamma$] \hfill  $\mu^+\to ({\rm e}^+)\gamma$ 
\item[Decay $e^+2\gamma$]  \hfill $\mu^+\to {\rm e}^+\gamma\gamma$ 
\item[Michel $e^+$] \hfill  $\mu^+\to e^+(\nu\overline \nu)$ \cite{kuno_2001}
\item[Radiative $e^+\gamma$] \hfill $\mu^+\to {\rm e}^+\gamma(\nu\overline \nu) $\cite{kuno_1996_prd,kuno_2001}
\item[Radiative $\gamma$] \hfill $\mu^+\to ({\rm e}^+)\gamma(\nu\overline \nu)$ \cite{kuno_1997_prd,kuno_2001}
\item[AIF $\gamma$]  \hfill $\mu^+\to e^+(\nu\overline \nu), {\rm e^+e^-}\to \gamma (\gamma)$ 
\item[Cosmic $\mu^+$]  \hfill $\mu^+$ from cosmic rays 
\item[Calibration $2\gamma$] \hfill  $\pi^-p \to \pi^0 (n)$, $\pi^0\to \gamma\gamma$
\item[Calibration Dalitz] \hfill  $\pi^-p \to \pi^0 (n)$, $\pi^0\to \gamma e^+e^-$
\item[Boron Calibration ] \hfill  $^{11} \mathrm{B}(p,2\gamma)^{12}\mathrm{C}$
\item[Lithium Calibration ] \hfill  $^{7}\mathrm{Li}(p,\gamma)^{8}\mathrm{Be}$
\item[Neutron Generator ]  \hfill $^{58}\mathrm{Ni}(n,\gamma)^{59}\mathrm{Ni}$
\item[Mott Scattering ]  \hfill $e^+ \mathrm{A \to} e^+  \mathrm{A}$
\end{basedescript}

When appropriate the generation includes first order radiative correction and $\mu^+$ polarisation effects
\cite{kuno_2001}.

The $\mu^+$ decay vertices may be forced to be in the target or each $\mu^+$ is propagated through the beam line
starting at the last quadrupole of Triplet II 
(see Fig.~\ref{fig:PiE5}) 5647~mm upstream of the centre of COBRA to the decay vertex.

\subsection{Detector simulation}

The beam propagation, interaction in the target and propagation of the decay
products in the detector are simulated with a FORTRAN 77 MC program ({\bf gem})
based on the GEANT3 package \cite{GEANT3}, which describes the 
sub-detector responses up to the read-out.

The program is heavily modularised using the {\bf rem} framework. That allows 
easy addition of new Modules for implementing the various parts of the simulation. 
Settings dependent on the run-period are copied 
from the MySQL database into the FFREAD cards, and used in {\bf gem}. The output is 
written in ZEBRA format \cite{zebra} and the code for dealing with the ZEBRA package is 
automatically generated starting from a bank description in DZDOC format.

\begin{figure}[htbp]
  \begin{center}
   \includegraphics[width=.95\linewidth] {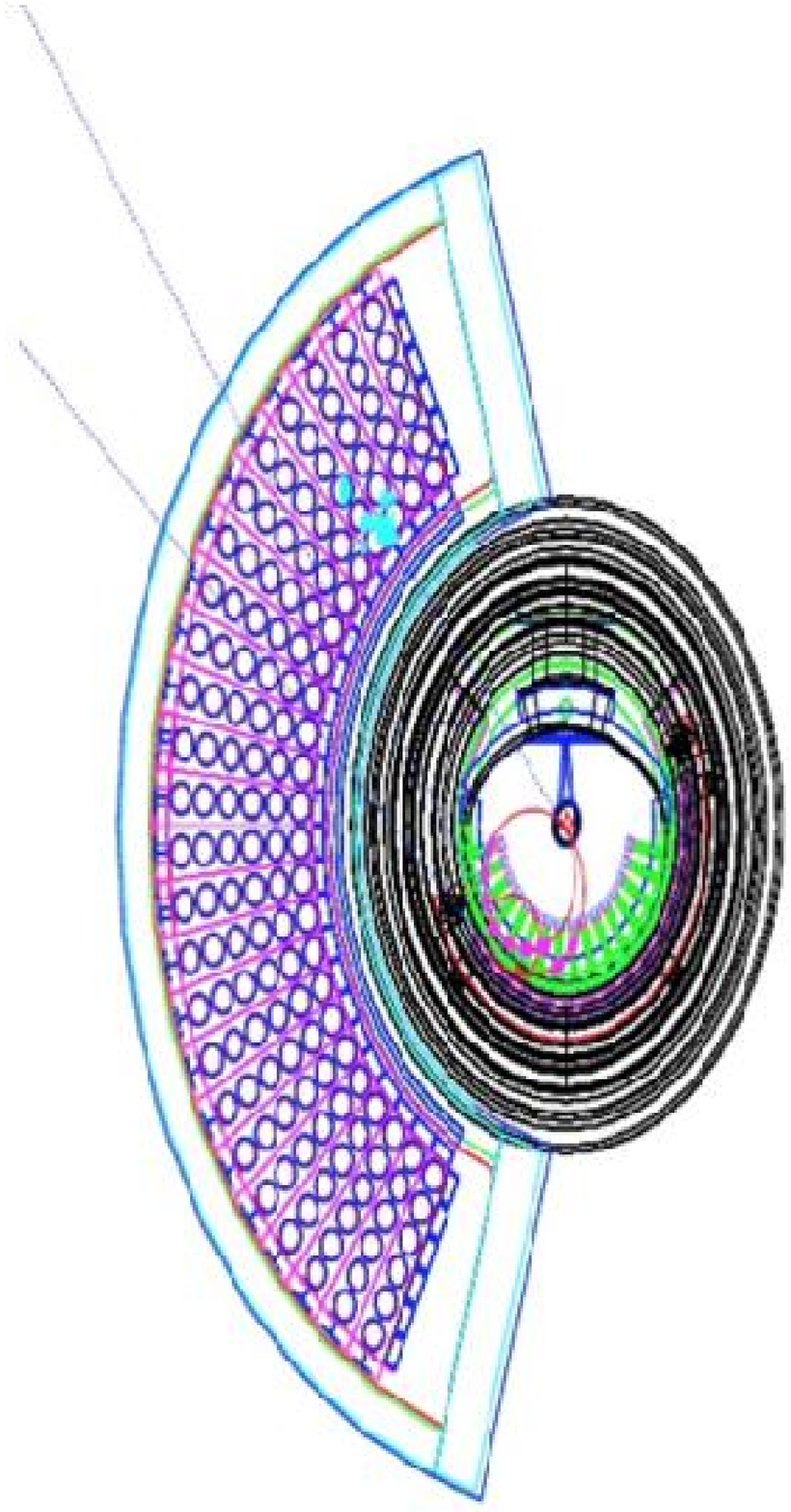}
    \caption{A simulated event \meg.}
    \label{gem_event}
  \end{center}
\end{figure}

Each sub-detectors is described by a Modules: DCH,  Longitudinal TC,
Transverse TC and LXe detector.

Additional Modules, e.g. COBRA, gas volume, $\mu$ target,
upstream and downstream elements outside COBRA, complete the 
description of the material distribution.

The GEANT3 routines simulate the energy release in the detectors, while the
conversion into electrons/photons and their propagation up to the read-out devices
are managed in {\bf gem} user routines.

The LXe detector Module includes a very detailed simulation of the 
propagation of each scintillation photon in the detector volume, accounting for 
reflection, refraction and polarisation effects. The output is the photoelectron 
current on the PMT photocathodes.
In the TC Modules (longitudinal and transverse sections) the photoelectron
currents on PMT (APD) photocathodes are calculated with an analytical expression for propagation
of the scintillation photons.
In the DCH Module the GEANT3 energy hits are split into clusters along the track. 
Then the electrons are drifted toward the sense wires following the position-time 
relation obtained with GARFIELD \cite{garfield}, including diffusion effects. The output is wire currents.
Figure.~\ref{gem_event} displays a simulated signal event.

The {\bf gem} package stops short of simulating the read-out, which is performed by Bartender.

\subsection{Bartender }

Between the detector simulation and the reconstruction program an
intermediate program, called Bartender, is required for the processing of MC data.

The Bartender serves different roles:

\begin{itemize}
\item Conversion of ZEBRA files into ROOT \cite{root} files
\item Readout simulation
\item Event mixing
\end{itemize}

\subsubsection{Format conversion }

Bartender reads several ZEBRA files written by {\bf gem} and converts them into ROOT
format. Simulation-specific data such as kinematics of generated particles,
true hit information etc. can be streamed to a ``sim'' Tree in separate ROOT
files for further studies.

\subsubsection{Readout simulation }

Bartender simulates the detector read-out electronics and makes waveforms. For
example, for the LXe detector, waveforms are made by convolution of the single
photoelectron response of a PMT with hit-time information of scintillation
photons simulated by {\bf gem}. PMT amplification, signal attenuation, saturation of the
read-out electronics, noise etc. are taken into account. Simulated waveforms are
encoded in the same manner as in the experiment and written to a ``raw'' Tree in ROOT
files.

\subsubsection{Event Mixing}

Bartender makes a mixture of several sub-events; rates of each event type
are set with a configuration file.
To study the combinatorial background events,
sub-events are mixed with adjustable relative timing with respect to each other
and with respect to the trigger; random and fixed timing can be selected.
This allows simulation of many different pile-up configurations with a limited number
of event samples simulated in the detector.

\subsection{Data Base}

Run-dependent parameters such as geometry, calibration and analysis
constants are stored in a relational database based on MySQL (DB). The DB is used by the DAQ
front-end, analysis and simulation. For simulation, the
dedicated program {\bf megdb2cards} reads the DB and writes the
FFREAD cards required by {\bf gem}. Therefore the DB is used consistently by all packages.

\begin{figure}[htbp]
  \begin{center}
    \includegraphics[width=.95\linewidth] {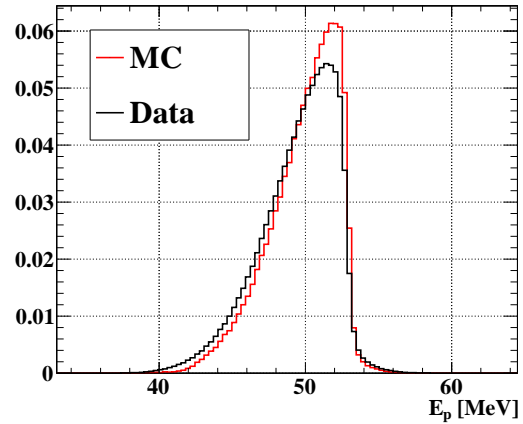}
    \caption{Normalised momentum spectrum measured in the spectrometer for data and MC simulation.}
    \label{mom_dch}
  \end{center}
\end{figure}

\subsection{Comparison with data}

\begin{figure}[htbp]
  \begin{center}
    \mbox{\begin{tabular}[t]{ll}
    \subfigure[]{\includegraphics[width=.95\linewidth] {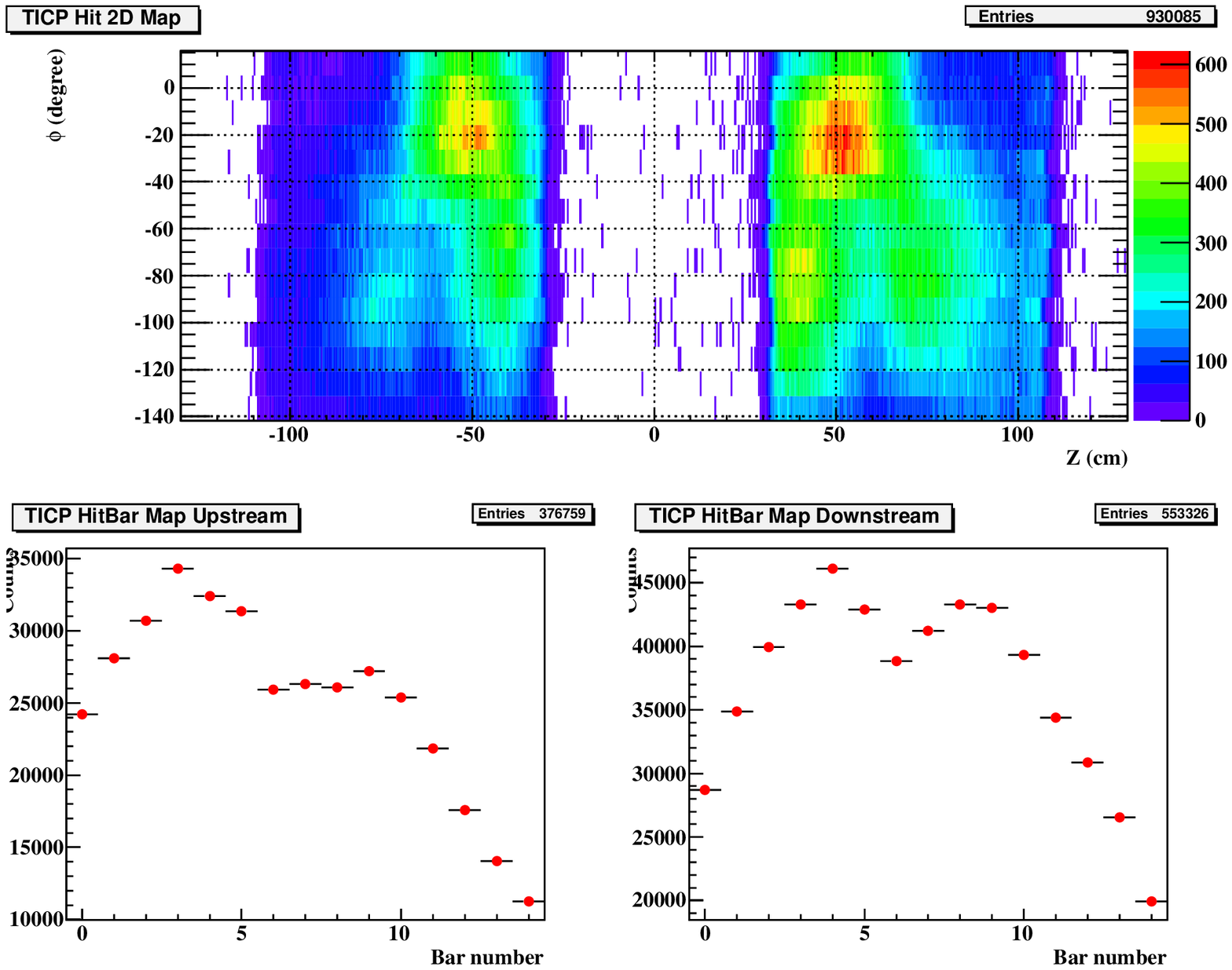}
    \label{hit_tpdata}
    }  \\
    \subfigure[]{\includegraphics[width=.95\linewidth] {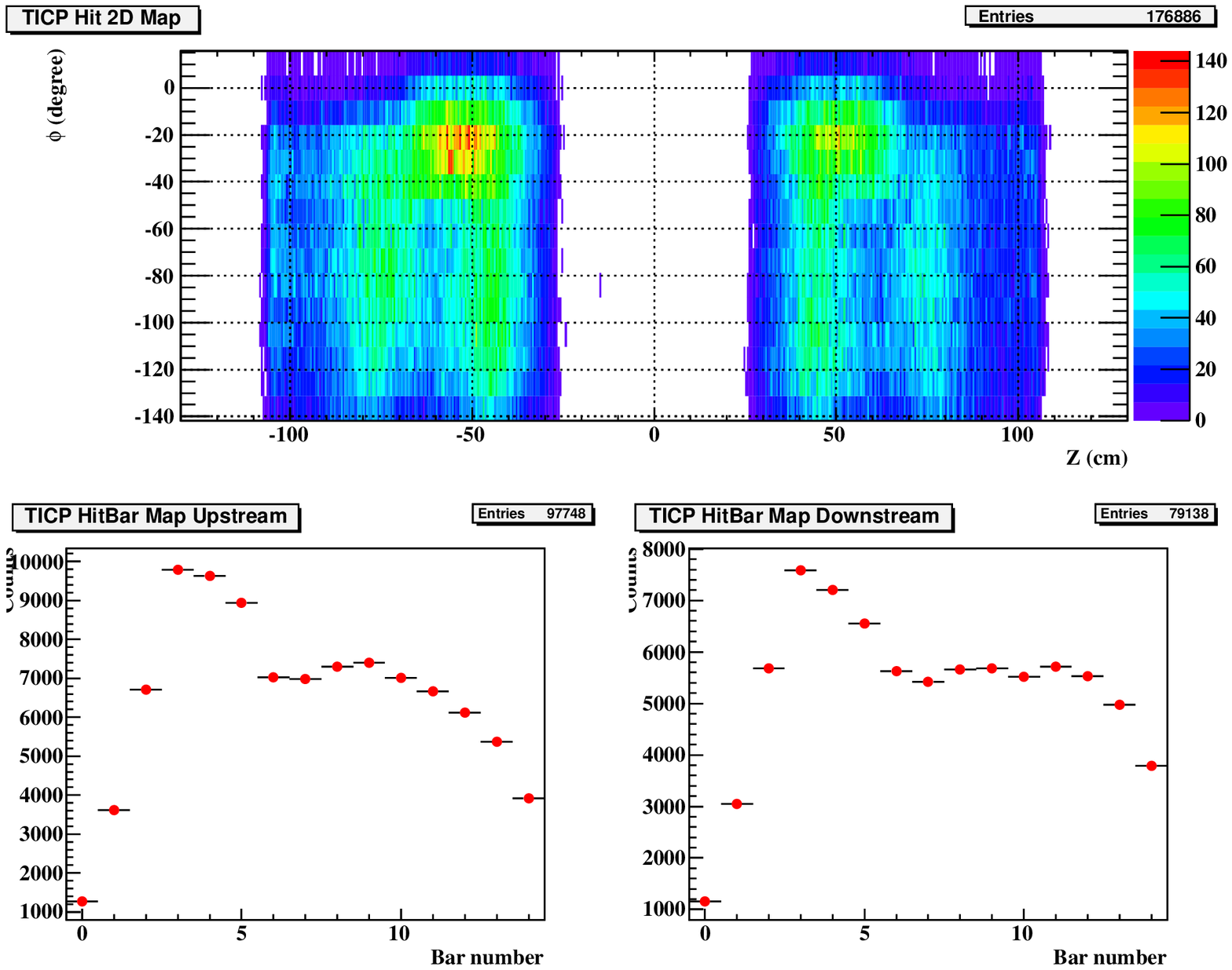}
    \label{hit_tpmc}
    } \\
    \end{tabular}}
    \caption{Hit distribution in longitudinal TC for data (above) and MC simulation (below).}
  \end{center}
\end{figure}

\begin{figure}[htbp]
  \begin{center}
    \includegraphics[width=.95\linewidth] {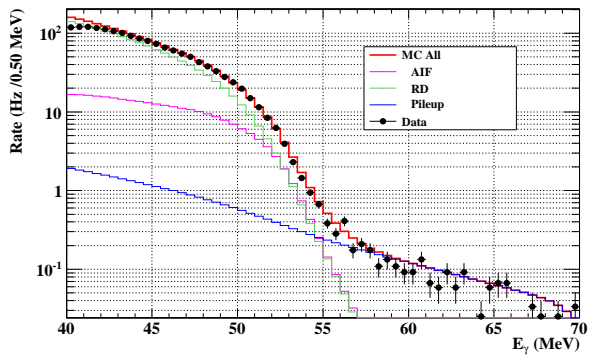}
    \caption{Inclusive $\gamma$-ray spectrum in LXe. Data (dots) and MC (lines).
     MC spectrum is separated into contributions from AIF,
    RMD and pile-up from multiple $\gamma$-rays.}
    \label{gamma_xec}
  \end{center}
\end{figure}

Figure~\ref{mom_dch} shows the normalised momentum spectrum of the $e^+$ measured in the spectrometer
for data and MC events.
Figure~\ref{hit_tpdata}-\ref{hit_tpmc} shows the hit distribution on the longitudinal Timing
Counter for data and MC events.
Figure~\ref{gamma_xec} shows the inclusive $\gamma$-ray spectrum measured in the LXe detector
for data and MC events. The MC events are separated into different sources.

%
%
\section{Summary}

The MEG detector described in this paper has been operational since 2008.
In the period 2009--2011, it has collected $\sim 3.63 \times 10^{14}$ muon decays.\\
In the Table~\ref{table:performance} the design, Monte Carlo and obtained resolutions are summarised.
The obtained resolutions are, for some observables, worse than the design values. The differences
are, at large extent, understood and the Monte Carlo resolutions match quite well the obtained
ones.

\begin{table}[hbct]
\caption{Resolutions expressed in Gaussian standard deviations (core or single side, 
when required) and efficiencies for all observables: comparison between expected and obtained.}
\begin{center}
\begin{tabular}{lrrr}
\\{\bf Variable} & Design & Monte Carlo & Obtained  \\[1mm]
\hline\\[-3mm]
Resolutions & & & \\
\hline\\[-3mm]
Positron(${\rm e}$) & & & \\
\hline\\[-3mm]
$\sigma_{E_{\rm e}}$ (keV)&  200  & 315 &306   \\
$\sigma_{\phi_{\rm e},\theta_{\rm e}}$ (mrad)& 
     5($\phi_{\rm e}$),5($\theta_{\rm e}$)  &
     8($\phi_{\rm e}$),9($\theta_{\rm e}$)  &
     9($\phi_{\rm e}$),9($\theta_{\rm e}$)  \\
$\sigma_{z_e,y_e}$ (mm) & 1.0 &  2.9($z_e$)/1.0($y_e$) & 2.4($z_e$)/1.2($y_e$)\\
$\sigma_{t_{\rm{e}}}$ (ps) & 50 & 65 & 102 \\
\hline\\[-3mm]
Photon ($\gamma$) & & & \\
\hline\\[-3mm]
$\sigma_{E_\gamma}$ (\%) & 1.2 & 1.2 &  1.7  \\
$\sigma_{t_\gamma}$ (ps)& 43  &  69 & 67 \\
$\sigma_{(u_\gamma,v_\gamma)}$ (mm) & 4 & 5 & 5\\
$\sigma_{w_\gamma}$ (mm) & 5 & 6 & 5\\
\hline\\[-3mm]
Combined(e-$\gamma$) & & & \\
\hline\\[-3mm]
$\sigma_{t_{{\rm e}\gamma}}$ (ps)& 66 & 95 & 122 \\
$\sigma_{\Theta_{{\rm e}\gamma}}$ (mrad)  & 11  & 16 &  17  \\
\hline\\[-3mm]
Efficiencies & &  & \\
\hline\\[-3mm]
$\epsilon_{\rm{e}}$  (\%) & 90 & 40 & 40\\
$\epsilon_\gamma$ (\%) & 60 & 63 & 63 \\
$\epsilon_{trg}$  (\%) & 100 & 99 & 99 \\
\hline \\

\label{table:performance}
\end{tabular}
\end{center}
\end{table}

The influence of the differences in resolutions on the analysis sensitivity 
is approximately evaluated firstly calculating the number
of expected signal events for a given branching ratio (B)
\begin{equation}
N_{sig} \propto \epsilon_\gamma \times \epsilon_{\rm{e}},
\end{equation}
from Table~\ref{table:performance}, the decrease in $N_{sig}$ is $\sim 2$ due 
to $\epsilon_{\rm{e}}$. Secondly, the number of accidental coincidences is 
\begin{equation}
N_{acc} \propto \sigma^2_{E_\gamma} \times \sigma_{E_{\rm{e}}} \times 
\sigma_{t_{{\rm e}\gamma}} \times
\sigma^2_{\Theta_{{\rm e}\gamma}}.
\end{equation}

This number increases a factor of $\sim 10$ compared to the design value.
Those effects combined implies that the sensitivity of the present MEG experiment
is $\sim 5$ compared to the design value.

The results from the analysis of data collected up to 2010 are already published 
and set the most stringent limit to date on the existence of 
the \meg\ decay. Data collected in 2011 with statistics comparable to
the 2009--2010 sample are being analysed.

The experiment run in 2012 collecting a data 
sample comparable in size to the 2011 sample.

\section*{Acknowledgements}

We are grateful for the support and co-operation provided 
by PSI as the host laboratory and to the technical and 
engineering staff of our institutes. This work is
supported by DOE DEFG02-91ER40679 (USA), INFN
(Italy) and MEXT KAKENHI 16081205 (Japan).

We thank Prof. David Stoker of the University of California, Irvine, for
his careful proofreading of the manuscript and its improvements.

\bibliographystyle{unsrt}
\bibliography{MEGDetectorPaper}
\end{document}